 \documentclass[a4paper,12pt]{article}
\usepackage{jheppub,esint,shuffle,psfrag}
 \usepackage[utf8]{inputenc}
 \usepackage{booktabs}

\usepackage{esint} 
\usepackage{breqn}
\usepackage{amsmath,bm}
\usepackage{hyperref}
\usepackage{mathrsfs} 

\definecolor{myred}{RGB}{233, 33, 45}

\makeatletter
\DeclareFontFamily{OMX}{MnSymbolE}{}
\DeclareSymbolFont{MnLargeSymbols}{OMX}{MnSymbolE}{m}{n}
\SetSymbolFont{MnLargeSymbols}{bold}{OMX}{MnSymbolE}{b}{n}
\DeclareFontShape{OMX}{MnSymbolE}{m}{n}{
    <-6>  MnSymbolE5
   <6-7>  MnSymbolE6
   <7-8>  MnSymbolE7
   <8-9>  MnSymbolE8
   <9-10> MnSymbolE9
  <10-12> MnSymbolE10
  <12->   MnSymbolE12
}{}
\DeclareFontShape{OMX}{MnSymbolE}{b}{n}{
    <-6>  MnSymbolE-Bold5
   <6-7>  MnSymbolE-Bold6
   <7-8>  MnSymbolE-Bold7
   <8-9>  MnSymbolE-Bold8
   <9-10> MnSymbolE-Bold9
  <10-12> MnSymbolE-Bold10
  <12->   MnSymbolE-Bold12
}{}

\let\llangle\@undefined
\let\rrangle\@undefined
\DeclareMathDelimiter{\llangle}{\mathopen}%
                     {MnLargeSymbols}{'164}{MnLargeSymbols}{'164}
\DeclareMathDelimiter{\rrangle}{\mathclose}%
                     {MnLargeSymbols}{'171}{MnLargeSymbols}{'171}
\makeatother

\newcommand{\bs}{\begin{shaded}}
\newcommand{\es}{\end{shaded}\noindent}
\def\ba#1\ea{\begin{align}#1\end{align}}		
\newcommand{\be}{\begin{equation}}
\newcommand{\ee}{\end{equation}}
\newcommand{\mc}{\mathcal }

\newcommand{\la}{\label}
\newcommand{\lp}{\notag \\ & }

\DeclareMathOperator{\tr}{\text{tr}}

\newcommand{\cf}{\textit{cf.} }
\newcommand{\ie}{\textit{i.e.} }
\newcommand{\eg}{\textit{e.g.}}
\newcommand{\N}{\mathcal N}

\newcommand\re[1]{(\ref{#1})}

\newcommand\lr[1]{{\left({#1}\right)}}

\newcommand \VEV [1] {\Big\langle{#1}\Big\rangle}
\newcommand \ket [1] {|{#1}\rangle}
\newcommand \bra [1] {\langle {#1}|}
\def \qqquad {\qquad\quad}
\def \qqqquad {\qquad\qquad}

\def\numberbysection{\@addtoreset{equation}{section}
                     \def\theequation{\thesection.\arabic{equation}}}

\newcommand{\gs}{g_{\small {\rm s}}}
\newcommand{\gym}{g_{_{\small {\rm YM}}}}
\newcommand{\foot}{\footnote}
\newcommand{\ci}{\cite}
\def\ov{\over}
\newcommand{\rf}[1]{(\ref{#1})}
\def\no{\nonumber}
\def\z{\zeta}
\def\OO{\mc O}
\def \adss {$\text{AdS}_{5}\times S^{5}$\ }
\def \vol {{\rm vol}}
\def\aa{{\alpha'}}

\def \no {\notag}
\def \iffa {\iffalse} 
\def \aa  {\alpha}
\def \Z {{\cal Z}}
\def \gs  {g_s} \def \adsss {$AdS_5 \times S^5$}
\def \adss {$AdS_5 \times S^5$ }

\newcommand{\sql}{\sqrt\l}
\renewcommand{\l}{\lambda}

\newcommand{\vev}[1]{\big\langle #1 \big\rangle}
\newcommand{\sF}{\mathsf{F}}
\newcommand{\sW}{\mathsf{W}}

\newcommand{\SA}{\mathsf{SA}}

\newcommand{\QQ}{\mathsf{Q}_{2}}
\newcommand{\Q}{\mathsf{Q}}

 \def \rI   {{\rm I}}
\def \rI   {{\sl I}}
\def \aa  {\alpha}
\def \Z {{\cal Z}}
\def \gs  {g_s}
\def \adsss {$AdS_5 \times S^5$} 
\def \adss {$AdS_5 \times S^5$ }
\def\hatlambda{(\tfrac{\l}{8\pi^{2}})}
\def\hatlambda{\bigg(\frac{\l}{8\pi^2}\bigg)}

\begin{document}

\vspace{-4cm }

\begin{flushleft}
 \hfill \parbox[c]{50mm}{
 IPhT--T23/015
 \\
 Imperial-TP-AT-2023-01}
\end{flushleft}
\author{M. Beccaria$^a$, G.P. Korchemsky$^{b,c}$ and A.A. Tseytlin$^{d,}$\footnote{Also on leave from  Inst. for Theoretical and Mathematical Physics (ITMP)  and Lebedev Inst.}}
\affiliation{
$\null$
$^a$Universit\`a del Salento, Dipartimento di Matematica e Fisica \textit{Ennio De Giorgi},\\ 
\phantom{a} and I.N.F.N. - sezione di Lecce, Via Arnesano, I-73100 Lecce, Italy
\\		
$\null$
$^b${Institut de Physique Th\'eorique\footnote{Unit\'e Mixte de Recherche 3681 du CNRS}, Universit\'e Paris Saclay, CNRS,\\ \phantom{a}  91191 Gif-sur-Yvette, France}  
\\
$\null$
$^c${Institut des Hautes \'Etudes Scientifiques, 91440 Bures-sur-Yvette, France}  
\\
$\null$
$^d${Blackett Laboratory, Imperial College London,  SW7 2AZ, U.K.}
}
\title{Non-planar  corrections in orbifold/orientifold  $\bm{\N=2}$  superconformal theories  from localization}
\abstract{
\small
We study non-planar corrections in two 
special $\N=2$ superconformal $SU(N)$ gauge theories that are planar-equivalent to $\N=4 $ SYM theory:  two-nodes quiver model with equal couplings 
 and  $\N=2$  vector multiplet coupled to two hypermultiplets in rank-2 symmetric and antisymmetric representations. We focus on two observables in these theories that admit representation  in terms of localization matrix model:  free energy on 4-sphere and the expectation value of half-BPS  circular Wilson loop. We extend the methods developed in arXiv:2207.11475 to derive a systematical expansion of non-planar corrections to these observables at strong  't Hooft coupling constant $\lambda$. We show that the leading non-planar corrections are given by a power series  in
 $\l^{3/2}/N^2$ with rational coefficients.  
Sending $N$ and the coupling constant $\lambda$ to infinity with $\l^{3/2}/N^2$ kept fixed
corresponds to the  familiar double scaling limit  in  matrix models. We find that 
in this limit  the observables in the two models are related  in a  remarkably simple way: the free energies  
differ by the factor of $2$, whereas the Wilson loop expectation values coincide. Surprisingly, these relations hold only  at strong coupling,
they
 are not valid in the weak coupling regime. We also  discuss a  dual   string theory interpretation of the leading corrections to the free energy in the double scaling limit suggesting their relation to  curvature corrections in type IIB string effective action.
}

\maketitle

\newpage

\section{Introduction and summary}

Localization \cite{Pestun:2007rz,Pestun:2016zxk} is a remarkable tool that allows us to compute exactly
various observables in  conformal $\N=2$ supersymmetric 4d gauge theories
(free energy on 4-sphere,  circular half-BPS  Wilson loop, correlation functions of chiral primary operators)  
in terms of special matrix models. It offers the possibility to study AdS/CFT correspondence beyond the planar limit and, in this way,  understand better  the structure of higher loop  corrections  in the dual  string theory.
 
According to the standard AdS/CFT dictionary, 
(see, \eg, \ci{Aharony:1999ti})\footnote{This dictionary is valid in maximally supersymmetric $SU(N)$ $\N=4$ SYM --  \adss  string  duality. In $\N=2$ cases some additional 
shifts of couplings   may be required.} 
\be \la{1.1}
\gs = {\l \ov 4\pi N} \,,\qqqquad  T= { L^2\ov 2 \pi \alpha'{}} = {\sql \ov 2 \pi}\,,
\ee 
the expansion of observables in gauge theory in $1/N$ and, then, in inverse powers of large 't Hooft coupling $\l\equiv  \gym^2N$ corresponds on the string side to
expanding 
in string coupling $g_s\sim 1/N$  and, then, in the  inverse string tension $T^{-1}\sim 1/\sqrt\l$.
 Within the localization approach, this expansion is well understood in maximally supersymmetric $SU(N)$ $\N=4$ Yang-Mills theory where the underlying 
matrix model is Gaussian  (see,  \eg, \ci{Erickson:2000af,Drukker:2000rr,Beccaria:2020ykg,Beccaria:2021alk}). 
In $\N=2$ superconformal    theories the localization 
matrix models contain nontrivial interaction potentials given by an infinite sum of single and  double trace terms  \cite{Pestun:2007rz,Fiol:2015mrp,Fiol:2020bhf,
  Fiol:2020ojn}. This makes the derivation of  large $N$, large $\l$ expansion in these theories  a non-trivial problem. 
  
For the special class of superconformal $\N=2$  theories that are planar-equivalent to $\N=4$ SYM,  the leading non-planar  corrections to free energy and circular  Wilson loop were studied using  localization in a number of recent papers  
\ci{Beccaria:2020ykg,Beccaria:2021ksw,Beccaria:2021vuc,Beccaria:2021ism,Beccaria:2021hvt,Beccaria:2022ypy,Bobev:2022grf,Beccaria:2022kxy}. 
The aim of the present  paper is to develop a systematical expansion of these observables beyond the leading order in $1/N$ 
and understand the properties of non-planar corrections at strong coupling. 
We shall consider two particular examples of $\N=2$  superconformal theories  that we denote as $\QQ$ and $\SA$ models. 

The $\QQ$ model is the $2-$node quiver $\N=2$  gauge  theory (hence the name $\QQ$) obtained as $\mathbb{Z}_{2}$ orbifold projection of  the $SU(2N)$  $\N=4$ SYM. It describes an adjoint $SU(N)$ vector multiplet coupled to two $SU(N) \times SU(N)$ bi-fundamental  $\N=2$ hypermultiplets with the same coupling constant. 
This model is dual to   type IIB superstring on the orbifold 
  ${\rm AdS}_{5}\times (S^{5}/\mathbb{Z}^{\rm orb}_{2})$ \cite{Kachru:1998ys,Lawrence:1998ja,Oz:1998hr,Gukov:1998kk}.

 The $\SA$ model describes a vector multiplet coupled to two $\N=2$ hypermultiplets in the rank-2 symmetric and antisymmetric representations of the $SU(N)$
 (hence the name $\SA$ for ``symmetric-antisymmetric'').\foot{This  model  is   also  sometimes referred to as ``{E}--theory" \cite{Billo:2019fbi}  
 being one  of the five (ABCDE) superconformal
4d  theories with gauge group $SU(N)$. } 
This theory may be viewed as an ``orientifold'' projection of the $\QQ$ model. 
 Its string theory dual is 
a special  orientifold   AdS$_{5}\times (S^{5}/\Gamma)$,  
where $\Gamma =\mathbb{Z}_{2}^{\rm orb}
\times \mathbb{Z}_{2}^{\rm orient}$ is the product of the orbifold  projection 
and an orientifold action that, besides inversions of target space  coordinates,  also  
involves a  product of  world-sheet parity and $(-1)^{F_{L}}$   \cite{Park:1998zh,Ennes:2000fu}.
   
We shall mostly focus on computing the two important observables in these 
 models -- the free energy on the unit 4-sphere  
  and the vacuum expectation value of a circular half-BPS Wilson loop. 
Due to the large $N$ planar equivalence,  at  leading order in $1/N$
these observables coincide  with those of  planar $\mathcal N=4$ SYM. The leading non-planar correction was found using the localization matrix model in \ci{Beccaria:2021ksw, Beccaria:2021vuc,Beccaria:2021hvt,Beccaria:2022ypy,Bobev:2022grf}. Here  we will compute the {\it subleading} non-planar corrections 
to the free energy and the circular Wilson loop in the $\SA$ and $\QQ$ models
and  also discuss their  interpretation on the dual string theory  side.  

Let  us   summarize our main results found from the corresponding localization matrix model. 

\subsection*{Strong coupling  expansion of  free energy}

\iffa 
One natural quantity that may be considered in our context is the 4-sphere free energy $F(\l; N)=-\log Z$, where $Z$ is the gauge theory partition function
on $S^{4}$.  Strictly speaking, it is not an observable since it contains in general a UV divergence proportional to the conformal a-anomaly.
In the framework of localization, the identification of $Z$ with the (finite) matrix model partition function is an implicit choice of 
renormalization scheme \cite{Pestun:2007rz,Russo:2012ay} that we will always assume in our analysis.
\fi

The localization yields a matrix model representation of the partition function $Z(\l,N)$ of the $\QQ$ and $\SA$ models on the 4-sphere.  As these models are planar equivalent to $\N=4$ SYM,  
it is convenient to split their free energy $F(\l,N)=-\log Z(\l,N)$ into the sum of the free energy of  the $SU(N)$  $\N=4$ SYM theory\foot{We assume the same 
definition of the matrix model measure and regularization as in \cite{Pestun:2007rz,Russo:2012ay} and omit 
a $\l$-independent constant.} and the difference function  
\begin{align}\notag
\la{1.2}
& F^\SA(\l; N) = F^{\N=4}(\l; N)+ \Delta F^\SA(\l; N)\,,  
\\[1.2mm]\notag
&  F^{\QQ}(\l; N)=  2 F^{\N=4}(\l; N)+\Delta F^{\QQ}(\l; N)  \ , 
\\ 
& F^{\N=4}(\l; N) = -\frac{1}{2}(N^{2}-1)\log\l \,. 
\end{align}
 In the $\QQ$ model,   
 the  $SU(N)$ $\N=4$  SYM contribution is  doubled as  in the planar limit each node of the quiver gives rise to $F^{\N=4}(\l; N)$.
  
In contrast to $F^{\N=4}(\l; N)$, the difference free energy $\Delta F(\l;N)$ remains finite at large $N$ and has the following expansion\foot{Note that in the $\N=2$ models   with hypermultiplets in the fundamental 
representation there are   also terms   with odd  powers of $1/N$ (see, e.g.,  \ci{Beccaria:2021ism,Beccaria:2022kxy}).}
  in powers of $1/N^{2}$
\be
\la{1.4}
\Delta F(\l; N) = \sF_{0}(\l)+\frac{1}{N^{2}}\sF_{1}(\l)+\frac{1}{N^{4}}\sF_{2}(\l)+\cdots\ .
\ee
As was mentioned above, the partition functions of the $\SA$ and $\QQ$ models are given by the $SU(N)$ matrix 
model integrals containing the interaction potential given by an infinite sum of double traces of powers of the $SU(N)$ matrices. A peculiar feature of the interaction potential is that the double traces are accompanied by powers of the coupling constant. As a consequence, the weak coupling expansion of the difference free energy $\Delta F(\l; N)$ can be obtained by expanding the
matrix integrals in powers of the interaction potential and evaluating them in a free  Gaussian   model. 

For  the same reason, the  evaluation of $\Delta F(\l; N)$ at strong coupling becomes an extremely nontrivial problem because it requires taking into account an infinite number of terms in the interaction potential. For the leading term in \re{1.4}, this leads to 
a representation of $\sF_{0}(\l)$ as the determinant of a certain (model-dependent) semi-infinite matrix $K(\l)$.
Early attempts to extract the strong coupling expansion of $\sF_0(\l)$ used various  approximations  or   numerical  approaches  in the  $\SA$ model \cite{Beccaria:2021vuc,Beccaria:2021hvt,Bobev:2022grf}  
and a numerical analysis  \cite{Beccaria:2021ksw} in the $\QQ$ model.
 
Recently,  it was observed  \cite{Beccaria:2022ypy} that the semi-infinite matrix $K(\l)$ in the $\SA$ model 
coincides with the matrix elements of the so called truncated (or temperature dependent) Bessel operator.
It is interesting to note that this operator has previously appeared in the study of level spacing distributions in matrix models \cite{Tracy:1993xj} and in the computation of   four-point correlation functions of infinitely heavy half-BPS operators 
in planar $\N = 4$ SYM~\cite{Belitsky:2019fan,Belitsky:2020qrm,Belitsky:2020qir}. 
Applying the methods developed in these papers, 
  made it 
possible not only to compute the strong coupling expansion of $\sF_{0}(\l)$ to any order in $1/\sql$ but also
determine non-perturbative (exponentially suppressed) $O(e^{-\sqrt\l})$ corrections
\cite{Beccaria:2022ypy}. 
 
 Similarly, it was shown that in the $\QQ$ model
the matrix $K(\l)$ can be split into two irreducible blocks (associated with the untwisted and twisted sectors of states), whose 
determinants are captured by  the extended Szeg\H{o}-Akhiezer-Kac
 formula for the Fredholm determinant of the Bessel operator~\cite{Beccaria:2022ypy}.
 
The resulting strong coupling expansion of the leading term in \rf{1.4}  was found to be  \cite{Beccaria:2022ypy}
\footnote{We omit non-perturbative  $O(e^{-\sql})$ contributions to the free energy in what follows.} 
\ba
\sF_{0}^{\SA}(\l) = 
& \frac18 \l^{1/2}-\frac{3}{8}\log\l
-3 \log\mathsf{A} +\frac{1}{4}-\frac{11}{12}\log 2  + {3\over 4} \log ( 4 \pi) \notag \\
& +\frac{3}{32} \log {\l'\ov \l}-\frac{15 \zeta (3)}{64\,
    {\l'}^{3/2}}-\frac{945 \zeta (5)}{512\, {\l'}^{5/2}}-\frac{765 \zeta (3)^2}{128\,
   {\l'}^{3}}+O({\l'}^{-7/2}), \la{1.5} 
\\[2mm]
\sF_{0}^{\QQ}(\l) =& \frac{1}{4} \l^{1/2}-{1\over 2} \log \l 
-6 \log\mathsf{A}+\frac{1}{2}-\frac{4}{3} \log 2  + \log (4 \pi)  \notag \\
& +\frac{1}{16} \log {\l'\ov \l}-\frac{3 \zeta (3)}{32\,{\l'}^{3/2}} -\frac{135 \zeta (5)}{256 \,
   {\l'}^{5/2}}-\frac{99\zeta (3)^2}{64 \,{\l'}^3}+O({\l'}^{-7/2}),\la{1.6} 
\ea
where $\mathsf{A}$ is the Glaisher constant and $\lambda'{}^{1/2} \equiv  \lambda^{1/2} -4\log 2$ is a shifted coupling constant. 
The rationale for redefining the expansion parameter $\l\to\l'$ is that it allows us to perform a 
resummation of all terms  with coefficients  containing powers of  $\log 2$. 
Let us note  that  in the $\QQ$  model \rf{1.6}  the coefficient of the leading $O(\l^{1/2})$ term is doubled  as  compared  to the  $\SA$ one in \rf{1.5} 
 (just like the planar $O(N^2)$ contribution to the free energies  $F^{\QQ}$ and $F^\SA$ in \rf{1.2}).

In this paper  we extend the analysis of \cite{Beccaria:2022ypy} and derive the strong coupling
expansion of subleading non-planar corrections to \re{1.4}. We show that in both models the functions $\sF_{1}(\l), \sF_{2}(\l), \dots$  in \rf{1.4}
are given by polynomials in the basic traces $\tr\big[RK(\l)/\lr{1-K(\l)}\big]$ involving again the leading-order matrix $K$ and
some specific coupling-independent semi-infinite matrices $R$. We demonstrate that these traces
can be expressed in terms of matrix elements of the resolvent  of the Bessel operator mentioned above. 
We develop a technique for computing these matrix elements at strong coupling in a systematic  way and, thus, find 
the corresponding expansions of coefficient functions in \rf{1.4}.
 
Explicitly,  we find for the functions $\sF_1$ and $\sF_2$ in the $\SA$  model 
\ba
\notag
\sF_{1}^{\SA}(\l)  ={}&
-\frac{\lambda ^{3/2}}{2048}-\frac{3 \lambda }{2048}+\lambda^{1/2}
   \Big(\frac{11}{2048}-\frac{\log 2 
   }{128}\Big)+\Big(\frac{3}{128}-\frac{\log ^2 2 }{32}-\frac{\log  2 }{512}\Big)
\\\notag &   
   +  {\frac{1}{\lambda^{1/2} }}\Big(-\frac{15 \zeta
   (3)}{2048}-\frac{\log ^3 2 }{8}-\frac{3 \log ^2 2 }{128}+\frac{279 \log 2 
   }{2048}\Big) \notag
\\ &   
   +{1\over \lambda
   } \lr{\frac{105 \zeta (3)}{8192}-\frac{15 \zeta (3) \log 2 }{128} 
   -\frac{\log ^4 2 }{2}-\frac{5 \log ^3 2 }{32}+\frac{441 \log ^2 2 }{512}} +\dots, \la{1.8}
\\[2mm]
\notag
\sF_{2}^{\SA}(\l) = {}&  \frac{\lambda ^{3}}{2949120}-\frac{\lambda ^{5/2}}{1310720}+\lambda ^2 \Big(\frac{\log 2
   }{61440}-\frac{251}{7864320}\Big) 
 \\\notag
 {}&  
   +\lambda ^{3/2}
   \Big(\frac{107}{3145728}+\frac{\log^{2}2}{15360}-\frac{47\log 2}{245760}\Big)
\\ &     
   +\lambda  \Big(\frac{409}{524288}+\frac{\zeta (3)}{65536}+\frac{\log
   ^32}{3840}-\frac{19 \log ^22}{20480}-\frac{191\log 2}{262144}\Big)+\dots,  \la{1.9}
 \ea
and for the function $\sF_1$   in  the $\QQ$ model   
\ba
\sF_{1}^{\QQ}(\l) =& -\frac{\l^{3/2}}{1024}-\frac{\l}{3072}+\l^{1/2}   \Big(\frac{5}{1024}-\frac{\log2}{192}\Big)
-\frac{\log   ^22}{48}+\frac{\log   2}{256}\notag\\
   &+\frac{1}{\l^{1/2}} \Big(-\frac{3 \zeta   (3)}{1024}-\frac{\log
   ^32}{12}+\frac{\log^22}{64}+\frac{35 \log2}{1024}\Big)+\dots.\la{1.10}
\ea
These expressions involve terms with powers of $\log 2$. Their  resummation  
is less  obvious than in the leading terms $\sF_{0}^{\SA}(\l)$ in \rf{1.5}   and $\sF_{0}^{\QQ}(\l)$ in  \rf{1.6}
  and will be discussed   below. 

\subsection*{Double scaling limit}

The obtained expressions for the non-planar corrections $\sF_{0}(\l)$, $\sF_{1}(\l)$ and $\sF_{2}(\l)$ reveal an interesting structure. Namely, keeping only the leading large $\l$ terms in  \re{1.5}--\re{1.10}, we get  for the corresponding $\Delta F$ in  \re{1.4}
\ba \notag
\Delta F^{\SA}(\l; N) &=  \frac{1}{8}\,\l^{1/2}  
-\frac{1}{N^{2}}\frac{\l^{3/2}}{2048} 
+\frac{1}{N^{4}}\frac{\l^{3}}{2949120} + \dots \,, 
\\
\Delta F^{\QQ}(\l; N) &=  \frac{1}{4}\,\l^{1/2} 
-\frac{1}{N^{2}}\frac{\l^{3/2}}{1024}+\dots\,,\la{1.11} 
\ea
where dots  stand for terms  with subleading powers of $\l$ at each order in $1/N^2$. 
 We observe that the coefficients of $1/N^2$ have  increasing power of $\lambda$. The relation \re{1.11} suggests 
that in this  limit (i.e. $N \to \infty$ and then $\l \to \infty$) 
the expansion of the free energy effectively runs in powers of $\lambda^{3/2}/N^2$.
Moreover, we again observe that, as  it happened at  the  $O(N^2)$ and $O(N^0)$ orders,
  the coefficients  of the subleading  $\l^{3/2}/N^2$  terms  in \rf{1.11} 
are again    related by the  factor of 2.

We shall argue that these properties can be understood as a consequence of the familiar 
double-scaling limit in the matrix models, see \eg ~\cite{DiFrancesco:1993cyw,Eynard:2015aea}. In the present context it  corresponds to the limit 
\be
\la{1.14}
 N\to\infty\,, \qquad \l \to \infty\,, \qquad \   \frac{\l^{3/2}}{N^{2}}= {\rm fixed}\,. 
\ee
Taking this limit directly in the localization matrix model representation of the gauge theory  
partition function  gives an efficient way of computing the
 coefficients  of the leading  large $\l$ terms in $\sF_n$ in  \rf{1.4}.
 By exploiting some recent results in the matrix models, we  find that in the $ \SA$  theory 
\ba
\Delta F^\SA 
\simeq {} &  \frac{1}{8}\,\sql-\frac{1}{2048}\,\frac{\lambda ^{3/2}}{N^2}+\frac{1}{2949120}\,\Big(\frac{\lambda ^{3/2}}{N^2}\Big)^{2}
-\frac{1}{1486356480}\,\Big(\frac{\lambda ^{3/2}}{N^2}\Big)^{3}  \no\\&
-\frac{1}{304405807104}\,\Big(\frac{\lambda ^{3/2}}{N^2}\Big)^{4}
+\frac{17}{365286968524800}\,\Big(\frac{\lambda ^{3/2}}{N^2}\Big)^{5}+\ldots\,,\la{1.15}
\ea
where the sign `$\simeq$' indicates that this relation is valid in the limit \re{1.14}.

In addition, we prove that in the double scaling limit the free energies  in the $\SA$ and $\QQ$ models are related to each other 
 as (to all orders in $1/N^2$)
\begin{align}\label{Fequal}
\Delta F^{\QQ} \simeq 2\Delta F^\SA\,.
\end{align}
We would like to emphasize that this relation is not satisfied at weak coupling. 
The   appearance of this  relation at strong coupling  admits a possible 
      interpretation   on the  string theory side as  being a consequence of 
       the fact that   the $\SA$ model  may be    obtained   from the $\QQ$ one by an extra  projection (see  section \ref{sec7}). 

The relation of \rf{1.15} suggests that, in the double scaling limit \re{1.14}, the difference free energy takes the following general form~\foot{Note that the leading $O(\sql)$ term in \rf{1.13}    has  a  special structure compared to 
  subleading terms. 
On the matrix model side,  this has to do with its 
 origin from  $ \log  \det(1-K(\l))$ of the  Bessel matrix, see  \ci{Beccaria:2022ypy} and \re{F-LO} below. 
} 
\be 
 \Delta F \simeq  c_0 \sql +   \sum_{n=1}^\infty   c_n \Big({ \l^{3/2}\ov N^2}\Big)^n  
 \la{1.13} =  2\pi c_0 T+   \sum_{n=1}^\infty   c_n \Big(8\pi { g^2_s\ov T}\Big)^n  
  \,, 
\ee
where $c_n$ are {\it rational} coefficients.
 Here in the second relation we switched to the expansion parameters \rf{1.1} of  the dual 
string theory.
  We shall  discuss the string theory interpretation of this expansion and coefficients $c_n$ 
    in section \ref{sec7}    below. 

 Remarkably, a similar scaling behaviour  was  observed  earlier in the strong coupling expansion 
of the  circular Wilson loop in the $ \N=4$   SYM  theory  in \ci{Drukker:2000rr}
 (its  dual   \adss  string-theory interpretation was  given  in \ci{Giombi:2020mhz}).
However, in contrast to that case where the  $(\l^{3/2}/N^2)^n $ corrections to the Wilson loop exponentiated 
into  $\exp(\l^{3/2}/(96N^2))=\exp ( \pi  { g^2_s/ (12T)})$,
here the series in \rf{1.15} and \re{1.13}  is likely to develop Borel singularities indicating the need  to include  non-perturbative, exponentially suppressed corrections.\footnote{Let us
 note that  an    explicit  form of 
a similar   strong coupling  (large $N$ and  large $\l$) scaling limit   may   depend 
on a particular model and also on a particular observable. 
For example, a correlator of the   circular Wilson loop with a  chiral primary operator
 has   an  expansion in powers of $ \l/N^2 \sim  g^2_s/ T^2 $ (which sums up to a simple square root expression)
  \ci{Beccaria:2020ykg}. 
Another 
model with reduced supersymmetry where a similar double scaling limit  exists  
 is the  4d $U(N)$ $\N=4$ SYM theory with 
 a $\frac{1}{2}$-BPS codimension-one defect (hosting a 3d $\mc N=4$ theory)
which is dual to a D3-D5 system without flux.
The associated localization matrix model  potential 
has an infinite number of single-trace terms and no double-trace terms.
The analysis of the free energy at strong coupling shows that it has
a well-defined limit   $N\to \infty, \ \l\to \infty$ with fixed $\l/N^{2}$
(compared to \rf{1.14}  above). It was  computed  in this limit 
in a closed
form in   \cite{Beccaria:2022bjo} (see Eq.~(5.20) there).
}

\subsection*{Strong coupling  expansion of circular Wilson loop}

The expectation value of the half-BPS circular Wilson loop in the $\SA$ and $\QQ$ models admits a representation in the localization matrix model similar to that  for  the free energy.\footnote{\la{f7}In the $\QQ$ model, the Wilson loop is defined in terms of  the fields of the $\N=2$ vector multiplet at one of the nodes.} Its large $N$ expansion in both models takes the form
\be\la{622}
N^{-1}  {W} = \sW_{0}+\frac{1}{N^{2}}\sW_{1}+\frac{1}{N^{4}}\sW_{2}+\cdots
\,.
\ee
As for  the free energy \rf{1.2}, the leading planar correction $\sW_{0}$ is the same in the $\SA$ and $\QQ$ models. It is equal to the planar term in the  $SU(N)$ 
$\N=4$ SYM expression $\sW_{0}= 2\rI_{1}(\sql)/\sql$ where $I_1$ is a Bessel function  \ci{Erickson:2000af,Drukker:2000rr}. %

The deviation of $W$ from the exact $\N=4$ SYM result  \ci{Drukker:2000rr} 
 in  the  $\SA$ and $\QQ$ models 
 starts at order $O(1/N^{2})$. It was observed in \cite{Beccaria:2021ksw,Beccaria:2021vuc} that at this order 
 it is 
proportional to a derivative of the free energy with respect to the coupling constant.
 We found that  a similar relation to the free energy 
  holds  also at  higher orders of $1/N^2$ expansion. Namely,  the ratios of the Wilson
   loop expectation values expanded in $1/N^2$  may be written as 
\ba\notag
\la{1.17} 
{ {W} ^\SA \ov  {W} ^{\N=4}} =&1 -\frac{\l^{2}}{4N^{2}}\,\sF_{0}'+\frac{1}{N^{4}}\,\Big(
-\frac{\l^{3}}{48}\,\sF_{0}'+\frac{\l^{4}}{96}\,\sF_{0}'{}^{2}-\frac{\l^{4}}{96}\sF_{0}''
-\frac{\l^{3/2}}{4}\frac{\rI_{2}(\sql)}{\rI_{1}(\sql)}\,\sF_{1}-\frac{\l^{2}}{4}\,\sF_{1}'\Big)+...
\\[2mm] 
 { {W} ^{\QQ} \ov  {W} ^{\N=4}}=& 1 -\frac{\l^{2}}{8N^{2}}\,\sF_{0}'+\frac{1}{N^{4}}\,\Big(
-\frac{\l^{3}}{192}\,\sF_{0}'+\frac{\l^{4}}{384}\,\sF_{0}'{}^{2}-\frac{\l^{4}}{384}\sF_{0}''
-\frac{\l^{3/2}}{8}\frac{\rI_{2}(\sql)}{\rI_{1}(\sql)}\,\sF_{1}-\frac{\l^{2}}{8}\,\sF_{1}'\Big)+\dots\,, 
\ea
where $\sF_{0}$ and $\sF_{1}$ are the coefficients of the $1/N^2$ expansion \re{1.4} of the free energy 
$\Delta F^{\SA}$ and $\Delta F^{\QQ}$ given by \rf{1.5} -- \rf{1.10} and 
prime denotes  derivative over $\l$. The leading $O(1/N^{2})$ terms in \re{1.17} 
were found  in \cite{Beccaria:2021ksw,Beccaria:2021vuc}.

The relations \re{1.17} hold for an arbitrary coupling $\l$. In the double scaling limit \re{1.14}, keeping the leading term at  strong coupling at each order in $1/N^2$,   
the above relations simplify as  
\be
\la{1.18}
{ {W} ^\SA \ov  {W} ^{\N=4}} \simeq { {W} ^{\QQ} \ov {W} ^{\N=4}}\simeq 
 1-{1 \ov 64} \frac{\l^{3/2}}{N^{2}}+{1\ov 6144}\Big( \frac{\l^{3/2}}{N^{2}}\Big)^2 +\cdots\,.
\ee
Thus  the Wilson loops in  the $\SA$ and $\QQ$ models  coincide in the double scaling
limit,  
\begin{align}\la{1.188}
{W} ^\SA\simeq {W} ^{\QQ}\,.
\end{align}
The  matrix model origin  of this  strong-coupling equality will be discussed below.

\subsection*{Structure of the paper}

The rest of the paper is organized as follows.

In section \ref{sec:2}  we  discuss  the localization matrix models for the $\SA$ and $\QQ$ models and
describe  the structure of the diagrammatic expansion of the free energy in $1/N$. 

In section \ref{sec:3}  we present explicit representations for the leading $O(N^0)$  term of the free energy \re{1.4}
and also  for the next two  $O(1/N^{2})$ and $O(1/N^{4})$   non-planar  corrections. 
The latter  are expressed    in terms of certain matrix elements of the resolvent of the truncated (finite temperature) Bessel operator. 

Section \ref{sect:res} 
is devoted to the explicit  evaluation of these matrix elements in the non-trivial strong coupling regime. It contains  the main results of the paper. 

In section \ref{sec:5}  we clarify the origin of the peculiar structure that the strong coupling  expansion of the free energy in the $\SA$ and $\QQ$ models
 takes 
when only the highest  power of $\l$ is kept at each order in the  $1/N^2$  expansion. This  limit 
is related to the familiar  double scaling limit in matrix models. The explicit results  for the coefficients of $(\l^{3/2}/N^2)^n$
terms in \rf{1.15} and \rf{1.13}  are 
obtained  up to  order  
 $O(1/N^{10})$. 

Section \ref{sec:6}  describes the computation of  the subleading  non-planar corrections 
to the circular half-BPS Wilson loop  in the $\SA$ and $\QQ$ models   and their form in the 
the double scaling limit.  

In section \ref{sec7}  we suggest the dual string theory interpretation of the leading strong coupling terms in the non-planar 
corrections  to free energy  in \rf{1.15} and \re{1.13}, relating the values of the coefficients $c_n$ in \rf{1.13}
to those of the few leading  higher-derivative $D^n R^4-$like     corrections  in the type IIB superstring effective action. 

There are also  four appendices  containing   derivations of  some of the results used in the  text. 

 \section{Matrix model representation}
 \la{sec:2}
 
In this section   we discuss the  matrix model representation for  the partition function of the $\SA$ and quiver $\QQ$ models with a gauge group $SU(N)$ on the unit sphere $S^{4}$.
 
The partition function of the $\SA$ model is given by a matrix 
 integral   \ci{Pestun:2007rz} (see \ci{Beccaria:2021vuc} for details)
\ba \la{21}
Z_{\SA} = \int\prod_{r=1}^{N}da_{r}\,\delta\Big(\sum_{r}a_{r}\Big)\,\Delta^2(\bm{a})\,e^{-S_\SA (\bm{a})}\,, 
\ea
where integration goes over eigenvalues $\bm{a}=\{a_1,\dots,a_N\}$ of a hermitian traceless $N\times N$ matrix $A$ describing zero modes of a scalar field on $S^4$.
Here $\Delta(\bm{a}) = \prod_{r<s}(a_{r}-a_{s})$ is a Vandermonde determinant and the potential  $S_\SA (\bm{a})$
has the following form 
\ba\label{S-SA}
S_\SA (\bm{a}) 
&= \frac{8\pi^{2}N}{\l} \sum_{r=1}^N a_{r}^{2}
+\sum_{r,s=1}^N \Big[\log H(a_{r}+a_{s})-\log H(a_{r}-a_{s})\Big].
\ea
It contains the function $H(x)$ given by the product of the Barnes $G-$function
\ba
H(x) & = e^{-(1+\gamma_{\rm E})\,x^{2}}
\,G(1+ix)G(1-ix)
= \exp\left( \sum_{n=1}^{\infty}\frac{(-1)^{n}}{n+1}\zeta(2n+1)\,x^{2n+2}\right)\,.
\la{23} 
\ea
The second relation yields its expansion at small $x$ and it involves the Riemann zeta values.

In a similar manner, the partition function of the quiver $\QQ$ model with equal coupling constants on the two nodes is given by
\begin{align}\label{Z-Q2}
Z_{\QQ}  =\int\prod_{r=1}^{N}da_{1,r}da_{2,r}\,\delta\Big(\sum_{r}a_{1,r}\Big)\delta\Big(\sum_{r}a_{2,r}\Big)\,\big[\Delta(\bm{a}_1)\Delta(\bm{a}_2)\big]^{2} \,e^{-S_{\QQ} (\bm{a}_1,\bm{a}_2)}\,,
\end{align}
where $a_{\alpha,i}$ are eigenvalues of the $SU(N)$ matrices $A_\alpha$ (with $\alpha=1,2$). The potential is given by
\begin{align}\notag\label{S-Q2}
S_{\QQ} (\bm{a}_1,\bm{a}_2) {}& = \frac{8\pi^{2}N}{\l} \sum_{r=1}^N (a_{1,r}^{2}+a_{2,r}^{2})
\\
{}&
+\sum_{r,s=1}^N \Big[2\log H(a_{1,r}-a_{2,s})-\log H(a_{1,r}-a_{1,s})-\log H(a_{2,r}-a_{2,s}) \Big],
\end{align}
where the function $H$ is defined in (\ref{23}). Writing down (\ref{21}) and (\ref{Z-Q2}), we neglected the instanton contribution to the partition function as it is exponentially small at large $N$.

It is convenient to express the potentials (\ref{S-SA}) and (\ref{S-Q2}) in terms of traces of the hermitian matrices
\begin{align}\label{O}
\OO_{i}(A)= \tr \Big({A\over  \sqrt N}\Big)^{i} = \sum_{r=1}^N \left(a_r\over \sqrt N\right)^{i}\,,
\end{align}
where $\OO_1(A)=0$ for the $SU(N)$ matrices.
Expanding the $H-$functions in (\ref{S-SA}) and (\ref{S-Q2}) in powers of eigenvalues $a_r$ and rescaling them as $a_r\to (8\pi^2 N/\lambda)^{-1/2} a_r$, we get
\begin{align}\notag
 S_{\SA}(\bm{a}) {}&= \tr A^2  - S_{\rm int}(A)\,,  
\la{24} 
\\[2mm]
S_{\QQ}(\bm{a}_1,\bm{a}_2) {}&= \tr A_1^2 +\tr A_2^2  - S_{\rm int}(A_1,A_2)\,.
\end{align}
The interaction terms in both models are given by infinite bilinear combinations of the single traces (\ref{O})  
\begin{align} \label{int-terms1}
S_{\rm int}(A) &= \frac12 \sum_{i,j\ge 1} C^-_{ij}(\lambda) \, \OO_{2i+1}(A) \OO_{2j+1}(A)\,,
\\ \notag\label{int-terms2}
S_{\rm int}(A_1,A_2) &= \frac14 \sum_{i,j\ge 1}C^-_{ij}(\lambda)  \left[ \OO_{2i+1}(A_1)-\OO_{2i+1}(A_2) \right]\left[ \OO_{2j+1}(A_1)-\OO_{2j+1}(A_2) \right]
\\
& +
 \frac14 \sum_{i,j\ge 1}C^+_{ij}(\lambda)  \left[ \OO_{2i}(A_1)-\OO_{2i}(A_2) \right]\left[ \OO_{2j}(A_1)-\OO_{2j}(A_2) \right]\,,
\end{align}
where the expansion coefficients $C^\pm_{ij}$ with $i,j\ge 1$ are 
\begin{align}\la{25}\notag
C^-_{ij} (\lambda)&= 8\,\Big(\frac{\l}{8\pi^{2}}\Big)^{i+j+1}\,(-1)^{i-j+1}\,\zeta(2(i+j)+1)\,\frac{\Gamma(2(i+j)+2)}{\Gamma(2i+2)\Gamma(2j+2)}\,,
\\
C^+_{ij} (\lambda)&= 8\,\Big(\frac{\l}{8\pi^{2}}\Big)^{i+j}\,(-1)^{i-j+1}\,\zeta(2(i+j)-1)\,\frac{\Gamma(2(i+j))}{\Gamma(2i+1)\Gamma(2j+1)}\,.
\end{align}
They define two semi-infinite matrices whose entries are proportional to a power of 't Hooft coupling and odd Riemann zeta values.
Notice that the interaction term in the $\SA$ model (\ref{int-terms1})  is given by the sum of double traces containing odd powers of matrices. At the same time, 
the interaction term in the $\QQ$ model (\ref{int-terms2}) involves the double traces with both even and odd powers of matrices. 
The superscript in $C_{ij}^\pm$ refers to the parity of the powers of matrices in the double traces.
 
The partition function of $\mathcal N=4$  $SU(N)$ super Yang-Mills theory $Z_{\mathcal N=4}$ is given by the same integral \re{21} but with the interaction term $S_\SA$ set to zero. 
Taking the ratio of the partition functions, we can express (\ref{21}) and (\ref{Z-Q2}) as the following matrix integrals
\begin{align}\notag\la{2.8}
& {Z_{\SA}\over  Z_{\mathcal N=4}} = \int DA \, e^{-\tr A^2 + S_{\rm int}(A)}  \equiv \vev{e^{S_{\rm int}(A)}}\,,
\\
& {Z_{\QQ} \over [Z_{\mathcal N=4}]^2} =\int DA_1 DA_2 \, e^{-\tr A_1^2-\tr A_2^2 + S_{\rm int}(A_1,A_2)}  \equiv  \vev{e^{S_{\rm int}(A_1,A_2)}}\,,
\end{align}
where the integration measure is normalized in such a way that $\int DA \, e^{-\tr A^2}=1$.  
As a consequence, the free energy $F=-\log Z$ may be written as \re{1.2}.

According to \re{2.8}, the difference free energy \re{1.4} in both models 
can be computed as expectation values of interaction terms \re{int-terms1} and \re{int-terms2} in a Gaussian matrix model
\begin{align}\label{F-vev}
e^{-\Delta F_{\SA}} =  \vev{e^{S_{\rm int}(A)}}\,,\qqqquad e^{-\Delta F_{\QQ}} =  \vev{e^{S_{\rm int}(A_1,A_2)}}\,,
\end{align}
where the average is computed using the measure defined in \re{2.8}.

At large $N$ and fixed $\lambda$, the matrix integrals in (\ref{2.8}) admit a topological expansion over the so-called touching surfaces \cite{Das:1989fq,Korchemsky:1992tt,Alvarez-Gaume:1992idg,Klebanov:1994pv}. A somewhat unusual feature of these surfaces, that follows from the double-trace form of the 
potentials \re{int-terms1} and  \re{int-terms2}, is that they are given by a collection of spherical bubbles that touch other bubbles at two isolated points at least. At large $N$ the leading contribution to \re{2.8} comes from the touching bubbles with neckless configuration and it scales as $O(N^0)$. This implies that
the difference free energy, $\Delta F_{\SA}$ and $\Delta F_{\QQ}$, stays finite in the large $N$ limit and it
takes the form \re{1.4}.
The leading $O(N^2)$ contribution to the free energy \re{1.2} in the two models coincides (up to a factor of $2$ in the $\QQ$ model) with that of $\mathcal N=4$  $SU(N)$ SYM theory. 
 
\subsection{Free energy at weak coupling}\label{sect:weak}

At weak coupling, for $\l \ll 1$, the free energy \re{1.4} can be computed by expanding the expectation values on the right-hand side of (\ref{F-vev}) in powers of $S_{\rm int}$ and doing Gaussian averages. 

This way we get, for instance,  the first three functions in \re{1.4}  in the $\SA$ model,
\ba\notag\label{SA-weak}
\sF_{0} &= 5 \zeta_5 \hatlambda^3-\frac{105}{2} \zeta_7\hatlambda^4+441 \zeta_9 \hatlambda^5+(-25 \zeta_5^2-3465 \zeta_{11}) 
\hatlambda^6+\cdots, \\
\notag
\sF_{1} &= -25 \zeta_5 \hatlambda^3+\frac{735 }{2}\zeta_7 \hatlambda^4-3780 \zeta_9 
\hatlambda^5+(-650 \zeta_5^2+32340 \zeta_{11}) \hatlambda^6+\cdots, \\
\sF_{2} &= 20 \zeta_5 \hatlambda^3-735 \zeta_7 \hatlambda^4+11907 
\zeta_9 \hatlambda^5+(9075 \zeta_5^2-127050 \zeta_{11}) \hatlambda^6+\cdots\,.
\ea
In the $\QQ$ model we have instead
 \ba\notag\label{Q2-weak}
\sF_{0} &=  3 \zeta _3 \bigg(\frac{\l}{8\pi^2}\bigg)^2-15 \zeta _5 \bigg(\frac{\l}{8\pi^2}\bigg)^3+\bigg(-9 \zeta _3^2+\frac{315 \zeta _7}{4}\bigg) 
\bigg(\frac{\l}{8\pi^2}\bigg)^4+(120 \zeta _3 \zeta _5-441 \zeta _9) \bigg(\frac{\l}{8\pi^2}\bigg)^5
+\cdots,  
\\\notag
\sF_{1} &=  
-3 \zeta _3 \bigg(\frac{\l}{8\pi^2}\bigg)^2+25 \zeta _5 \bigg(\frac{\l}{8\pi^2}\bigg)^3+\bigg(-9 \zeta _3^2-\frac{735 \zeta 
_7}{4}\bigg) \bigg(\frac{\l}{8\pi^2}\bigg)^4+(120 \zeta _3 \zeta _5+1260 \zeta _9) \bigg(\frac{\l}{8\pi^2}\bigg)^5
+\cdots, \\
\sF_{2} &=  -10 \zeta _5 \bigg(\frac{\l}{8\pi^2}\bigg)^3+\bigg(18 \zeta _3^2+\frac{315 \zeta _7}{2}\bigg) \bigg(\frac{\l}{8\pi^2}\bigg)^4+(-420 
\zeta _3 \zeta _5-1701 \zeta _9) \bigg(\frac{\l}{8\pi^2}\bigg)^5
+\cdots\, .
\ea
Higher order terms of the expansion have increasing complexity and involve multilinear combinations of the Riemann 
$\zeta$-values  $\zeta_n \equiv \zeta(n)$.  

Notice that the expansion of $\sF_{0}$ and $\sF_{1}$ in \re{SA-weak} and \re{Q2-weak}
starts at different order in $\lambda$ and there is no obvious relation between the functions $\Delta F_{\SA}$ and $\Delta F_{\QQ}$ at weak coupling. As we will see below, the situation is different at strong coupling.

\subsection{Hubbard--Stratonovich transformation}\label{sect:HS}

As was already mentioned, the interaction terms (\ref{int-terms1}) and  (\ref{int-terms2}) are bilinear in single traces (\ref{O}). We can linearize $S_{\rm int}$ by introducing auxiliary fields coupled to the traces (\ref{O}).
For instance, in the $\SA$ model we use (\ref{int-terms1}) to get
\begin{align}\label{HS}
e^{S_{\rm int}(A)} = (\det C^-)^{-1/2} \int  dJ^- \exp\Big(\sum_{i\ge 1} J_i^- \OO_{2i+1}(A) - \frac12\sum_{i,j\ge 1} J_i^- J_j^- (C^-)_{ij}^{-1} \Big),
\end{align}
where the integration measure is $d J^-=\prod_{i\ge 1} dJ_i^-/\sqrt{2\pi}$.
Substituting this identity into the first relation in (\ref{2.8}),  the matrix integral over $A$ takes the form
\begin{align}\label{Z}
Z(J^-) =\vev{e^{\sum_{i\ge 1} J_i^- \OO_{2i+1}(A)}} =   \int DA \, e^{-\tr A^2 + \sum_{i\ge 1} J_i^- \OO_{2i+1}(A)}  \,.
\end{align}
It coincides with  a generating function of the correlators of single traces (\ref{O}) in a Gaussian unitary ensemble. In this way,
we arrive at the following representation for the difference free energy \re{F-vev} in the $\SA$ model \footnote{Here and in what follows the summation over repeated indices $i,j\ge 1$ is tacitly assumed.}
\begin{align}\label{F-SA}
e^{-\Delta F_{\SA}} =  (\det C^-)^{-1/2} \int  dJ^- Z(J^-) \, e^{- \frac12J_i^- J_j^- (C^-)_{ij}^{-1}}\,.
\end{align}
Repeating the same calculation for the $\QQ$ model we obtain a similar representation
\begin{align}\label{F-Q2}
e^{-\Delta F_{\QQ}} = (\det C^-  \det C^+)^{-1/2}\int  dJ^- dJ^+ Z(J^-,J^+) Z(-J^-,-J^+) \, e^{-  J_i^- J_j^- (C^-)_{ij}^{-1}-  J_i^+ J_j^+ (C^+)_{ij}^{-1}}\,,
\end{align}
where $d J^\pm=\prod_{i\ge 1} dJ_i^\pm/\sqrt{2\pi}$. Here the factors of $Z(J^-,J^+)$ and $Z(-J^-,-J^+) $ come from integration over $A_1$ and $A_2$, respectively. They are given by
\begin{align}\label{ZZ}
Z(J^-,J^+) =\vev{e^{\sum_{i\ge 1} (J_i^+ \OO_{2i}(A)+J_i^- \OO_{2i+1}(A))}} =   \int DA \, e^{-\tr A^2 + J_i^+ \OO_{2i}(A)+ J_i^- \OO_{2i+1}(A)} \,,
\end{align}
and similarly  for $Z(-J^-,-J^+)$.  

As we show below, the relations \re{F-SA} and \re{F-Q2} can be effectively used to derive the strong coupling expansion of the free energy \re{1.4} to any order in $1/N^2$.
Notice that the dependence of  \re{F-SA} and \re{F-Q2} on the coupling constant comes from the semi-infinite matrices $C^\pm_{ij}$ defined in \re{25}. At the same time, the dependence on $1/N^2$ comes from the correlators \re{Z} and \re{ZZ} in a Gaussian unitary ensemble. 

\subsection{Correlators in a Gaussian unitary ensemble}

In this subsection, we examine the properties of the function $Z(J^-,J^+)$ defined in \re{ZZ}. 
The function $Z(J^-)$ introduced in \re{Z}  can be considered as its special value for $J_i^+=0$
\begin{align}\label{Z-ZZ}
Z(J^-) =Z(J^-,0) \,,\qquad\qquad Z(J^-,J^+)=Z(-J^-,J^+)\,.
\end{align}
The second relation follows from invariance of (\ref{ZZ}) under $A\to -A$. It implies that the product $Z(J^-,J^+) Z(-J^-,-J^+)$ that enters \re{F-Q2}  is an even function of $J^-$ and $J^+$ separately.

According to its definition \re{ZZ}, $Z(J^-,J^+)$ is a generating function of correlators in a Gaussian matrix model
\begin{align}\label{3.3}
Z(J^-,J^+) = \vev{e^{\sum_{i\ge 2} J_i \OO_{i}}} 
 =e^{\sum  J_i  G_{i} +\frac12 \sum  J_iJ_j G_{ij} +\frac1{3!} \sum J_iJ_j J_k G_{ijk} 
 +\dots}\,.
\end{align}
Here $J_i$ coincides with $J_i^+$ or $J_i^-$ depending on the parity of index, \ie $J^+_i=J_{2i}$ and $J^-_i=J_{2i+1}$,
and $G_{i_1\dots i_L}$ is a connected $L-$point correlator (see Appendix~\ref{App:gauss})
\begin{align} 
G_{i} =\vev{\OO_{i}}\,,
\qqquad 
 G_{ij} =\vev{\OO_{i} \OO_{j} }_c\,,
\qqquad 
 G_{ijk}=\vev{\OO_{i} \OO_{j} \OO_k}_c\,, \quad  ...  \ .
 \la{3.4}
\end{align}
Recall that $\OO_1=0$ for  the $SU(N)$ group and, therefore, $G_{i_1\dots i_L}$ is different from zero for $i_p\ge 2$.

At large $N$, the correlators \re{3.4} admit an expansion in powers of $1/N^2$  
\begin{align}\label{G-gen} 
G_{i_1\dots i_{L}} = {\beta_{i_1}\dots\beta_{i_{L}}\over N^{L-2}}\Big[P_{L-3} +{1\over N^2} P_{L} 
+{1\over N^4} P_{L+3}  +\dots \Big],
\end{align}
where $\beta_i$ is given by a ratio of $\Gamma-$functions and depends on the parity of $i$, see \re{beta} below. 

The correlator \re{G-gen} scales as $O(1/N^{L-2})$. The leading term $P_{L-3}$ is a polynomial in $i_1,\dots,i_L$ of degree $L-3$ for $L\ge 3$.  
For $L=2$ we have $P_{-1} \sim 1/(i_1+i_2+c)$ with the constant $c$ depending on the parity of indices (see \re{Q2}). The subleading corrections to \re{G-gen} involve polynomials in $i$'s of increasing degree. 
To the next order in $1/N^2$ their degree increases by $3$. Notice that for $i_p=O(N^{2/3})$ each term inside the brackets in \re{G-gen} scales as $O(N^{2(L-3)/3})$ leading to $G_{i_1\dots i_{L}}=O(1/N^{2L/3})$. This property will play an important role in what follows.
 
The explicit expressions for the polynomials $P_{L-3}, P_L, P_{L+3},\dots$ in \re{G-gen} depend on the parity of indices $i_1,\dots,i_L$. For all indices even and for all but two indices even, the leading polynomials $P_{L-3}$ are known in a literature for a long time~\cite{tutte_1962}. For arbitrary values of indices and any $L$, a general expression for $P_{L-3}$ was derived only recently \cite{Bouttier}. A general expression for the subleading polynomials in \re{G-gen} remains unknown. 
Luckily, for the purpose of computing the first few terms of $1/N^2$ expansion of the free energy \re{1.4}, we only need the correlators \re{G-gen} of finite length $L\le 6$. The corresponding expressions are summarized in Appendix~\ref{App:gauss}.

Replacing the correlators with their expressions \re{G-gen}, we find that the terms  in \re{3.3} containing the product $J_{i_1} \dots J_{i_p}$ are suppressed by the factor of $1/N^{p-2}$. Therefore, computing the free energy \re{F-SA} and \re{F-Q2} to order $O(1/N^{2(g-1)})$ we are allowed to retain in the exponent of \re{3.3} only the first $2g$ terms. 

An additional simplification occurs after we take into account the relation \re{Z-ZZ}. For the function $Z(J^-)=Z(-J^-)$ it leads to
\begin{align}\label{Z-Q}
Z(J^-) = e^{\frac12  J_i^-J_j^-  \Q_{ij}^- + \frac1{4!}   J_i^- J_j^- J_k^- J_l^- \Q_{ijkl}^-+\frac1{6!} J_i^- J_j^- J_k^- J_l ^- J_n^- J_m^- \Q_{ijklnm}^-+O(1/N^6)}\,,
\end{align}
where the exponent only involves even powers of $J^-_i=J_{2i+1}$. Here the notation was introduced for the correlators with odd indices
\begin{align}\label{Q-}
\Q_{i_1\dots i_n}^- = G_{2i_1+1,\dots, 2i_n+1} = \vev{\OO_{2i_1+1}\dots \OO_{2i_n+1}}_c\,.
\end{align}
For the product of $Z-$functions in \re{F-Q2} we get in a similar manner
\begin{align}\notag\label{ZZ-Q}
& Z(J^-,J^+) Z(-J^-,-J^+) 
\\[2mm]
& =e^{J_i^-J_j^-  \Q_{ij}^- +J_i^+ J_j^+  \Q_{ij}^+ + \frac1{12}   J_i^- J_j^- J_k^- J_l^- \Q_{ijkl}^-+ \frac1{12}   J_i^+ J_j^+ J_k^+ J_l^+ \Q_{ijkl}^++ \frac1{2} J_i^+ J_j^+ J_k^- J_l^- \Q_{ijkl}^{+ -}+O(1/N^4)}\,,
\end{align}
where $J^+_i=J_{2i}$ and the following notation was introduced 
\begin{align}\notag\label{Q+}
& \Q_{i_1\dots i_n}^+ = G_{2i_1,\dots, 2i_n} = \vev{\OO_{2i_1}\dots \OO_{2i_n}}_c\,,
\\[2mm]
& \Q_{ijkl}^{+ -} = G_{2i,2j, 2k+1,2l+1} = \vev{\OO_{2i}\OO_{2j} \OO_{2k+1}\OO_{2l+1}}_c\,.
\end{align}
As follows from their definition \re{Q-} and \re{Q+}, the functions $ \Q_{i_1\dots i_n}^+$ and $ \Q_{i_1\dots i_n}^-$ are completely symmetric in their indices, whereas $ \Q_{ijkl}^{+ -}$ is symmetric with respect to the first and second pair of indices.

\subsection{Diagrammatic technique}

Substituting \re{Z-Q} into \re{F-SA} we can express the free energy in the $\SA$ model as an integral over the auxiliary $J-$fields. This integral can be evaluated using a Feynman diagram technique. 

\begin{figure}[t!]
\begin{center}
\includegraphics[width=0.9\textwidth]{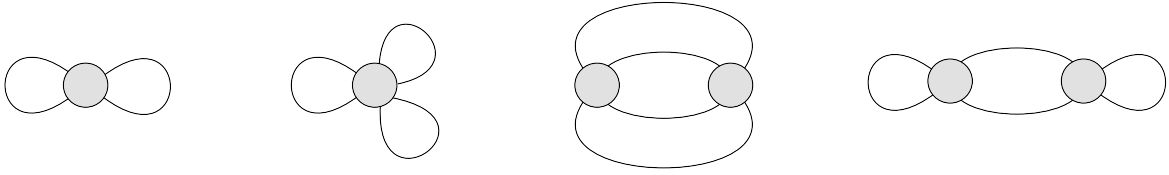}
\end{center}
\caption{Diagrammatic representation of various terms in \re{3.9} involving the product of $X^-$ and $\Q$ matrices. Solid lines denote the propagators  \re{X-} and grey disks represent the $\Q-$matrices.}\label{fig}
\end{figure}

To this end, we combine together the two factors in the integrand of \re{F-SA} and identify the quadratic form in the exponent as defining a propagator of the auxiliary field
\begin{align}\label{X-}
X_{ij}^-\equiv \vev{J_i^- J_j^-} = \left[C^-(1-\Q^-C^-)^{-1}\right]_{ij}\,.
\end{align}
Here semi-infinite matrix $C^-_{ij}$ is defined in \re{25} and $\Q^-_{ij}$ is the two-point correlator in a Gaussian matrix model \re{Q-}.
The remaining terms in the exponent of \re{Z-Q}, proportional to the matrix model correlators $\Q_{i_1\dots i_n}^-$, define interaction vertices with $n\ge 4$ outgoing lines. Then, the free energy in the $\SA$ model is given by the sum of vacuum diagrams involving an arbitrary number of interaction vertices as shown in Figure~\ref{fig}. 
Their contribution to the difference free energy \re{F-SA} is given by
\begin{align}\notag\label{3.9}
\Delta F_{\SA} {}& =  \frac12 \log \det(1-\Q^-C^-) - \frac1{8} X_{ij}^-X_{kl}^- \, \Q_{ijkl}^- - \frac1{48}X_{ij}^-X_{kl}^-X_{nm}^-\,\Q_{ijklnm} ^-
\\ \notag
& -\frac1{48}  X_{i_1 i_2}^- X_{j_1 j_2}^-X_{k_1 k_2}^- X_{l_1 l_2}^- \Q_{i_1j_1k_1l_1}^- \Q_{i_2j_2k_2l_2} ^-
\\
&  -\frac1{16}  X_{i_1j_1}^- X_{i_2 j_2}^- X_{k_1 k_2}^- X_{l_1l_2}^- \Q_{i_1j_1k_1l_1}^-  \Q_{i_2j_2k_2l_2}^-  +O(1/N^{6}) \, .
\end{align}
Notice that the semi-infinite matrices $X^-$ and $\Q^-$ have an expansion in powers of $1/N^2$ and, therefore, each term in \re{3.9} generates a series in $1/N^2$.

In the $\QQ$ model, the coefficient in front of the quadratic term in the exponent of \re{ZZ-Q} is two times bigger as compared with that in \re{Z-Q}. To use the 
same propagator as in \re{X-} it is convenient to rescale auxiliary fields in \re{F-Q2} as $J^-_i\to J^-_i/\sqrt{2}$ and $J^+_i\to J^+_i/\sqrt{2}$. Then, the propagator of 
$J_+$ is defined similar to \re{X-}
\begin{align}\label{X+}
X_{ij}^+\equiv \vev{J_i^+ J_j^+} =\left[ C^+(1 -\Q^+C^+)^{-1}\right]_{ij}\,.
\end{align}
The resulting expression for the difference free energy in the $\QQ$ model \re{F-Q2} is
\begin{align}\notag\label{dF-sum}
\Delta F_{\QQ} {}& = \frac12 \log \det(1-\Q^-C^-) + \frac12 \log \det(1-\Q^+ C^+) 
\\
{} & - \frac1{16} X_{ij}^+X^+_{kl} \Q_{ijkl}^{+}- \frac1{16} X_{ij}^-X^-_{kl} \Q_{ijkl}^{-}
- \frac1{8} X_{ij}^+X^-_{kl} \Q_{ijkl}^{+-}
+ O(1/N^4)\,.
\end{align}
The first four terms in this relation are analogous to those in \re{3.9}. The last term proportional to $\Q_{ijkl}^{+-}$ takes into account the interaction of $J^+$ and $J^-$ fields.

In the next section, we first evaluate the individual terms in \re{3.9} and \re{dF-sum} and, then, compute the difference free energy \rf{1.4} in the two models. 

\section{Large $N$ expansion of the free energy}
\la{sec:3}

The relations \re{3.9} and \re{dF-sum} express the difference free energy in the $\SA$ and $\QQ$ models in terms of two sets of semi-infinite matrices $\Q^\pm$ and $C^\pm$ defined in \re{Q-}, \re{Q+} and \re{25}, respectively. The former are given by  correlators in a Gaussian matrix model that are independent of the coupling constant. The latter are fixed by the localization to be nontrivial functions of the coupling constant and are independent of $N$. 

The relations \re{3.9} and \re{dF-sum} are valid for any coupling constant.  
It follows from \re{25} that $C_{ij}^\pm =O(\lambda^{i+j+1})$ at weak coupling and, therefore, the matrices $C^\pm$ can be replaced in \re{3.9} and \re{dF-sum} by their finite dimensional minors. Expanding \re{3.9} and \re{dF-sum}  in powers of $C^\pm$ one can reproduce the weak coupling expansion \re{SA-weak} and \re{Q2-weak}.

 At strong coupling, we have to find an efficient way to deal with the product of semi-infinite matrices in \re{3.9} and \re{dF-sum}. It proves convenient to think about semi-infinite matrices $C^\pm$ and $X^\pm$ as matrix elements of certain integral operators. In this way, the product of these matrices can be cast into a convolution of the corresponding integral kernels. We show in this section that the integral operators defined by $C^\pm$ and $X^\pm$ coincide with the temperature dependent (or truncated) Bessel operator. We exploit this property in Section~\ref{sect:res} to work out the strong coupling expansion  
 of the difference free energy \re{3.9} and \re{dF-sum}.
 
\subsection{Leading order} 

At large $N$, the leading contribution to \re{3.9} and \re{dF-sum} comes from terms involving $\log\det(1-\Q^\pm C^\pm)$. 
The contribution of the remaining terms  
is suppressed by powers of $1/N^2$
in virtue of \re{Q-}, \re{Q+} and \re{G-gen}. 

Let us first summarize the properties of the matrices $\Q_{ij}^\pm$ that enter the leading term in \re{3.9} and \re{dF-sum}. According to 
\re{Q-} and \re{Q+}, they are given by two-point correlators in a Gaussian matrix model
\begin{align}
\Q^-_{ij} = \vev{\OO_{2i+1}\OO_{2j+1}}\,,\qqqquad
\Q^+_{ij} = \vev{\OO_{2i}\OO_{2j}}-\vev{\OO_{2i}}\vev{\OO_{2j}}\,.
\end{align}
To the first few orders in $1/N^2$ they take the following form
\begin{align}\label{Q2}\notag
\Q^-_{ij} & = \beta_i^- \beta_j^- \left[{2\over i+j+1}+ \frac{1}{6 N^2} (i^2+i j+j^2-5    (i+j)-13) +O(1/N^4)\right]\,,
\\
\Q^+_{ij} & = \beta_i^+ \beta_j^+ \left[{2\over i+j}+ \frac{1}{6 N^2} (i^2+i j+j^2-8    (i+j)+ 7) +O(1/N^4)\right]\,,
\end{align}
where normalization factors are
\begin{align}\label{beta}
\beta_i^+=2^{1/2-i} {\Gamma(2i)\over \Gamma^2(i)} \,,\qqqquad
\beta_i^-=2^{-1-i}  {\Gamma(2i+2)\over \Gamma(i+2)\Gamma(i)}\,.
\end{align}
These relations match \re{G-gen} for $L=2$. 

The leading $O(N^0)$ term in \re{Q2} admits the following representation
\begin{align}\label{Q-UU}
\Q^\pm = U^\pm (U^\pm)^t + O(1/N^2) \,, 
\end{align}
where $U^\pm$ are lower triangular matrices and $(U^\pm)^t $ denote transposed matrices. 
The explicit expression  for  these matrices can be found in \re{U-mat}.~\footnote{The matrices $U^\pm$ have a simple interpretation in the Gaussian matrix model as they allow us to construct the basis of orthonormal traces $\widehat \OO_{i}$ satisfying $\vev{\widehat \OO_{i} \widehat \OO_{j} }=\delta_{ij}+O(1/N^2)$ (see \re{ortho}).}
We apply the matrices $U^\pm$ to define 
\begin{align}\label{K-C}
K^\pm = (U^\pm)^t C^\pm U^\pm \,.
\end{align}
A distinguished feature of these matrices is that they admit an integral representation in terms of Bessel functions, see \re{K-B} and \re{chi} below. Combining together \re{Q-UU} and \re{K-C} we obtain
\begin{align}\label{logdet-LO}
\log\det(1-\Q^\pm C^\pm) = \log\det(1-K^\pm) + O(1/N^2)\,.
\end{align}
 As we will see in a moment, the expression on the right-hand side coincides with a Fredholm determinant of the Bessel operator.  

To summarize, the leading corrections to the difference free energy \re{3.9} and \re{dF-sum}  in the $\SA$ and $\QQ$ models are given by
\begin{align}\notag\label{F-LO}
& \sF_{0}^{\SA} = \frac12 \log\det(1-K^-)\,,
\\ 
& \sF_{0}^{\QQ} = \frac12 \log\det(1-K^-)+\frac12 \log\det(1-K^+)\,.
\end{align}
 
\subsection{Subleading corrections}\label{sect:sub} 

To compute subleading corrections to \re{logdet-LO} in $1/N^2$, we generalize \re{Q-UU} as
\begin{align}\label{Q-R}
\Q^\pm = U^\pm \mathsf R^\pm (U^\pm)^t  \,,
\end{align}
where $\mathsf R^\pm= 1 + O(1/N^2)$. 
To determine the (matrix) coefficients in the expansion of $\mathsf R^\pm$ in powers of $1/N^2$, we apply an inverse transformation to the matrices \re{Q2}. 

To any order in $1/N^2$, the correlators \re{Q2} are given by a sum of terms of the form $\beta_i \beta_j  i^n j^m $
\begin{align}\label{Q-c}
\Q^\pm= U^\pm (U^\pm)^t + {1\over N^2} \beta_i \beta_j  \sum_{n,m\ge 0} c_{nm}^\pm   i^n j^m\,,
\end{align}
where the expansion coefficients $c_{nm}^\pm$ are series in $1/N^2$ with rational coefficients. They can be found by matching \re{Q-c} to the expression \re{Q2} of the two-point correlators $\Q^\pm$. For instance, for $N\to\infty$ the only nonzero entries are
\begin{align}\label{cc}
c_{02}^\pm = c_{11}^\pm =\tfrac16\,, \qquad
c_{01}^+=-\tfrac43 \,,\qquad c_{01}^-=-\tfrac56 \,,\qquad c_{00}^+=\tfrac76\,,\qquad c_{00}^-=-\tfrac{13}6\,.
\end{align}
and $c_{nm}^\pm=c_{mn}^\pm$.

Combining together \re{Q-R} and \re{Q-c}, we find    
\begin{align}\label{R}
\mathsf R^\pm = 1 + {1\over N^2} \sum_{n,m} c_{nm}^\pm R_{nm}^\pm \,.
\end{align}
As compared 
to  \re{Q-c}, each term of the sum gets replaced by a matrix $R_{n,m} =  R_n \otimes R_m$ defined as
\begin{align}\label{RxR}
 (R_{nm})_{ij} = (R_n)_i (R_m)_j\,,\qqqquad (R_n)_i = \sum_{k\ge 1} U_{ik}^{-1} k^n\beta_k\,.
\end{align}
To simplify the formula, we do not display here the superscript `$\pm$'. The explicit expressions for the vectors $R_n^\pm$ can be found in Appendix~\ref{app:mat}. 

It follows from \re{R} and \re{cc} that the semi-infinite matrices $\mathsf R^\pm$ are given to order $O(1/N^2)$ by
\begin{align}\notag
&\mathsf R^+ = 1 + {1\over 6N^2} \left[ R_{0,2}^+ +R_{2,0}^+ + R_{1,1}^+ - 8 R_{1,0}^+ - 8 R_{0,1}^+ +7 R_{0,0}^+\right] + O(1/N^4)\,,
\\
& \mathsf R^- = 1 + {1\over 6N^2} \left[ R_{0,2}^- +R_{2,0}^- + R_{1,1}^- -5 R_{1,0}^--5 R_{0,1}^- -13 R_{0,0}^-\right] + O(1/N^4)\,.
\end{align}
The main advantage of dealing with matrices $\mathsf R^\pm$ is that they allow us to express the subleading corrections to the free energy \re{3.9} and \re{dF-sum} in terms of the same matrices $K^\pm$ that appeared  at  the leading order \re{F-LO}. 

To show this we 
take into account \re{K-C}, \re{Q-R} and \re{R} to obtain  
\begin{align}\notag
\log\det(1-\Q^\pm C^\pm) &=  \log\det(1-\mathsf R^\pm K^\pm)
\\
& =  \log\det(1-K^\pm)
+\log\det\Big(1-{1\over N^2}\sum_{n,m} c^\pm_{nm} R_{n,m}^\pm {K^\pm\over 1-K^\pm}\Big)\,.
\end{align}
Expanding the second term on the right-hand side in powers of $R^\pm$ and using a factorized form of this matrix, $R_{nm} = R_n\otimes R_m$, we get 
\begin{align}\label{logdet-fin}
\log\det(1-\Q^\pm C^\pm) =  \log\det(1-K^\pm) -\sum_{L\ge 1} {1\over L N^{2L}} c^\pm _{n_1,m_1} W^\pm_{m_1,n_2} c^\pm_{n_2,m_2} \dots  c^\pm_{n_L,m_L}  W^\pm_{m_L,n_1}\,,
\end{align}
where the notation was introduced for the  scalar quantity
\begin{align}\label{W}
W^\pm_{nm} = (R^\pm_n{)}_i \lr{K^\pm\over 1-K^\pm}_{ij} (R^\pm_m{)}_j\,,\qqqquad (n,m\ge 0)\,.
\end{align}
It is symmetric in indices, $W^\pm_{nm}=W^\pm_{mn}$, depends on the  't Hooft coupling $\lambda$ but is independent of $N$. 
Expansion of $W^\pm_{nm}$ at strong coupling is derived in Section~\ref{sect:res}.

The relation \re{logdet-fin} can be used to systematically expand $\log\det(1-\Q^\pm C^\pm)$ in powers of $1/N^2$. In particular, the $O(1/N^2)$ correction to \re{logdet-fin} is given by
\begin{align}\notag \label{1/N2}
&  \frac12 \log\det(1-\Q^+ C^+) \Big|_{O(1/N^2)}=   -\frac{7  }{12}W^+_{0,0} +\frac{4
    }{3}W^+_{0,1}-\frac{1}{6}W^+_{0,2}-\frac{1}{12}W^+_{1,1}\,,
    \\
& \frac12 \log\det(1-\Q^- C^-) \Big|_{O(1/N^2)}=\frac{13  }{12}W^-_{0,0}+\frac{5
    }{6}W^-_{0,1}-\frac{ 1}{6}W^-_{0,2}-\frac{1}{12}W^-_{1,1}\,.
\end{align}

Let us now examine the remaining terms in \re{3.9} and \re{dF-sum}. They contain the product of matrices $X^\pm$ whose indices are contracted with
the $\Q-$tensors, \eg~$X_{ij}^-X^-_{kl} \Q_{ijkl}^{-}$ and $X_{ij}^+X^-_{kl} \Q_{ijkl}^{+-}$.
Applying \re{K-C} and \re{Q-R}, we can express the matrices $X^\pm$ defined in \re{X-} and \re{X+}, in terms of the Bessel matrices \re{K-pm}
\begin{align}\label{UXU}
 X^\pm = (U^{\pm\, t})^{-1}  K^\pm(1-\mathsf  R^\pm K^\pm)^{-1} (U^\pm)^{-1}\,.
\end{align}
According to \re{G-gen}, \re{Q-} and \re{Q+}, the $\Q-$tensors are proportional to the product of $\beta-$factors and
certain polynomials in indices. As a result, $X_{ij}^-X^-_{kl} \Q_{ijkl}^{-}$ and $X_{ij}^+X^-_{kl} \Q_{ijkl}^{+-}$ are given by a sum of terms each of which factorizes into a product of 
terms of the form
\begin{align}\notag\label{Xbb}
i^n j^m \beta_i^\pm \beta_j^\pm X^\pm_{ij} {}&= (R_n)_i  (K^\pm(1-\mathsf  R^\pm K^\pm)^{-1} )_{ij} (R_m)_j
\\
{}&=W^\pm_{nm} + \sum_{L\ge 1}{1\over N^{2L}} W^\pm_{nn_1} c^\pm _{n_1m_1}W^\pm_{m_1n_2} \dots c^\pm_{n_L m_L} W^\pm_{m_L m}\,.
\end{align}
Here in the first relation we took into account \re{RxR} and in the second relation expanded the right-hand side in powers of $\mathsf R^\pm$
and applied \re{R} and \re{W}.  

The relation \re{Xbb} allows us to express various terms in \re{3.9} and \re{dF-sum} in terms of the matrix elements \re{W}.
As an example, the leading $O(1/N^2)$ correction to the second term in \re{3.9} can be computed as
\begin{align}\notag
- \frac1{8} X_{ij}^-X_{kl}^- \, \Q_{ijkl}^- {}&= -(2+i+j) \beta_i^-\beta_j^- X^-_{ij} \beta_k^-\beta_l^- X^-_{kl} + O(1/N^4)
\\[2mm]
{}&
= -2 (W_{0,0}^-+W^-_{0,1}) W^-_{0,0}+ O(1/N^4)\,,
\end{align}
where we replaced $\Q_{ijkl}^-$ with its expression \re{Q-4ind} and applied \re{Xbb}.
 
Combining together the above relations we obtain from \re{3.9} and \re{dF-sum} the $O(1/N^2)$ correction to the difference free energy \re{1.4} in the $\SA$ and $\QQ$ models
\begin{align}\notag\label{F-NLO}
\sF_1^{\SA} ={}& -2 W_{0,0}^-(W_{0,0}^-+W^-_{0,1})+\frac{13  }{12}W^-_{0,0}+\frac{5
    }{6}W^-_{0,1}-\frac{ 1}{6}W^-_{0,2}-\frac{1}{12}W^-_{1,1}\,,
\\[2mm]\notag
\sF_1^{\QQ} ={}& -(W^+_{0,0}+W^-_{0,0})(W^+ _{0,1}+W^-_{0,1})
   -(W^-_{0,0})^2+\frac{1}{4} (W^+
   _{0,0})^2 
 \\ &  
   +\frac{13
    }{12}W^-_{0,0}+\frac{5
    }{6}W^-_{0,1}-\frac{1}{6}W^-_{0,2}-\frac{1}{12}W^-_{1,1} -\frac{7  }{12}W^+_{0,0} +\frac{4
    }{3}W^+_{0,1}-\frac{1}{6}W^+_{0,2}-\frac{1}{12}W^+_{1,1}\,.
\end{align} 
Notice that the expression for $\sF_1^{\QQ}$ contains mixed terms proportional to the product of $W^+$ and $W^-$. They come from  
$X_{ij}^+X^-_{kl} \Q_{ijkl}^{+-}$ term in \re{dF-sum}.

The same technique can be used to determine subleading $O(1/N^4)$ correction to the free energy \re{1.4}. 
Going through the calculation we obtain in the $\SA$ model
\begin{align}\notag\label{F-NNLO}
\sF_2^{\SA}  &=-\frac{64}{3} W_{0,0}^4-\frac{128}{3} W_{0,1} W_{0,0}^3-\frac{10}{3}
   W_{1,1} W_{0,0}^3-18 W_{0,1}^2 W_{0,0}^2  
   + \frac{11}{3}W_{0,0}^3-\frac{86}{3} W_{0,1} W_{0,0}^2-\frac{49}{3} W_{0,2} W_{0,0}^2
\\ \notag {}&    
   -W_{0,3}
   W_{0,0}^2-\frac{31}{3} W_{1,1} W_{0,0}^2-\frac{10}{3} W_{1,2} W_{0,0}^2-\frac{131}{3}
   W_{0,1}^2 W_{0,0}-13 W_{0,1} W_{0,2} W_{0,0}-\frac{37}{3} W_{0,1} W_{1,1}
   W_{0,0}
\\\notag {}&    
   -\frac{25}{3}W_{0,1}^3    
   -\frac{133}{144}
   W_{0,0}^2+\frac{499}{36} W_{0,1} W_{0,0}+\frac{307}{36} W_{0,2} W_{0,0}-\frac{1}{6}
   W_{0,3} W_{0,0}-\frac{1}{6} W_{0,4} W_{0,0}+\frac{443}{72} W_{1,1}
   W_{0,0}
\\\notag {}&    
   +\frac{41}{36} W_{1,2} W_{0,0}-\frac{1}{2} W_{1,3} W_{0,0}-\frac{25}{72}
   W_{2,2} W_{0,0}+\frac{77}{6}W_{0,1}^2-\frac{49}{72}W_{0,2}^2-\frac{37
   }{144}W_{1,1}^2+\frac{77}{36} W_{0,1} W_{0,2} 
\\\notag {}&    
   -W_{0,1} W_{0,3}+\frac{131}{36} W_{0,1}
   W_{1,1}-W_{0,2} W_{1,1}-\frac{73}{36} W_{0,1} W_{1,2}   
   + \frac{37}{60} W_{0,0}+\frac{41}{180}W_{0,1}-\frac{151}{120}W_{0,2}-\frac{31
   }{240} W_{0,3}
\\\notag {}&    
   +\frac{1}{10}W_{0,4}-\frac{1}{144}W_{0,5}-\frac{151}{240}W_{1,1}-\frac{251
   }{360}W_{1,2}+\frac{71 }{240}W_{1,3}-\frac{1}{48}W_{1,4}+\frac{5}{24}W_{2,2}-\frac{29
   }{720}W_{2,3} \,,
\\ {}&   
\end{align}
where $W_{n,m}\equiv W_{n,m}^-$.

We would like to emphasize that the relations \re{F-LO}, \re{F-NLO} and \re{F-NNLO} hold for an arbitrary 't Hooft coupling. We verified that at weak coupling they are in agreement with \re{SA-weak} and \re{Q2-weak}.

\subsection{Relation to the Bessel operator}
In what follows we shall use the notation 
\be \la{ggg}
g \equiv {\sql \ov 4 \pi}  \ .
\ee
Let us show that  the semi-infinite matrices \re{K-C} are given by matrix elements of the truncated Bessel operator $\bm K_\ell$. This operator is defined as
\begin{align}\label{Bes} 
\bm K_\ell  f(x) &= \int_0^\infty dy\, K_\ell(x,y) \chi\lr{\sqrt y\over 2g} f(y)\,,\end{align}
where $f(x)$ is a test function and  
 $K_\ell(x,y)$ is expressed in terms of the Bessel functions (hence the name of the operator)
\begin{align}\label{K-sum} 
K_\ell(x,y) &=\sum_{i\ge 1} \psi_i(x) \psi_i(y) 
= {\sqrt{x} J_{\ell+1}(\sqrt x)J_\ell(\sqrt y)-\sqrt{y} J_{\ell+1}(\sqrt y)J_\ell(\sqrt x)\over 2(x-y)}\,.
\end{align}
Here $\ell$ is an arbitrary positive real parameter and $\psi_i(x)$ is an orthonormal basis of functions
\begin{align}\notag\label{psi}
& \psi_i(x) = (-1)^i \sqrt{2i+\ell-1} {J_{2i+\ell-1}(\sqrt x)\over \sqrt x}\,,
\\
& \vev{\psi_i |\psi_j} = \int_0^\infty dx\, \psi_i(x) \psi_j(x) =\delta_{ij}\,.
\end{align}
The function $\chi(x)$ is conventionally called the ``symbol'' of the Bessel operator.
In what follows we choose this function as
\begin{align}\label{chi}
\chi(x) = - {1\over\sinh^2(x/2)}\,.
\end{align}
It vanishes at infinity and truncates the integral in \re{Bes} at $y=O(g^2)$. 

The Bessel operator \re{Bes} can be realized as a semi-infinite matrix on the space of functions spanned by $\psi_i(x)$ 
\begin{align}
\bm K_\ell \, \psi_i(x) = (K_\ell)_{ij} \psi_j(x)\,.
\end{align}
Its matrix elements $ (K_\ell)_{ij} = \vev{\psi_i|\bm K_\ell |\psi_j} $  are given by 
\begin{align}\label{K-B}
 (K_\ell)_{ij} = (-1)^{i+j}\sqrt{2i+\ell-1}\sqrt{2j+\ell-1}\,\int_{0}^{\infty}\frac{dt}{t} \,J_{2i+\ell-1}(\sqrt{t})\,J_{2j+\ell-1}(\sqrt{t})\chi\lr{\sqrt t\over 2g} \,.
\end{align}
where $i,j \ge 1$.

The reason for the choice of the symbol function \re{chi} is that the resulting matrix $K_\ell$ coincides  for $\ell=1$ and $\ell=2$ with the semi-infinite matrices $K^\pm$ defining the leading contribution to the free energy \re{F-LO} ,
\begin{align}\label{K-pm}
 K^+=K_{\ell=1}\,,\qqqquad K^-=K_{\ell=2}\,.
\end{align}
Being combined with  \re{F-LO}, this relation implies that the $O(N^0)$ contribution to the difference free energy in $\SA$ and $\QQ$ models is given by a Fredholm determinant of the Bessel operator
\begin{align}\notag \label{F-Fr}
& \sF_{0}^{\SA} = \frac12 \log\det(1-\bm K_{\ell=2})\,,
\\ 
& \sF_{0}^{\QQ} = \frac12 \log\det(1-\bm K_{\ell=1})+\frac12 \log\det(1-\bm K_{\ell=2})\,.
\end{align}
For the special choice of the symbol $\chi(x)=\theta(1-x)$, the Fredholm determinant of the Bessel operator coincides with 
the  celebrated Tracy-Widom distribution describing statistics of the spacing of the eigenvalues in Laguerre ensemble \cite{Tracy:1993xj}. In application to the $\SA$ and $\QQ$ models, we encounter the symbol of the form \re{chi}. 

Strong coupling expansion of \re{F-Fr} was derived in \cite{Beccaria:2022ypy} using the technique developed in \cite{Belitsky:2019fan,Belitsky:2020qrm,Belitsky:2020qir}. It relies on the relation
\begin{align}\label{logdet-str}
\log {}& \det  (1-\bm K_{\ell}) = \pi g - \frac12 \lr{2\ell-1} \log g + B_\ell + \frac18(2\ell-3)(2\ell-1) \log (g'/g) 
\\[2mm]\notag
& + (2\ell-5)(2\ell-3)(4\ell^2-1) {\zeta(3)\over 2048 \pi^3 g'{}^3} - (2\ell-7)(2\ell-5)(4\ell^2-9)(4\ell^2-1) {3\zeta(5)\over 262144 \pi^5 g'{}^5} + \dots\ . 
\end{align}
Here $g$ is defined in \re{ggg},  \  $g'=g-{\log 2/\pi}$  and dots denote subleading corrections suppressed by powers of $1/g$ as well as exponentially small $O(e^{-4\pi g})$ corrections.
Expanding again the series in powers of $1/g$, one can produce terms proportional to powers of $\log 2/\pi$. The constant term $B_\ell$ in \re{logdet-str}, conventionally called the Widom-Dyson constant, is given by
\begin{align}
B_\ell=-6 \log \mathsf A +\frac12 +\frac16\log 2 -\ell \log 2 + \log \Gamma(\ell)\,, 
\end{align}
where $\mathsf A$ is the Glaisher's constant. 
Substituting \re{logdet-str} into \re{F-Fr} we arrive at 
\begin{align}\notag\label{F0s}
& \sF_{0}^{\SA} = \frac{\pi  g}{2} -\frac{3
   \log g}{4}-3 \log \mathsf A+\frac{1}{4}-\frac{11 \log 2\, }{12} +\frac{3}{16} \log \frac{ g'}{g}  -\frac{15 \zeta (3)}{4096
   (\pi g')^3}-\frac{945 \zeta (5)}{524288 (\pi g')^5} +\dots\,,
\\[2mm]
& \sF_{0}^{\QQ}= \pi  g-\log g-6 \log \mathsf A +\frac{1}{2}-\frac{4 \log 2\, }{3}+\frac{1}{8} \log \frac{g'}{g}  -\frac{3 \zeta (3)}{2048 (\pi g')^3}-\frac{135
   \zeta (5)}{262144 (\pi g')^5} +\dots\,,
\end{align}
where 
$g'$ is defined in \re{logdet-str}.  For $g={\sqrt{\lambda}/ 4\pi}$ these relations coincide with \re{1.5} and \re{1.6}.

Subleading corrections to the free energy \re{F-NLO} and \re{F-NNLO} involve the quantities $W_{nm}^\pm$ defined in \re{W}. To establish the relation between $W_{nm}^\pm$ and the Bessel operator, it is convenient to 
introduce auxiliary functions $\phi_n^\pm  (x)$ with $n\ge 0$
\begin{align}\label{phi-n}
\phi_n^\pm  (x) = \sum_{i\ge 1} (R^\pm_n{)}_i \, \psi_i^\pm (x)\,,
\end{align}
where $\psi_i^+(x)$ and $\psi_i^-(x)$ coincide with \re{psi} for $\ell=1$ and $\ell=2$, respectively, and the expansion coefficients $(R^\pm_n{)}_i$ are defined in \re{RxR}.
The matrices \re{K-pm} and \re{K-B} admit the representation
\begin{align}
K^\pm_{ij} = \int_0^\infty dt\,  \psi_i^\pm (t) \chi\lr{\sqrt t\over 2g} \psi_j^\pm (t)\,.
\end{align}
Their product can be expressed using \re{K-sum} as a convolution of the Bessel kernels
\begin{align} \notag
[(K^\pm)^L]_{ij} = \int_0^\infty  {}& dt_1\dots dt_L\, \psi_i^\pm (t_1)\chi\lr{\sqrt t_1\over 2g}
\\
{}&\times  K^\pm(t_1,t_2)
   \chi\lr{\sqrt t_{2}\over 2g} \dots    K^\pm(t_L,t_{L-1})  \chi\lr{\sqrt t_{L}\over 2g}\psi_j^\pm (t_L)\,,
\end{align}
where $K^\pm(x,y)$ is the kernel \re{K-sum} evaluated at $\ell=1$ and $\ell=2$. This relation can be written in a compact form as
\begin{align}\label{K^L}
[(K^\pm)^L]_{ij} = \vev{\psi_i^\pm |\bm{\chi} \,(\bm K^\pm)^{L-1} | \psi_j^\pm}\ , 
\end{align}
where the operator $\bm{\chi}$ has a diagonal kernel $\delta(x-y)\chi(\sqrt x/(2g))$.

We apply the relation \re{K^L} to obtain the following representation of \re{W}
\begin{align} \label{W-K}
W^\pm_{nm} {}& =  (R^\pm_n{)}_i 
  \vev{\psi_i^\pm |\bm{\chi} \,{1\over 1-\bm K^\pm} | \psi_j^\pm}(R^\pm_m{)}_j
   = \vev{\phi_n^\pm |\bm{\chi} {1\over 1-\bm K^\pm} | \phi_m^\pm} \,,
\end{align} 
where in the second relation we used \re{phi-n}.
Thus, the quantities $W_{nm}^\pm$ are given by  matrix elements of the resolvent of the Bessel operator with respect to the special states $\phi_n^\pm(x)$ defined in \re{phi-n}. 

According to \re{phi-n}, the functions $\phi_n^\pm(x)$ are given by infinite sums of the Bessel functions \re{psi} with the expansion coefficients \re{RxR}. These sums can evaluated in a closed form leading to (see Appendix~\ref{app:mat})
\begin{align}\notag\label{phi-J}
& \phi_n^-(x) = -\frac1{2\sqrt2} (x \partial_x)^n J_2(\sqrt x)\,,
\\
& \phi_n^+(x) = -\frac1{2\sqrt2}\sum_{i=0}^n 2^{i-n} \lr{n\atop i}(x \partial_x)^iJ_1(\sqrt x)\,.
\end{align}

To summarize, we demonstrated in this section that non-planar corrections to the free energy admit 
a compact representation \re{F-NLO} and \re{F-NNLO}  in terms of the matrix elements $W^\pm_{nm}$  of the resolvent \re{W-K} of the truncated Bessel operator.
In the next section, we develop a technique for computing $W^\pm_{nm}$ and, then, apply it to 
derive the strong coupling expansion of the free energy \re{F-NLO} and \re{F-NNLO}.
 
\section{Resolvent of the Bessel operator}\label{sect:res}

The matrix elements \re{W-K} involve the Bessel operator \re{Bes} for $\ell=1$ and $\ell=2$ (see \re{K-pm}). To treat them in a unified manner, we generalize \re{W-K} to arbitrary $\ell$ and define the matrix elements of the resolvent of the Bessel operator \re{Bes}
\begin{align}\label{w-nm}
w_{nm}= \vev{\phi_n|\bm{\chi} {1\over 1-\bm K_\ell} |\phi_m}\,,
\end{align}
where the functions $\phi_n(x)$ (with $n\ge 0$) are given by
\begin{align}\label{phi}
\phi_0(x) = J_\ell(\sqrt x)\,,\qqqquad 
\phi_n(x)= (x\partial_x)^n \phi_0(x)\,.
\end{align}
The operator $\bm{\chi}$ is defined in \re{K^L},  it acts on a test function as $\bm{\chi} f(x) = \chi(\sqrt x/(2g))f(x)$. 

In the previous section, we encountered different matrix elements \re{W-K}.  As follows from \re{W-K} and \re{phi-J}, they are given by a linear combination of $w_{nm}$ evaluated for $\ell=1$ and $\ell=2$
\begin{align}\label{W-w}\notag
{}& W_{nm}^-=\frac18 w_{nm}\Big|_{\ell=2}
\,,
\\[2mm]
{}& W_{nm}^+=\frac18 \sum_{i=0}^n\sum_{j=0}^m 2^{i+j-n-m} \lr{n\atop i}\lr{m\atop j}w_{ij}\Big|_{\ell=1}\,.
\end{align} 

The matrix elements \re{w-nm} depend on the integer $\ell$ and the coupling constant $g=\sqrt\lambda/(4\pi)$. They 
have the following important properties. Expanding \re{w-nm} in powers of $\bm K_\ell$ and taking into account the definition \re{Bes} of the Bessel operator, we find that $w_{nm}$ are symmetric in indices  
\begin{align}
w_{mn}=w_{nm}\,.
\end{align}
Besides, the matrix elements \re{w-nm} are not independent. We show in Appendix~\ref{app:method} 
that $w_{nm}$ satisfy a functional equation 
\begin{align}\label{fun}
\lr{\frac12 g\partial_g-1}w_{nm} = \frac14w_{0n} w_{0m} +w_{n+1,m} + w_{n,m+1}\,.
\end{align}
For lowest values of $n,m\ge 0$ it leads to
\begin{align}\notag\label{w01}
w_{01} {}&= -\frac18 w_{00}^2 + \lr{\frac14 g\partial_g-\frac12} w_{00}\,,
\\\notag
w_{11} {}&= -w_{02} -\frac14 w_{01}w_{00} + \lr{\frac12 g\partial_g-1} w_{01}\,,
\\
w_{12} {}&= -\frac{1}{8} w_{01}^2+ \lr{\frac{1}{4} g \partial_g -\frac12 }w_{11}\,, \qquad \dots \,.
\end{align}
These relations hold for an arbitrary coupling. They allow us to express $w_{01}$, $w_{11}$ and $w_{12}$ in terms of $w_{00}$ and $w_{02}$. 
Examining the relation \re{fun} for arbitrary $m$ and $n$, we find that the matrix elements $w_{nm}$  can be expressed in terms of independent quantities $w_{00},w_{02},w_{04},\dots $. We discuss them in the next subsection.

\subsection{Method of differential equations}

The matrix element $w_{00}$ is related to the Fredholm determinant of the Bessel operator (see  \cite{Belitsky:2019fan,Belitsky:2020qrm,Belitsky:2020qir})
\begin{align}\label{w00}
w_{00} = -2 g\partial_g \log \det (1-\bm K_\ell)\,.
\end{align}
Its expansion at strong coupling follows from \re{logdet-str}
\begin{align}\notag\label{w00-str}
w_{00} = {}& -2 \pi  g+(2\ell-1)-\frac{(2 \ell -3) (2 \ell -1) \log 2}{4 \pi  g}-\frac{(2 \ell -3) (2 \ell
   -1) \log^2 2}{4 (\pi g)^2}
\\   
 {}&  +\frac{(2 \ell -3) (2 \ell -1) \left(3 \zeta (3) (2 \ell -5) (2 \ell +1)-256 \log^3 2\right)}{1024 (\pi
   g)^3}+\dots\,,
\end{align}
 where $g=\sqrt{\lambda}/(4\pi)$. Substituting this expression into the first relation in \re{w01}, we can obtain the strong coupling expansion of $w_{01}$. 
 
The strong coupling expansion of $w_{0n}$ can be found by applying the method of differential equations~\cite{Korepin:1993kvr,Tracy:1993xj}. 
It is based on the following identify \cite{Belitsky:2019fan} (see \re{delta-w} in Appendix~\ref{app:method})
\begin{align}\label{dw} 
\partial_g w_{0n} {}&= \int_0^\infty dx\, Q_0(x) Q_n(x) \partial_g \chi\lr{\sqrt x\over 2g}
\,,
\end{align}
where the functions $Q_n(x)$ (with $n\ge 0$) are matrix elements of the resolvent of the Bessel operator
\begin{align}\label{Qn-def}
Q_n(x) = \vev{x|{1\over 1-\bm K_\ell} |\phi_n }\,.
\end{align}
It is tacitly assumed that $Q_n(x)$ also depends on the coupling $g$.

We show in Appendix~\ref{app:method} that the functions $Q_n(x)$ satisfy recurrence relations 
\begin{align}\label{Q-rec}
Q_{n+1}(x) = -\frac14 Q_0(x) w_{0n}(g) + \frac12 \lr{g\partial_g +2 x\partial_x} Q_n(x)\,.
\end{align}
They can be used to express $Q_n(x)$ for $n\ge 1$ in terms of $Q_0(x)$. In its turn, the function $Q_0(x)$
satisfies a second-order partial differential equation~\cite{Belitsky:2020qrm,Belitsky:2020qir}
\begin{align}\label{pde}
\Big[(g\partial_g + 2 x\partial_x)^2 +
x-\ell^2+(1-g \partial_g)w_{00}\Big] Q_0(x)=0\,.
\end{align}
It involves a nontrivial function of the coupling $w_{00}$ given by \re{w00} and \re{w00-str}. 
A solution to the differential equation \re{pde} at strong coupling   is described in Appendix~\ref{app:method}.
Being combined with \re{Q-rec}, it allows us to expand the integral on the right-hand side of \re{dw} in powers of $1/g$ and, then,  
obtain the strong coupling expansion of $w_{0n}$.

In this way we obtain  
\begin{align} \notag  
w_{02}{}& =\frac{(\pi g)^3}{4}+\frac{1}{8} (\pi g)^2 (2 \ell -1)+\pi g \left(-\frac{  \ell
   ^2}{2}-\frac{1}{32}   (2 \ell -3) (2 \ell -1) \log
   2 \right)
\\ \notag {}&   
   +\left(\frac{1}{32} (2 \ell -1) \left(4 \ell ^2+4 \ell +3\right)+\frac{1}{32} (2 \ell -3) (2 \ell -1) \log 2\,  (2
   \ell +1-\log 2\, )\right)+O(1/g)
\,, 
\\[2mm] 
\label{w04-as}\notag
w_{04} {}&=   
   -\frac{23 (\pi g)^5}{64}-\frac{5(\pi g)^4}{128}(2 \ell -9)+{(\pi g)^3\over 8} \left(\left(\ell ^2-2 \ell +2\right)+\frac{5}{64} (2 \ell -3) (2 \ell -1) \log 2\right)
\\&   
   +{(\pi g)^2\over 256} \left( \left(12 \ell ^2-12 \ell +11\right) (2 \ell -1)
   -\frac{1}{2}(2 \ell -3) (2 \ell -1)(2(2\ell-7) \log 2- 5\log^22)\right)+O(g)\,.
\end{align}
We recall that all other matrix elements $w_{nm}$ follow unambiguously from the functional relations \re{fun}.

Examining the relations \re{w00-str}  and \re{w04-as}, we observe that the matrix elements $w_{0n}$ behave at strong coupling as a power of $g$
\begin{align}\label{w0-dsl}
w_{0n} = \omega_{0n}\, g ^{n+1}+ O(g^{n})\,.
\end{align}
It follows from \re{fun} that $w_{nm}$ have a similar behaviour
\begin{align}\label{w1-dsl}
w_{nm} = \omega_{nm}\,  g ^{n+m+1}+ O(g^{n+m})\,.
\end{align}
The explicit expressions for the leading coefficients $\omega_{nm}$ are given by relations \re{gen-fun} -- \re{G(x,y)} below. 
Most importantly they are independent of $\ell$.~\footnote{A logarithm of the Fredholm determinant of the Bessel operator \re{logdet-str} has the same property.}  

We recall  that the free energy in the $\SA$ and $\QQ$ models is expressed in terms of the matrix elements 
\re{W-w} evaluated at $\ell=1$ and $\ell=2$. The fact that the leading asymptotic behaviour of $w_{nm}$ at strong coupling is independent of $\ell$
suggests that the free energy in the two models should be related to each other. Indeed, we establish such a relation in Section~\ref{sect:proof} below.

\subsection{Next-to-leading corrections to the free energy}

We are now ready to compute the subleading corrections to the (difference) free energy \re{F-NLO}. 

Applying the relations \re{W-w} we can express $F_1^{\SA}$ and $F_1^{\QQ}$ in terms of the matrix elements $w_{nm}$ 
\begin{align}\label{F1-1}\notag
F_1^{\SA} ={}&   -{1\over 32} w_{00}^-(w_{00}^-+w_{01}^-)+ \frac{13 }{96}w_{00}^-+\frac{5 }{48}w_{01}^--\frac{1}{48}w_{02}^--\frac{1}{96}w_{11}^-\,,
\\[2mm] \notag
F_1^{\QQ}={}&  -\frac{1}{64}
   (w^+_{01}+w^-_{01})(w^+_{0,0}+
   w^-_{00}) -\frac{1}{128} w^+_{00}w^-_{00} -\frac{1}{64}(w^-_{00})^2-\frac{1}{256} (w^+_{00})^2   
\\ &   
+\frac{w^+_{00}}{384}
+\frac{13
  }{96} w^+_{01}
   -\frac{1}{48}w^+_{02}-\frac{1}{96}w^+_{11} 
+\frac{
   13}{96}w^-_{00} +\frac{5
  }{48}w^-_{01}-\frac{1}{48}w^-_{02}-\frac{1}{96}w^-_{11} \,,
\end{align}
where $w_{nm}^+$ and $w_{nm}^-$ are given by $w_{nm}$ for $\ell=1$ and $\ell=2$, respectively.

Replacing the matrix elements with their expressions \re{w00-str}, \re{w04-as} and \re{w01}, we obtain the strong coupling expansion of the difference free energy in the two models 
  \begin{align}\notag\label{F1-SA}
F^{\SA}_1
& = -\frac{(\pi g)^3}{32}-\frac{3 (\pi g)^2}{128}+\pi g \left(\frac{11}{512}-\frac{\log 2\, }{32}
   \right)    +\left(\frac{3}{128}-\frac{\log ^22}{32}-\frac{\log 2\, }{512}\right)
\\\notag &   
   +\lr{-\frac{15 \zeta (3)}{8192}-\frac{\log ^32}{32}-\frac{3 \log ^22}{512}+\frac{279 \log 2\, }{8192}}{1\over \pi g}
\\  &   
   +\lr{\frac{105 \zeta (3)}{131072}-\frac{15 \zeta (3) \log 2\, }{2048}-\frac{\log ^42}{32}-\frac{5 \log ^32}{512}+\frac{441 \log
   ^22}{8192}}{1\over (\pi g)^2}+ O(1/g^3)\,,
\\[2mm]
\notag\label{F1-Q2}
F_1^{\QQ}&=-\frac{(\pi g)^3}{16}-\frac{(\pi g)^2}{192}+\pi g \left(\frac{5}{256}-\frac{\log 2\, }{48}
     \right)+\left(\frac{\log 2\, }{256}-\frac{\log
   ^22}{48}\right)
\\\notag &   
   +\lr{-\frac{3 \zeta (3)}{4096 }-\frac{\log ^32}{48}+\frac{\log ^22}{256}+\frac{35 \log 2\, }{4096}}{1\over \pi g}
\\ &   
   +\lr{\frac{21 \zeta
   (3)}{65536}-\frac{3 \zeta (3) \log 2\, }{1024}-\frac{\log ^42}{48}+\frac{\log ^32}{256}+\frac{53 \log ^22}{4096}}{1\over (\pi g)^2}+ O(1/g^3)\,.
\end{align}
For $g=\sqrt{\lambda}/(4\pi)$ these relations coincide with \re{1.8} and \re{1.10}.

The following comments are in order.

A close examination of \re{F0s} shows that the leading $O(N^0)$ terms of the strong coupling expansion \re{1.4} of the free energy in the $\SA$ and $\QQ$ models differ by the factor of $2$. It follows from \re{F1-SA} and \re{F1-Q2} that the same relation holds at order $O(1/N^2)$
\begin{align}\label{fact-2}
{F_0^{\SA}\over F_0^{\QQ}}= \frac12 + O(1/g)\,,\qqqquad 
{F_1^{\SA}\over F_1^{\QQ}}= \frac12 + O(1/g)\,.
\end{align}
This relation is yet another manifestation of universality of matrix elements mentioned in the previous subsection. We show in Section~\ref{sect:proof} that it holds to any order of $1/N^2$ expansion \re{1.4}.

It is important to emphasize that the relations \re{fact-2} only hold at strong coupling. At weak coupling, one uses \re{SA-weak} and \re{Q2-weak} to verify that the ratios of functions differ already at order $O(g^2)$
\begin{align}
{F_0^{\SA}\over F_0^{\QQ}}= {10\zeta(5)\over 3\zeta(3)} g^2  + O(g^4)\,,\qqqquad 
{F_1^{\SA}\over F_1^{\QQ}}= {50\zeta(5)\over 3\zeta(3)} g^2  + O(g^4)\,.
\end{align}

Higher order corrections to \re{F1-SA} and \re{F1-Q2} involve powers of $\log2$. The leading order $O(N^0)$ functions 
$F_0^{\SA}$ and $F_0^{\QQ}$ given by
\re{F0s} have the same property. In this case, terms containing $\log 2\, $ can be absorbed into redefinition of the coupling constant
$g'=g-{\log 2/ \pi}$. It turns out that the functions $F_1^{\SA}$ and $F_1^{\QQ}$ have the same property. 

Indeed, it is easy to see from \re{F1-SA} and \re{F1-Q2} that terms with the maximal power of $\log2$ to each order in $1/g$ form a geometrical progression. As a consequence, the relations 
 \re{F1-SA} and \re{F1-Q2} can be written in the following form
 \begin{align}\notag
F^{\SA}_1
& =- \lr{\frac{(\pi g)^3}{32}-\frac{(\pi g)^2}{128}} -{(\pi g)^2\over 32}{g\over g'} +\dots\,,
\\[2mm]
F_1^{\QQ}&=- \lr{\frac{(\pi g)^3}{16}-\frac{(\pi g)^2}{64}}-{(\pi g)^2\over 48}{g\over g'}+ \dots\,,
\end{align}
where dots denote the remaining terms. Notice that the expressions inside the curly brackets satisfy the relation \re{fact-2}.

\subsection{Next-to-next-to-leading correction to the free energy}

The $O(1/N^4)$ correction to the difference free energy \re{1.4} in the $\SA$ model is given by \re{F-NNLO}. We use the first relation in \re{W-w} 
and replace the matrix elements $w_{nm}$ with their expressions to get
\begin{align}\notag\label{F2-SA}
F^{\SA}_2{}& =\frac{(\pi g)^6}{720}-\frac{(\pi g)^5}{1280}+(\pi g)^4 \left(\frac{\log
   2}{240} -\frac{251}{30720}\right)
\\   \notag
{}&   +(\pi g)^3 \left(\frac{\log ^22}{240} -\frac{47 \log 2\, }{3840}+\frac{107}{49152}\right)
\\
{}&   
   +(\pi g)^2 \left(\frac{ \zeta
   (3)}{4096}+\frac{\log ^32}{240}-\frac{19
   \log ^22}{1280}-\frac{191\log
   2}{16384}+\frac{409}{32768}\right)+O(g)\,.
\end{align}
For $g=\sqrt{\lambda}/(4\pi)$ this relation coincides with \re{1.9}.

We observe that the terms with the maximal power of $\log2$ can be again eliminated through the redefinition of the coupling
\begin{align}
F^{\SA}_2=\frac{(\pi g)^6}{720}-\frac{19(\pi g)^5}{3840}+\frac{(\pi g)^5}{240}{g\over g'}+\dots\,,
\end{align}
where $g'=g-{\log2 / \pi}$.

The relations \re{F0s}, \re{F1-SA} and \re{F2-SA} define the corrections to the difference free energy \re{1.4} in the $\SA$ and $\QQ$ models at strong coupling.

\section{Double scaling limit}
\la{sec:5}

As follows from \re{F1-SA} and \re{F2-SA}, the subleading corrections to the free energy \re{1.4} in the $\SA$ model exhibit an interesting scaling behaviour at strong coupling, $F_n^{\SA} = O(g^{3n})$ for $n=1,2$. 

This suggests to consider the double scaling limit 
\begin{align}\label{dsl}
 N\to\infty\,, \qquad
g\to\infty,\qquad
g^3/N^2=\text{fixed}\,.
\end{align}
In this limit  we retain the leading $O(g^{3n})$ terms in the expression for $F_n^{\SA}$ to arrive at the following remarkably simple result
\begin{align}\notag\label{F-dsl}
\Delta F^{\SA} &=F^{\SA}_0+{1\over N^2} F^{\SA}_1 + {1\over N^4} F^{\SA}_2 + \dots
\\
& \simeq \frac12 \pi g -{1\over N^2}\frac{(\pi g)^3}{32}+{1\over N^4}   \frac{(\pi g)^6}{720}+ \dots\,,
\end{align}
where `$\simeq$' denotes the limit \re{dsl}.

Moreover, taking into account the relation \re{fact-2}, we expect that, in the double scaling limit, the free energy in the $\QQ$ model differs from \re{F-dsl} by the factor of $2$
\begin{align}\label{rat2}
\Delta F^{\QQ}\simeq 2 \Delta F^{\SA}\,.
\end{align}
In this section  we elucidate the meaning of the double scaling limit \re{dsl} and prove the relation \re{rat2}.

\subsection{Gaussian correlators in the double scaling limit}\label{sect:proof} 

We recall that, in the matrix model representation \re{F-SA} and \re{F-Q2}, the dependence of the free energy \re{1.4} on $1/N^2$ is generated by non-planar corrections to the correlators \re{Z} and \re{ZZ} in a Gaussian matrix model. 
The main observation is that in the double scaling limit \re{dsl}, the dominant contribution to the free energy 
\re{3.9} and \re{dF-sum} comes from the correlators \re{G-gen} with large indices.
Indeed, for $i_p=O(N^{2/3})$ (with $p=1,\dots,L$) all terms inside the brackets in \re{G-gen} scale at large $N$ as  $O(N^{2(L-3)/3})$.
As we show below, it is this property that leads to the scaling behaviour of the free energy \re{F-dsl} for $g=O(N^{2/3})$.

As was mentioned above, the correlators \re{G-gen} with even and odd indices are described by two different functions, see \eg~\re{Q-} and \re{Q+}. It turns out that for large values of indices these functions coincide. Indeed, one can see from \re{Q2} that this is true for 
the functions $\Q_{ij}^+$ and $\Q_{ij}^-$  for $i,j=O(N^{2/3})$ as $N\to\infty$. The same property holds for $L-$point correlators. 

It can be understood by writing the correlators \re{G-gen} as integrals over eigenvalues in the Gaussian matrix model (see \eg~\re{OO-int}). It is well-known that the distribution density of the eigenvalues in this model has a finite support. In the limit of large indices, the dominant contribution to the correlator \re{G-gen} comes from integration close to the edge of the spectrum. The distribution density of eigenvalues has remarkable universal properties in this region. This allows one to determine the correlators \re{G-gen} for $i_p=O(N^{2/3})$ in a closed form.  For instance, the two-point correlators are given 
in this limit by, see e.g. \cite{Dijkgraaf:1990qw,Eynard:2021zcj}
\begin{align} \label{Q-ee}
\Q^-_{ij} = 2\beta^-_{i} \beta^-_{j}  e^{i^3+j^3\over 12 N^2}\sum_{k=0}^\infty 
{1\over (2k+1)!!}{(i+j)^{k-1} (ij)^k \over 2^k N^{2k}}\,.
\end{align}
One can verify that $O(N^0)$ and $O(1/N^2)$ terms in this relation are in agreement with \re{Q2} for $i,j =O(N^{2/3})$.
For $L-$point correlator, the analogous expressions are known to the first four orders of $1/N^2$ expansion \cite{Eynard:2021zcj,eynard2022new}. As we show below, they can be used to compute the subleading corrections to the free energy \re{F-dsl} in the double scaling limit at order $O(1/N^{10})$. 

We can exploit a universality of the correlators $\Q^+_{i_1,\dots,i_L}$ and $\Q^-_{i_1,\dots,i_L}$ for $i_p=O(N^{2/3})$ to establish the relation between the free energy in the $\SA$ and $\QQ$ models. Let us examine the relations \re{F-SA} and \re{F-Q2} in the limit when all matrix indices are large. According to \re{25}, the matrix elements $C_{ij}^-$ can be obtained from $C_{ij}^+$ by replacing the indices $i\to i+\frac12$ and $j\to j+\frac12$. In the limit of large $i$ and $j$ we can neglect this shift and identity the two matrices. In the similar manner, we identify $\Q^+$ and $\Q^-$ tensors in \re{Z-Q} and \re{ZZ-Q} to get from \re{F-SA} and \re{F-Q2}
\begin{align}\notag\label{V-repr}
& e^{-\Delta F^{\SA}} =  (\det C)^{-1/2} \int  dJ^- \, e^{- \frac12J_i^- J_j^- (C^{-1}-\Q)_{ij}+V(J^-)}\,,
\\
& e^{-\Delta F^{\QQ}} = (\det C)^{-1}\int  dJ^- dJ^+ \, e^{- (J_i^- J_j^-+J_i^+ J_j^+) (C^{-1}-\Q)_{ij}+V(J^++J^-)+V(J^+-J^-)}\,.
\end{align}
Going from \re{F-SA} and \re{F-Q2}, we separated  the terms quadratic in $J$'s in the exponent of  \re{Z-Q} and \re{ZZ-Q}   and absorbed the remaining terms into the potential 
\begin{align}\label{V}
V(J) =   \frac1{4!}   J_i J_j J_k J_l \Q_{ijkl}+\frac1{6!} J_i J_j J_k J_l  J_n J_m \Q_{ijklnm}+O(1/N^6)\,.
\end{align}
Changing the integration variables in the second relation \re{V-repr} as $J^\pm {}'=(J^+\pm J^-)$ we observe that the integrals over $J^+{}'$ and $J^-{}'$ factorize leading to
\begin{align}
e^{-\Delta F^{\QQ}} \simeq e^{-2\Delta F^{\SA}}\,.
\end{align}
We would like to emphasize that this relation only holds in the double scaling limit \re{dsl}. 

\subsection{Free energy in the double scaling limit}

As follows from the calculation presented in the previous section, the free energy in the $\SA$ and $\QQ$ models takes a remarkably
simple form \re{F-dsl} and \re{rat2} in the double scaling limit. In this subsection we use the representation \re{V-repr} of the partition function in this limit to extend the relations \re{F-dsl} and \re{rat2} to order $O(1/N^{10})$. 

The calculation of \re{V-repr} can be significantly simplified by replacing the correlators $\Q_{i_1i_2\dots}$ with their leading expressions in the double scaling limit (see \re{A.27}).
In addition, the corrections to the free energy \re{F-NLO} and \re{F-NNLO} depend on matrices $W_{nm}^\pm$ defined in \re{W-K}. According to  \re{W-w}, these matrices are given by linear combinations  of matrix elements $w_{nm}$ which have the scaling behaviour \re{w0-dsl} and \re{w1-dsl} at strong coupling. As a consequence,
in the double scaling limit \re{dsl} the relations \re{W-w} and \re{w1-dsl} can be simplified as
\begin{align}\label{W-w1}
 W_{nm}^-=W_{nm}^+ \simeq \frac18 w_{nm}  = \frac18 \omega_{nm}g^{n+m+1} + O(g^{n+m})\,,
\end{align}
where we took into account that the leading coefficients $ \omega_{nm}$ are independent of $\ell$.

Applying the relation \re{W-w1}, we observe that various terms in \re{F-NLO} and \re{F-NNLO} have different behaviour in $g$. In the double scaling limit, we can retain the terms with the maximal power of $g$ and neglect remaining ones. For instance, the first relation in \re{F-NLO} simplifies as
\begin{align}
\sF_1^{\SA} ={}& -2 W_{0,0} W_{0,1}-\frac{ 1}{6}W_{0,2}-\frac{1}{12}W_{1,1} + O(g^2)\,,
\end{align}
where we suppressed the superscript of $W_{nm}^\pm$ in virtue of \re{W-w1}. 

As follows from \re{F-NLO}, \re{F-NNLO} and \re{W-w1}, the subleading $O(1/N^{2L})$ corrections to the free energy \re{1.4} are given in the double-scaling limit by a multilinear combination of $w$'s, schematically 
\begin{align}\label{F1-sum-W} 
\sF_L=  \sum  f_{n_1\dots n_{2p}} \,w_{n_1n_2} \dots w_{n_{2p-1} n_{2p}} = O(g^{3L}) \,,
\end{align}
where the sum runs over non-negative integers $n_i\ge 0$ satisfying the condition $\sum_i (n_i+\frac12)=3L$.

To illustrate \re{F1-sum-W} we use the first relation in \re{V-repr} to obtain the following representation of $\Delta F_{\SA}$ in the double scaling limit
\begin{align}\label{F-split}
\Delta F^{\SA}  = \Delta F^{\SA,(0)} +  \Delta F^{\SA,\rm (int)} \,.
\end{align}
Here $\Delta F^{\SA,(0)}$ takes into account the contribution of quadratic in $J$ terms in the exponent of \re{V-repr}. It is given by the integral \re{V-repr} with the potential $V(J^-)$ put to zero
\begin{align}\label{F0-split}
\Delta F^{\SA,(0)}=\frac12 \log \det(1-\Q C)\,.
\end{align}
The second term in \re{F-split} yields the contribution of the interaction terms described by the potential \re{V}  
\begin{align}\notag\label{DeltaF-V}
\Delta F_{\SA} ^{\rm (int)}  ={}& -\frac{\vev{J^4}}{24 N^2}-{1\over N^4}\Big({\frac{ \vev{J^4J^4}- \vev{J^4}^2}{1152} + \frac{\vev{J^6}}{720}}\Big)  
\\[1.5mm]
{}&-{1\over N^6} \Big( { \frac{
\vev{J^4J^4J^4}-3\vev{J^4}\vev{J^4 J^4}+2\vev{J^4}^3 }{82944}+\frac{ \vev{J^4 J^6}-\vev{J^4}\vev{J^6}}{17280}+\frac{\vev{J^8}}{40320}}\Big) + \dots, 
\end{align}
where $\vev{J^{2p_1}J^{2p_2}\dots}$ denotes an expectation value of $J^{2p}= N^{2p-2} J_{i_1} \dots J_{i_{2p}} \Q_{i_1\dots i_{2p}}$ (with $p=2,3,\dots$) with respect to a Gaussian measure in \re{V-repr}. The factor of $N^{2p-2}$ was inserted to ensure that $J^{2p}=O(N^0)$ at large $N$. In a close analogy with \re{3.9}, the right-hand side of 
\re{DeltaF-V} can be expanded over the product of `propagators' 
\begin{align}
X_{ij} \equiv \vev{J_i J_j} =\left[ C(1 -\Q C)^{-1}\right]_{ij}\,,
\end{align}
whose indices are
contracted with the $\Q-$tensors. Going through the same steps as in Section~\ref{sect:sub}, we can express \re{DeltaF-V} in terms of matrix elements \re{W-w1}.

The relation \re{F0-split} admits an expansion \re{logdet-fin} in powers of $1/N^2$. As was explained above, the matrices with `$\pm$' superscripts coincide in the double scaling limit and for this reason we suppress this superscript in what follows. The relation \re{logdet-fin} involves the coefficients $c_{nm}$ defined in \re{Q-c}. Their values can be found to any order in $1/N^2$ by matching \re{Q-c} to the exact expression for the correlator \re{Q-ee}. To leading order in $1/N^2$ they are given by \re{cc}. Going through the calculation and taking into account \re{W-w1} and \re{F-LO} we obtain from \re{logdet-fin} 
\begin{align}\label{F-logdet}\notag
\Delta F^{\SA,(0)}{}&=   \sF_{0}^{\SA}  -\frac{1}{N^2}\lr{\frac{w_{0,2}}{48}+\frac{w_{1,1}}{96}}
\\
{}& -
\frac{1}{N^4}\Big(\frac{w_{0,2}^2}{4608}+\frac{w_{1,1}^2}{9216}+\frac{w_{0,5}}{1152}+\frac{w_{0,1}
   w_{1,2}}{2304}+\frac{w_{1,4}}{384}+\frac{w_{0,0} w_{2,2}}{4608}+\frac{29
   w_{2,3}}{5760}\Big)+ \dots
\end{align}
where $\sF_{0}^{\SA}$ is given by \re{F0s}. It is straightforward to expand \re{F-logdet} to any order in $1/N^2$.  The resulting expressions are lengthy and we do not present them here to save space. It is easy to see that 
the coefficients in front of powers of $1/N^2$ in \re{F-logdet} have the expected form \re{F1-sum-W}. The coefficient of $1/N^2$ in 
\re{F-logdet} agrees with \re{1/N2} up to subleading correction in $1/g$. 
 
The expressions for the matrix elements $w_{nm}$ at strong coupling \re{W-w1} are defined by the set of parameters $\omega_{nm}$. It is convenient to define their generating functions 
\begin{align}\label{gen-fun}
G(x)=\sum_{n\ge  0} \omega_{0n}\, x^{-n}\,,\qqqquad 
G(x,y) = \sum_{n,m\ge  0} \omega_{nm}\, x^{-n} y^{-m}\,,
\end{align}
where $G(x,y)=G(y,x)$.  
As we argue  in Appendix~\ref{app:method},  the function $G(x)$   should   be given by 
\begin{align}\label{G(x)}
G(x) & = 8\pi \bigg[  {\Gamma^2(\frac12 -\frac{x}{2\pi}) \over \Gamma^2(-\frac{x}{2\pi})}+{x\over 2\pi}\bigg].
\end{align}
It is well defined for $x<0$ and its expansion at large negative $x$ generates the coefficients $\omega_{0n}$.
It follows from \re{fun} that the two functions in \re{gen-fun} are related to each other as
\begin{align} \label{G(x,y)}
 (x+y) G(x,y) = -\frac14 G(x) G(y)+  x G(y)+ y G(x)\,.
\end{align}
Being combined together the relations \re{gen-fun}, \re{G(x)} and \re{G(x,y)} allows us to determine the parameters $\omega_{nm}$ and, as a consequence, the matrix elements \re{w0-dsl} and \re{w1-dsl}. For instance,
\begin{align}
w_{00} = -2\pi g \,,\qquad w_{01} = -\frac12 (\pi g)^2\,,\qquad w_{11} = -\frac12 (\pi g)^3\,,\qquad
w_{02} = \frac14 (\pi g)^3\,.
\end{align}
These relations are valid up to corrections suppressed by powers of $1/g$. We check that they are in agreement with \re{w00-str} and \re{w04-as}.

Replacing $w_{nm}$ in \re{F-logdet} with their expressions we find after some algebra
\begin{align} \label{logdet-2}
\Delta  F^{\SA,(0)}={} & \frac{1}{2}\pi g -\frac{7 (\pi g)^6 }{92160 N^4}-\frac{221 (\pi  g)^9 }{743178240 N^6}+\frac{21253 (\pi g)^{12}}{23781703680 N^8}+\frac{18670639 (\pi g)^{15}}{45660871065600 N^{10}} +O(1/N^{12})\,,
\end{align}
where we added the additional terms in $1/N^2$ expansion as compared with \re{F-logdet}.
This relation holds in the double scaling limit \re{dsl}. The $O(1/N^2)$ term is absent in \re{logdet-2} due to the relation $w_{02}=-w_{11}/2+ O(g^2)$.

In a similar manner, repeating the calculation of \re{DeltaF-V} we get in the double scaling limit
\begin{align}\label{F-V}
  \Delta F^{\SA,\rm (int)}=-\frac{(\pi g)^{3}}{32N^{2}}+\frac{3(\pi g)^{6}}{2048N^{4}}-\frac{2077(\pi g)^{9}}{11796480N^{6}}-\frac{147997(\pi g)^{12}}{2642411520N^{8}}
+\frac{754343579 (\pi g)^{15}}{15220290355200N^{10}}+O({1/N^{12}})\,,
\end{align}
where compared to  \re{logdet-2}  the expansion starts at order $O(1/N^2)$.

Finally, adding together \re{logdet-2} and \re{F-V},
 we obtain the following expression for the difference free energy in the double-scaling limit
\begin{align}\label{F-fin} 
\Delta F^{\SA} &=  \frac{1}{2}\pi g-\frac{(\pi g)^{3}}{32N^{2}}+\frac{(\pi g)^{6}}{720N^{4}}-\frac{(\pi g)^{9}}{5670N^{6}}-\frac{(\pi g)^{12}}{18144N^{8}}
+\frac{17 (\pi g)^{15}}{340200N^{10}}+O(1/N^{12})\,.
\end{align}
This relation is one of the main results of this paper.
Compared to  \re{F-dsl}, it contains three additional terms. 

Notice that
the expansion coefficients of the series \re{logdet-2} and \re{F-V} are rather complicated rational numbers but their sum is remarkably simple. We believe that this property is not accidental and hints at the existence of hidden properties of the free energy \re{F-fin} in the double scaling limit.

 In a close analogy with 
the known solution of two-dimensional quantum gravity and $c<1$ noncritical strings 
(for a review, see e.g. \cite{DiFrancesco:1993cyw}),  
one might expect that the free energy in the double scaling limit satisfies a certain nonlinear differential equation. This will open up an  exciting possibility to compute the free energy of the $\SA$ and $\QQ$ models non-perturbatively, to any order in $1/N$.

\section{Circular Wilson loop} 
\la{sec:6}

Let us now apply the technique developed in the previous sections to compute non-planar corrections to expectation value of the circular  half-BPS Wilson loop in the $\SA$ and $\QQ$ models.

In the $\SA$ model the Wilson loop is defined as  (see also \ci{Beccaria:2021vuc})
\be\label{W-def}
W^{\SA} = \VEV{\tr\,\mc P\,\exp\Big\{\gym\,\oint ds\Big[i\,A_{\mu}(x) \dot  x^{\mu}(s)+\frac{1}{\sqrt 2}(\phi(x)+\phi^{*}(x)) \Big]\Big\}},
\ee
where the gauge field $A_{\mu}$ and scalar field  $\phi(x)$ from the $\N=2$ vector multiplet
 are integrated along a circle of unit radius, $x_\mu^2(s)=1$ and $\dot x_\mu^2(s)=1$. 
In the $\QQ$ model, the Wilson loop $W^{\QQ}$ is given by the same expression \re{W-def} 
where $A_\mu$ and $\phi$  correspond to  one of the two $SU(N)$ $\N=2$ vector multiplets. 

The  large $N$ expansion of  the Wilson loop in the $\SA$ and $\QQ$ models  has  the form  \re{622}. 
Due to the planar equivalence of these models with $\mathcal N=4$ SYM theory, the leading term of the expansion $\sW_{0}$ coincides with the analogous expression in the latter theory.
Below we present the  results for the subleading corrections $\sW_{1}(\l)$ and $\sW_{2}(\l)$ in \re{622}.

In the localization approach, the Wilson loop in both models can be represented as the matrix model expectation values
\begin{align}
\la{8.2}
 W^{\SA} = {\vev{\tr e^{\sqrt{\lambda\over 2N}A} e^{S_{\rm int}(A)}}\over\vev{e^{S_{\rm int}(A)}}} \,,\qqqquad 
 W^{\QQ} = {\vev{\tr e^{\sqrt{\lambda\over 2N}A_1}e^{S_{\rm int}(A_1,A_2)}}\over \vev{e^{S_{\rm int}(A_1,A_2)}}} \,,
\end{align}
where the average is taken with the same Gaussian measure as in (\ref{2.8}).
Expanding the exponential functions in powers of the  matrices, we get 
\begin{align}
W= N + \sum_{n\ge 1} {1\over (2n)!} \lr{\lambda\over 2 }^n{\vev{ \OO_{2n}e^{S_{\rm int}}}\over\vev{e^{S_{\rm int}}}},
\end{align}
where the single-trace operators  $\OO_{2n}$ are defined in \re{O}. 

\subsection{Non-planar corrections}

The expectation value $\vev{ \OO_{2i}e^{S_{\rm int}}}$ can be evaluated in the same manner as the difference free energy  $\vev{e^{S_{\rm int}}}=e^{-\Delta F}$, see Section~\ref{sect:HS}.
To accommodate for $\OO_{2i}$ inside the expectation value, we can differentiate the generating function \re{ZZ} with respect to the $J_i^+$. In this way, we get from 
\re{F-SA} and \re{F-Q2}
\begin{align}\label{W-uni}
&W^{\SA} =\mathcal N_{\SA} \int  dJ^-  \, e^{- \frac12J_i^- J_j^- (C^-)_{ij}^{-1}}\Big[1+  \sum_{n\ge 1} {1\over (2n)!} \lr{\lambda\over 2 }^n {\partial\over\partial J_n^+} \Big] Z(J^-,J^+)\Big|_{J^+=0}\,,
\\
&W^{\QQ} =\mathcal N_{\QQ}  \int  dJ^- dJ^+   \, e^{-  J_i^- J_j^- (C^-)_{ij}^{-1}-  J_i^+ J_j^+ (C^+)_{ij}^{-1}} Z(-J^-,-J^+)\Big[1+  \sum_{n\ge 1} {1\over (2n)!} \lr{\lambda\over 2 }^n {\partial\over\partial J_n^+} \Big] Z(J^-,J^+)\,, \no 
\end{align}
where the normalization factors $\mathcal N_{\SA}$ and $\mathcal N_{\QQ}$ are such that both integrals equal $N$ after one neglects the derivatives inside the brackets. Here in the first relation one has to put $J_n^+=0$ after applying the derivative because the interaction potential in the $\SA$ model \re{int-terms1} only involves $\OO_i$ with odd indices. In the second relation, the derivatives act on the generating function corresponding to one of the nodes of the $\QQ$ model. 
 
Replacing the generating function $Z(J^-,J^+)$ with its expression \re{3.3} in terms of the connected correlation function \re{3.4} we get from \re{W-uni}
\begin{align}\label{W-gen}
 {W} = N+ \sum_{n\ge 1} {1\over (2n)!} \lr{\lambda\over 2 }^n & \Big[G_{2n} + \frac1{2!}   \langle J_{i}J_{j}\rangle G_{2n, ij }+ \frac1{3!}   \langle J_{i}J_{j}J_k \rangle G_{2n,ijk}
+ \frac1{4!}  \langle J_{i}J_{j}J_{k}J_{l} \rangle G_{2n,ijkl} +\dots \Big],
\end{align}
where $\vev{J_{i}J_{j}\dots }$ denotes the average with respect to the measure \re{F-SA} or \re{F-Q2} with $J_{2k}=J_k^+$ and $J_{2k+1}=J_k^-$. Due to the symmetry of the integration measure under $J_i\to -J_i$, the terms in \re{W-gen} with odd number of $J$'s vanish. At large $N$, the connected correlators scale as $G_{2n, i_1\dots i_L}=O(1/N^{L-1})$ and their contribution to \re{W-gen} takes the expected form \re{622}. 

We recall that the auxiliary fields $J_i$ were introduced to linearize the double-trace interaction term \re{HS}. The first term inside the brackets in 
\re{W-gen} is independent of these fields. It arises from evaluating \re{8.2} for $S_{\rm int}=0$ and coincides with the expectation value of the circular Wilson loop in $\mathcal N=4$ SYM theory  
\begin{align}\label{8.10}
 W^{\mathcal N=4} = N+ \sum_{n\ge 1} {1\over (2n)!} \lr{\lambda\over 2 }^n  G_{2n} \,. 
\end{align}
 Using known results for the correlators $G_{2n}$ in the  Gaussian matrix model, $ W^{\mathcal N=4}$ can be found exactly for arbitrary $\lambda$ and $N$ in terms of a Laguerre polynomial \cite{Drukker:2000rr}.   
 For our purposes we need its large $N$ expansion 
 \be
\la{8.17}
 W^{\N=4} = N\sW_{0}+\frac{1}{N}\Big(-\frac{\l}{8}\sW_{0}+\frac{\l^{2}}{48}\partial_\lambda\sW_{0} \Big)+\frac{1}{N^{3}}\Big(
\frac{\l ^2 (744+5\l)}{92160}\,\sW_0-\frac{\l^{2}(2+3\l)}{960}\,\partial_\lambda\sW_{0} \Big)+\cdots,
\ee
The leading term is given  in terms of the $\rI_1$ Bessel function \ci{Erickson:2000af}
\begin{align}\label{W0}
\sW_{0}={2 \over \sqrt\lambda}\rI_1(\sqrt\lambda)= \sqrt{2\over\pi}\,\l^{-3/4}\,e^{\sql}\lr{1-{3\over 8\sqrt\lambda} +\cdots}\,, 
\end{align}
where dots denote subleading corrections at strong coupling.

Subtracting the $\mathcal N=4$ result \re{8.10} from \re{W-gen}, we obtain the leading correction to the difference of Wilson loops in the $\SA$ and $\QQ$ models as 
\begin{align}\notag\label{W-W}
& W^{\SA} - W^{\N=4} =\sum_{n\ge 1} {1\over 2(2n)!} \lr{\lambda\over 2 }^n \langle J^-_{i}J^-_{j}\rangle G_{2n, 2i+1,2j+1} +O(1/N^3)\,,
\\
& W^{\QQ} - W^{\N=4} =\sum_{n\ge 1} {1\over 4(2n)!} \lr{\lambda\over 2 }^n  \left[  \langle J^+_{i}J^+_{j}\rangle G_{2n, 2i,2j}+ \langle J^-_{i}J^-_{j}\rangle G_{2n, 2i+1,2j+1}\right]+O(1/N^3).
\end{align}
In the second relation we inserted an additional factor of $1/2$  to take into account the difference between the two-point functions $\vev{J_i J_j}$ in the two models. It arises because
 the coefficient in front of the quadratic term in the exponent of \re{ZZ-Q} is two times  larger  as compared with that in \re{Z-Q}.
Going from \re{W-gen} to \re{W-W}, we separated the sum over even and odd indices and  put $J_i^+=J_{2i}$ to zero in the $\SA$ model.

The connected three-point Gaussian correlators in \re{W-W} are given by
\begin{align}\notag
& G_{2n,2i+1,2j+1}= {1\over N} \beta_i^- \beta_j^- n(n+1)G_{2n} +O(1/N^3) \,,
\\[1.2mm]
& G_{2n,2i,2j}= {1\over N}  \beta_i^+ \beta_j^+ n(n+1)G_{2n}+O(1/N^3) \,,
\end{align}
where $G_{2n}$ is the one-point correlator and  $\beta^\pm_i$ are defined in \re{beta}. Replacing the two-point functions $\langle J^\pm _{i}J^\pm _{j}\rangle$ with their expressions \re{X-} and \re{X+}, we find that both relations in \re{W-W} involve the quantities
\begin{align}
\beta_i^\pm  \beta_j^\pm \langle J^\pm_{i}J^\pm_{j}\rangle =  \beta_i^\pm  \beta_j^\pm X^\pm_{ij} = W_{00}^\pm + O(1/N^2)\,,
\end{align}
where in the last relation we applied \re{Xbb}.
 
Taking into account \re{W-w} we obtain from \re{W-W}
\begin{align}\notag\label{W-W1}
& W^{\SA} - W^{\N=4} ={1\over 8N}  S\,  w_{00}\big|_{\ell=2} +O(1/N^3)\,,
\\
& W^{\QQ} - W^{\N=4} ={1\over 16N}  S \left(w_{00}\big|_{\ell=1}+ w_{00}\big|_{\ell=2} \right)+O(1/N^3)\, , 
\end{align}
where we introduced the notation 
\begin{align}
S=\sum_{n\ge 1} {1\over (2n)!} \lr{\lambda\over 2 }^n n(n+1)G_{2n} =  \lambda \partial_\lambda^2 \lr{\lambda \sW_0} + O(1/N^2) = \frac14 \lambda \sW_0 + O(1/N^2) \,.
\end{align}
Here we applied \re{8.10} and \re{8.17} and used the properties of the leading term \re{W0}.

According to \re{w00} and \re{F-LO}, the matrix elements $w_{00}$ in \re{W-W1} are related to derivatives of the difference free energy. As a consequence, combining the above relations we find from \re{W-W1} that the leading  non-planar 
correction to the difference of the Wilson loops takes the same universal form in the two models
\begin{align}
& W^{\mathsf A} - {W}^{\N=4} =-\frac{\kappa_{\mathsf A} }{4N} \sW_0 \lambda^2\partial_\lambda \sF_{0}^{\mathsf A} +O(1/N^3)\,,\qqqquad {\mathsf A}=\{\SA, \QQ\}\,,
\end{align}
where $\kappa_{\SA}=1$ and $\kappa_{\QQ}=1/2$. Replacing $W^{\N=4}$ with its expansion  \re{8.17}, we determine the leading non-planar correction to the Wilson loop \re{622}
\be\la{619}
\sW_{1}^{\mathsf A} = -\frac{\l}{8}\sW_{0}+\frac{\l^{2}}{48}\partial_\lambda\sW_{0} -\frac{\kappa_{\mathsf A} }{4} \sW_{0} \lambda^2\partial_\lambda\sF_{0}^{\mathsf A}  \,, 
\ee
where $\sW_{0}$ is defined in  \re{W0} and $\sF_{0}^{\mathsf A}$ is the leading correction to the difference free energy \re{1.4} in the $\SA$ and $\QQ$ models. 
We use the strong coupling expansion \re{F1-SA} and \re{F1-Q2}  to get
\begin{align}\notag\label{W-1}
& \sW_{1}^{\SA} = -\sqrt{2\over\pi} \,\frac{\l^{3/4}}{192}\,e^{\sql} \,\lr{1+{69\over 8\sqrt\lambda} + \dots}\,,
\\
& \sW_{1}^{\QQ} = -\sqrt{2\over\pi} \,\frac{\l^{3/4}}{192}\,e^{\sql} \,\lr{1+{117\over 8\sqrt\lambda} + \dots}\,.
\end{align}
Notice that the leading terms in the two expressions coincide leading to 
\begin{align}
{W^{\SA} \over W^{\QQ}} =1 + {\lambda\over 32 N^2} + \dots \,.
\end{align}
This relation should be compared with the analogous relation \re{fact-2} for the free energy. We show below that the Wilson loops in the $\SA$ and $\QQ$ models coincide in the double scaling limit \re{dsl}.
 
At the next order in the $1/N$ expansion, the Wilson loop \re{W-gen}  in the $\SA$ model is given by
\begin{align}\notag\label{8.19}
 W^\SA- W^{\mathcal N=4} =&   \sum_{n\ge 1} {1\over (2n)!} \lr{\lambda\over 2 }^n 
\Big[\frac1{2 }\vev{J_{i}^-J_{j}^-} G_{2n,2i+1,2j+1}  +\frac1{24}   \vev{ J_{i}^-J_{j}^-J_{k}^-J_{l}^-}  G_{2n,2i+1,2j+1,2k+1,2l+1}
\\
&+ \frac1{48} \sum_{i,j,k,l} \vev{J_{p}^-J_{q}^-J_i^- J_j ^-J_k^- J_l^-} \Q_{ijkl}^-G_{2n,2p+1,2q+1}
+\dots \Big],
\end{align}
where $\Q^-_{ijkl}=\OO(1/N^2)$ and the average is taken in the  Gaussian model with the propagator (\ref{X-}). A long but straightforward calculation 
shows that $O(1/N^3)$ correction to \re{8.19} can be represented as 
\ba
\la{8.20}
\sW_{2}^{\SA} ={}& \frac{1}{96} \l ^4 \sW_0\, \sF_0'{}^2
-\frac{1}{4} \l ^2 \sW_0\, \sF_1'-\frac{1}{2} \l ^2 \sW_0'\, \sF_1 
-\frac{1}{96} \l ^4 \sW_0 \sF_0''\lp
+\frac{\l^{3}}{96}\Big(\sW_{0}-\frac{\l}{2}\sW_{0}'\Big)\,\sF_0'
+\frac{\l ^2 (744+5\l)}{92160}\,\sW_0-\frac{\l^{2}(2+3\l)}{960}\,\sW_{0}' \,,
\ea
where $\sF_0$ and $\sF_1$ are corrections to the difference free energy in the $\SA$ model and  prime denotes a derivative over $\lambda$. 
We apply the relations \re{F1-SA} and \re{F2-SA} to derive the  strong coupling  expansion of \re{8.20}
\ba\label{W2}
\sW_{2}^{\SA} &= \frac{\l^{9/4}}{\sqrt{2\pi}}e^{\sql}\,\Big[\frac{ 1}{9216}+\frac{121 }{122880 \sqrt\l}+\Big(\frac{1133}{655360}+\frac{3\log 2}{1024}\Big){1\over \l}+\cdots\Big].
 \ea
The relations \re{W0}, \re{619} and \re{8.20} allows us to compute the first few terms of the large $N$ expansion of the Wilson loop \re{622} in the $\SA$ and $\QQ$ models.
 
We use the obtained results to derive the ratio of the $\SA$   and $\N=4$ SYM Wilson loops
\ba\label{rat-W}
 \frac{ W^\SA}{ W^{\N=4}} = & 1-\frac{\l^{2}}{4N^{2}}\,\sF_{0}'+\frac{1}{N^{4}}\,\Big(
-\frac{\l^{3}}{48}\,\sF_{0}'+\frac{\l^{4}}{96}\,\sF_{0}'{}^{2}-\frac{\l^{4}}{96}\sF_{0}''
-\frac{\l^{3/2}}{4}\frac{\rI_{2}(\sql)}{\rI_{1}(\sql)}\,\sF_{1}-\frac{\l^{2}}{4}\,\sF_{1}'\Big)+O(1/N^6)\,.
\ea
It depends on the functions $\sF_{0}$ and $\sF_{1}$ defining corrections to the free energy \re{1.4} in the $\SA$ model.
The relation \re{rat-W} holds for an arbitrary coupling $\l$. 

\subsection{Double scaling limit}

At strong coupling,  keeping only the leading large $\l$  terms at each order in $1/N^2$, we get from \re{W0}, \re{W-1} and \re{W2}
\be\la{625}
W^\SA \simeq \frac{2N}{\sqrt{2\pi}}\,\l^{-3/4}\,e^{\sql}\,\Big[1-\frac{\l^{3/2}}{192N^{2}}+\frac{\l^{3}}{18432N^{4}}+\cdots\Big].
\ee
Surprisingly, the coefficient of $\l^3/N^4$ happens to be the same as in the $\N=4$  SYM theory \ci{Drukker:2000rr}
 \be
  {W}^{\N=4} \simeq \frac{2N}{\sqrt{2\pi}}\,\l^{-3/4}\,e^{\sql}\,\Big[1+\frac{\l^{3/2}}{96N^{2}}+\frac{\l^{3}}{18432N^{4}}+\cdots\Big].
\ee
 Taking the ratio of these  expressions   
 we get
\begin{align}\label{rat-W1} 
 \frac{ W^\SA}{ W^{\N=4}} \simeq 1-\frac{\l^{3/2}}{64N^{2}}+\frac{\l^{3}}{6144N^{4}}+\dots
\end{align}
 We observe that, similarly  to the free energy \re{F-fin},
  the strong coupling expansion of the ratio of the Wilson loops runs in powers of $\lambda^{3/2}/N^2=(4\pi)^3 g^3/N^2$.  
 
We have shown in the previous section that the free energy in the $\SA$ and $\QQ$ models differ in the double scaling limit \re{dsl} by the factor of $2$, see \re{rat2}. 
As was mentioned above, the ratio of the leading non-planar corrections $\sW_{1}^{\SA} / \sW_{1}^{\QQ} $ approaches $1$ in this limit. 
This suggests that the Wilson loops in the two models coincide in the double scaling limit
\be\label{W=W}
 {W}^{\QQ}\simeq {W}^{\SA}\,.
\ee
To leading order in $1/N^2$, this relation follows immediately from \re{W-W} after one takes into account that the two terms inside the brackets in the second relation in \re{W-W} coincide in the double-scaling limit. 

To any order in $1/N^2$, the relation \re{W=W} follows from the representation \re{V-repr} of the partition function of the $\SA$ and $\QQ$ models. We recall that in the  double-scaling  limit the partition function of the $\QQ$ model (see the second relation in \re{V-repr})
 factorizes into the product of two factors, one per $SU(N)$ node. Each of these factors coincides with the partition function of 
 the $\SA$ model, i.e. $Z_{\QQ}(J)\simeq [Z_\SA(J)]^2$. Applying \re{W-uni} we find that, in the double scaling limit, the Wilson loop in the $\SA$ and $\QQ$ models is obtained by applying the same differential operator to $Z_\SA(J)$. As a consequence, the free energies in the two models differ by the factor of $2$ whereas the Wilson loops coincide. 
 
Finding the exact expression for the ratio of the Wilson loops \re{rat-W1} in the double scaling limit \re{dsl} is an interesting challenging problem.
 
\section{String theory  interpretation  \la{sec7}}  

Given the explicit strong coupling results  obtained in this paper and summarized in the Introduction, it is 
important to attempt to  interpret  them on the dual  string  theory side.

In general,  in AdS/CFT context  the free energy $F$ of a    conformal theory on $S^4$  
should  correspond to  the type IIB  string partition function $\Z(g_s, T)$ evaluated  on the corresponding  $AdS_5 \times X^5$    background.  In  the perturbative string theory regime, it
should be   given by an  infinite  series   in   the string coupling constant  $g_s$  with coefficients $\Z_n(T)$
that are functions  of the effective string tension  $T\sim {L^2/ \alpha'{} } $  (with $L$  being the  curvature scale, \cf\rf{1.1})
\be \la{1.19} 
\Z(g_s, T)= {g^{-2}_s} \Z_{-1} ( T) + \Z_0 (T) + \sum^\infty_{n=1}  g_s^{2n}  \Z_n(T)  \ . \ee
Here the  (properly defined)  contribution of the 2-sphere $\Z_{-1} $   should  correspond to the planar term in \rf{1.2}, 
the   contribution   $\Z_0$ of the 2-torus -- to  the leading non-planar correction in \rf{1.4}, etc. 

In the maximally supersymmetric case of $\N=4$ SYM dual  to type IIB string on \adss 
the only non-trivial contributions  to $F$ should  come   from the sphere  and torus parts -- 
in fact, just from the  leading type IIB supergravity term in the tree-level effective action \ci{Russo:2012ay}
and the  massless mode (supergravity) 1-loop correction \ci{Beccaria:2014xda}. 
Together they   reproduce  indeed the $(N^2-1)\log \l$ term in \rf{1.2} 
(see  also  a  discussion  in  \cite{Beccaria:2021ksw,Beccaria:2021vuc}). 

Computing  the $\Z_n$ functions   in the superstring theory   defined on 
 half-supersymmetric   orbifold or orientifold  of \adss   
 appears to be a challenging task. 
 Being interested  here only in the 
leading  large tension limit ${\alpha'{}/ L^2} \to 0$  of each  $\Z_n$  in \rf{1.19},   
we shall make  a bold assumption that  the resulting  leading terms   can be found   
 just from the  corresponding  leading $\alpha'{}$   terms in the {\it local}  part
  of the string  low-energy effective action.\foot{Similar idea   goes back to \cite{Gubser:1998nz} where the  leading
    strong-coupling 
 corrections to  the  $O(N^2)$ and  $O(N^0)$ orders in the planar expansion of the 
  finite-temperature  free energy of the $\N=4$ SYM theory  were  discussed.} 
  
 In general, the   string    partition   function 
 on   a  curved  background may  contain  special local  terms of particular order in $\alpha'{}$ 
(i.e. in  derivatives, curvature, etc.)  that may  
receive    contributions  only  from 
 few leading orders in the  small $g_s$ perturbation theory.  They   may  then   play  a central role in the  ${\alpha'{}/ L^2} \to 0$  limit. 
 
 A support   to   this conjecture comes from the observation  \cite{Beccaria:2021ksw,Beccaria:2021vuc,Beccaria:2022ypy}
  that   given that the  leading 1-loop (torus) 
 term in the type IIB   10d  effective action  starts with the well-known  $S_{R^4}={1\ov \alpha'{}} \int d^{10} x\, \sqrt {-G}  R^4$  quartic curvature   term  
which on dimensional grounds  should scale as $S_{R^4}\sim {L^2\ov \alpha'{}}\sim \sql$,  this then 
reproduces  the  large $\l$  scaling 
of the leading terms in $\sF_0$ in \rf{1.5} and \rf{1.6}. 
The remaining   problem, however,  is  to explain  why this term (supplemented  by other  flux-dependent terms, etc.)
that   should  vanish on the \adss   background   
may give a  non-zero  contribution when evaluated  on the orbifold/orientifold of \adsss.\foot{That may require   understanding the role of  resolution of the orbifold singularity in such effective action computation.}

\subsection{Type IIB effective action  and  strong-coupling expansion }

Our aim  below will be  to  go beyond the 1-loop matching $S_{R^4}\sim\sql$ and 
demonstrate   that the current  knowledge   about the structure of    similar  higher  order terms  in the 
type IIB low-energy effective action is remarkably consistent with the strong-coupling scaling  
\be 
\Big({\alpha'{}\ov L^2} g_s^2\Big)^n\sim  \Big({ g^2_s \ov T}\Big)^n \sim  \Big({\l^{3/2} \ov N^2}\Big)^n \,,
\ee   
which we previously observed for the higher-genus terms in \rf{1.13} on the gauge theory side.
 
The  structure of type IIB effective action  can be symbolically summarized   as follows
(keeping  only  curvature  dependent terms   with 
powers of $\aa'$ that are required   on dimensional   grounds)\foot{Here   
$D^{2n} R^4$ ($n=0,2,3, ...)$  stand for the corresponding invariants (containing also other  terms with 5-form,  etc.,  fields) 
of the same dimension  that can be reconstructed from supersymmetry  considerations 
 and are implied by the structure of the low-energy expansion of type IIB string scattering amplitudes.
Some overall rational factors  are assumed to be  absorbed into these symbols.} 
\begin{align}
S_{\rm eff}= {1 \ov (2 \pi)^7} \int d^{10} x \sqrt {-G}   {}&   \Big[\alpha'{}^{-4}  \gs^{-2} R   + \alpha'{}^{-1} f_0(\gs)  R^4 \no 
\\  
{}& + 
\alpha'{}  f_1(\gs)  D^4 R^4  + \alpha'{}^2  f_2 (\gs)   D^6 R^4  + \alpha'{}^3  f_3 (\gs)   D^8 R^4  +\dots
 \Big] \ .  \la{1.20}
\end{align}
Note that the  $\aa'$-independent (dimension 10)  term   $D^2 R^4$  
does not appear   due to  maximal    supersymmetry  of type IIB  theory. 
The functions $f_{0},f_1,f_2$ (given by  Eisenstein series)
have  a {\it finite} number of 
perturbative terms  plus a tail of non-perturbative 
$O(e^{-1/\gs^2})$  corrections
(see  \ci{Green:1982sw,Gross:1986iv,Green:1999pu,Green:2005ba,Green:2006gt,Basu:2007ck,Gomez:2013sla,
DHoker:2014oxd,Green:2008uj,DHoker:2015gmr,DHoker:2017pvk,Basu:2019fim})
\begin{align} \la{1.21}\notag
&f_0 = {1\ov 16} \Big( 2 \z(3)  \gs^{-2}  + 4 \z(2) \Big)  
+ O(e^{-1/\gs^2}) \ ,  
\\\notag
&f_1 = {1\ov 32}  \Big( 2  \z(5)  \gs^{-2}    +    \tfrac{8  }{3} \z(4) \gs^2 \Big)+  O(e^{-1/\gs^2}) \ , 
\\
&f_2 = {1 \ov 48 }  \Big(  \z(3) ^2  \gs^{-2} + \z(3) \z(2)  +  6 \z(4)    \gs^2  +   \tfrac{2}{9} \z(6)    \gs^4\Big) +   O(e^{-1/\gs^2}) \ , 
\end{align}
while  $f_3$  contains an infinite series in $g^2_s$~\foot{The  $\log( - \aa' D^2)  $  term in $f_3$  indicates the presence of the corresponding $p^{16} \log p^2$ 
term  in the 4-graviton amplitude on flat background.
The  presence of an  infinite tail of perturbative  terms in $f_3$ appears to  be an open question 
(we thank J. Russo for this remark).} 
\begin{align}
f_3 = {1\ov 64}  \z(9) \gs^{-2}   + k_0 \z(3) \log( - \aa' D^2)  + O(g_s^2) + O(e^{-1/\gs^2}) \ . \la{1.24} 
\end{align}
Collecting the leading   $\aa'$ terms  at each order in $g^2_s$   in \rf{1.20} 
 corresponds to  including only 
 the    last  (supersymmetry-protected) perturbative terms      in $f_0,f_1,f_2$  in \rf{1.21}. 
 
Evaluating the action \rf{1.20} on the corresponding 10d   background  with curvature scale $L$ 
and separating   the tree-level $R$ term contribution\foot{\la{f1}The leading supergravity  term has  the 
expected planar scaling:  $L^8 \alpha'{}^{-4}  \gs^{-2}\sim   \l^2  {(4 \pi N)^2\ov \l^2} = \pi^2 N^2$.
The  factors of $\pi$   cancel  against the overall $1\ov \pi^7$ in \rf{1.20}:  indeed, for \adss we have 
  $\vol(S^5) =L^5 \pi^3$ and $\vol(AdS_5) =L^5 \pi^2 \log (L/a)$
where $a\to 0$ is an IR cutoff.  
Assuming a particular  regularization $a \sim { \sqrt{\aa'}}$ as in \ci{Russo:2012ay}  that leads to the tree-level term  being  $\sim N^2 \log  {L^2\ov \aa'} $   in agreement with  \rf{1.2}.}
we then expect  (on dimensional grounds, $R\sim D^2 \sim  L^{-2}$)    to  get from \rf{1.20}  
 the following  leading  non-planar contribution to the free energy 
\be \la{1.25}
\Delta F 
= {1\over \pi^2} \left[a_0 \zeta(2) {L^2\ov \aa'}    + a_1\zeta(4) {\aa' \ov L^2}  g^2_s + a_2  \zeta(6)\Big({\aa' \ov L^2}  g^2_s\Big)^2 + \dots\right]\,.
\ee
Here the terms proportional to $a_0$, $a_1$, $a_2$  originate the terms in \rf{1.21} containing $\zeta(2)$, $\zeta(4)\,g_s^2$ and $\zeta(6)\,g_s^4$, respectively.
 The overall factor of ${1\ov \pi^7} \times \pi^5 = {1\ov \pi^2}$ in \re{1.25} is
 dictated by the normalization of the planar term (see footnote \ref{f1}).
 The coefficients $a_0$, $a_1$, $a_2$    coming from curvature contractions should be rational, 
  i.e. should not contain extra factors of $\pi$. 
  One may conjecture that this pattern may extend also to higher-order terms  in \rf{1.25}.\foot{Let us recall in this connection  that  starting from the 
  structure of the  one-loop 4-graviton  amplitude  in 11d supergravity 
on $S^1$   it was suggested in   \ci{Russo:1997mk}   that  genus $k$ correction to the
$ D^{2k}R^4$  term in type IIA  theory   should be proportional to $\zeta(2k)$.
As there  may be several superinvariants ($ D^{2k}R^4, \, D^{2k-2}R^5, ..., R^{4+k}$) 
 of the same   dimension, that may actually apply  to some of them that  are non-vanishing on a given background. 
} 

 
Using that  ${L^2/\aa'} = 2\pi T$  we find that the expansion \rf{1.25} has precisely the same structure as was found from localization  in \rf{1.13}. 
Remarkably, we also check  that, in agreement with \rf{1.11} and \rf{1.14}, 
 the corresponding coefficients $c_n$ should be indeed {\it rational\/}  
\begin{align}
c_n= a_n {\zeta(2n+2)\over 16^n \pi^{2n+2}} =  a_n r_n \,,
\end{align}
where $r_n$ are rational numbers proportional to Bernoulli numbers. 
 
 \subsection{Comments}
 
   Let us now add  a few reservations  and comments.
The above argument  was based on the assumption  that  the  higher-order   curvature corrections in \rf{1.20},  that should vanish 
in the   maximally symmetric  \adss   case,  become  non-zero  once supersymmetry is reduced   as in the orbifold/orientifold case. As already mentioned above,  to show  this explicitly remains an open  problem. 
 Another  puzzle  is that since the   orbifold/orientifold  projection applies to $S^5$ only,   one  would expect that the
   IR divergent  factor of the 
$AdS_5$   volume   should  still remain and thus, as in  the  leading tree-level part, 
 should then  produce a $\log \l$  contribution  after  introducing an IR 
  regularization. However,  such $\log \l$ terms  should   be absent  in  non-planar corrections 
 to $F$ beyond the torus order  as seen from \rf{1.8}--\rf{1.11}.  
  It is  possible that there is a subtle  ``$0 \times \infty$'' cancellation mechanism  at work  that gives a finite contribution. 
 
 The   above discussion involved   the  special $f_0,f_1,f_2$ terms   in \rf{1.20}   that get contributions only from a  finite  number 
 of perturbative $g^{2n}_s$ terms; this made it   possible to isolate the leading  $\aa'$ correction at  each  of the  lowest 
  $g^{2n}_s$ ($n=0,1,2$)
 orders. 
 This  pattern    appears to change   starting with the  string 4-loop   $f_3 {\aa'}^3  D^8 R^4 $ term.
 For example, 
 the 1-loop $\z(3) \log (-\aa' D^2) $ term  in \rf{1.24}  would naively lead to a $\zeta(3) \log \l$ term in \rf{1.5} and \rf{1.6}
 but the $\log \l$ term there has a rational coefficient (which  should come from  the torus  partition function).\foot{Still, it is interesting to note that 
 a constant term that may  accompany this 1-loop  $\z(3) \log (-\aa' D^2) $ term  in \rf{1.24} 
 as a coefficient of the $\aa'{}^3  D^8 R^4$ term in \rf{1.20} 
 could   be a string counterpart of the $\z(3) \l^{-3/2}$   term in $\sF_0$ in \rf{1.5} and \rf{1.6}.}
Thus  extracting the leading  $\aa'$ term  at each of 
 higher  order $g^{2n}_s$ contributions   may  require  a detailed information  about 
  the structure of $f_n$ with $n\geq 3$. 
 
 It is possible   that  the  above  discussion  based on low-energy effective action  is  only a  short-cut  
 to understand
  the leading large tension  scaling of  string loop corrections   that appear  in 
  the  full string partition  function \rf{1.19}. 
 The latter   should include, in particular,  also the contribution  of the twisted sector  states  present in  orbifold  theories
  which are not  included  in the generic 10d  action effective action \rf{1.20}
  (where, e.g.,  the contributions of light twisted states should be added separately, cf. \ci{Gukov:1998kk}).
 
 Thus in general 
     instead of  starting 
      with the  effective action \rf{1.20}  (found  by first expanding string scattering amplitudes  near flat   space  in $\aa'$  
 and then  reconstructing  the  corresponding action for a generic background)
 and evaluating  it on the corresponding  background
 one is   first   to    compute  higher order  terms in the $g_s^2$ expansion 
 of  the partition function \rf{1.19}   of the  orbifold/orientifold string  theory  
 and then take the limit $\aa' /L^2 \sim T^{-1} \to 0$.

 We conjecture that   the  structure of the resulting  expansion  of the string partition function computed 
 for  the    $AdS_5 \times X^5$ orbifold/orientifold corresponding to $\QQ$ and $\SA$ models 
 will  remain the same  as in \rf{1.25}, i.e.  it will be given by the  sum of 
 powers of $  T^{-1} g^2_s \sim {\l^{3/2} / N^2}$   matching  the localization result    in \rf{1.13}. 
 
 This  would imply that the double-scaling limit \rf{1.14}  should have a string-theory counterpart:
 the leading $T^{-1} \to 0$  terms  at each order in $g^2_s$  may be  captured by taking the limit 
 $g^2_s\to 0$ and $T^{-1} \to 0$ with $T^{-1} g^2_s$ kept fixed.
 Thus  adding a  handle 
 to a genus $n$   surface   should   result in an extra factor of $ { g^2_s / T}\sim g^2_s{ \aa' / L^2}  $.\foot{An argument   of why    that should happen (at leading order in ${\aa'\ov  L^{2}} \to 0$)
  was attempted in \ci{Drukker:2000rr} in the case  of the  computation of the expectation value 
  of the  circular  Wilson loop  in  the maximally supersymmetric \adss  string theory. 
  This case may be   somewhat analogous to  the one of the free energy in the $\N=2$ supersymmetric theory 
  as this  Wilson loop   breaks   half of supersymmetry and gets corrections at  all orders in genus expansion. 
  The suggestion in  \ci{Drukker:2000rr}  was    that for   $\aa' \to 0$  only the  lightest (supergravity) 
  modes  should  propagate  along  thin  handle.  Inserting the latter  is then   like attaching a  ``massless''
   propagator  which in the 
  \adss   case may lead to a  factor of $\aa' /L^{2} $. 
  It  appears  to be hard, however,  to make this argument precise due to various ad hoc cutoffs  required
  (see  a discussion in appendix A of \ci{Beccaria:2020ykg}).
  An alternative  argument specific to the Wilson loop   case was given in \ci{Giombi:2020mhz}.
  We thank S. Giombi for a  discussion of this issue.}
 
 Let us   now comment on   the  string theory   interpretation of the 
 curious   prediction   of the localization  matrix model 
 that the leading strong  coupling coefficients at each order of  $1/N^2$ expansion of free energy 
   in the $\SA$ and $\QQ$ models 
 are related by a factor of  $1/2$   (\cf \rf{1.5}--\rf{1.10}  and  \rf{Fequal}). The  idea is to relate this fact to an extra  $ \mathbb{Z}_{2}$
 orientifold projection  required  to obtain the $\SA$ model  from the $\QQ$ one.
 
 If we parametrize the $S^5$  directions    by 3  complex coordinates  $(z_1,z_2, z_3)$  with $|z_i|^2=1$   then  the 
   $ \mathbb{Z}_{2}$ orbifold  model   corresponds to modding out the \adss  theory by  the action $z'_1= - z_1, \
 z'_2 = - z_2$ or  by the inversion of the 4 out of 6  real embedding coordinates.   
 The $\SA$  theory   (see, e.g., \ci{Ennes:2000fu})   is found 
  by an  additional  orientifold projection that involves 
   the product  of the  inversion in the  2 remaining  coordinates transverse to the 
   original  stack of D3-branes in flat  space or the $AdS_5$ boundary, i.e. $z_3'=-z_3$
   and also the  world-sheet parity $\Omega$  and $(-1)^{F_L}$  that changes the sign of the 
   Ramond sector of left-moving modes in  the NSR description in flat space. 
 
 Assuming, as we discussed  above, that  the  leading   in ${\aa'/ L^2} \sim {1/ \sql} \to 0 $   terms 
   at  least   low  orders in $g^2_s$ can be captured  just at the level of  the  effective action, 
    then 
    only the inversion  part of this extra  projection should matter.  It   implies 
   restricting  the angle  in the corresponding plane to half of its value    and this then    halves the volume of the corresponding 
   internal 5-space  thus producing an extra 1/2 factor in the coefficients in \rf{1.25}. 
  
  As for the Wilson loop equality  \rf{1.188}, 
  on the  string theory side its origin may be  related to   the fact that  in both orbifold and orientifold  cases  the  circular 
Wilson loop expectation value  is given by a semiclassical expansion near the same $AdS_2$  minimal surface that lies  in $AdS_5$   only
and then   the leading in $\aa'/L^2$  corrections  at each order $g^2_s$   expansion 
 may not  be   sensitive to extra orientifolding projection.

\subsubsection*{Acknowledgments}

We would like to thank Bertrand Eynard and  Emmanuele Guitter for very useful discussions. 
MB was supported by the INFN grant GSS (Gauge Theories, Strings and Supergravity). AAT was supported by the STFC grant ST/T000791/1.

\appendix
 
\section{Correlators in  Gaussian matrix model}\label{App:gauss}

In this appendix  we summarize the properties of correlation functions of single-trace operators in  $SU(N)$ Gaussian matrix model
\begin{align}\label{Gc}
G_{i_1\dots i_{L}} =\vev{\OO_{i_1} \dots \OO_{i_{L}}}_c\,,\qqqquad \OO_i=\tr\lr{A\over \sqrt N}^i\,,
\end{align}
where the subscript `$c$' denotes the connected part. 
The  partition function of this model is defined as
\begin{align}\label{app:Z}
Z[J] = \int DA \, e^{-\tr A^2+\sum_i J_i \OO_i}\,, 
\end{align}
where $A=\sum_{a=1}^{N^2-1} A^a T^a$ are hermitian traceless $N\times N$ matrices and the $SU(N)$ generators $T^a$ are normalized as $\tr(T^a T^b)=\frac12 \delta^{ab}$. The integration measure is $D A= \prod_a d A^a/\sqrt{2\pi}$. 

The correlation function \re{Gc} can be found as
\begin{align}
G_{i_1\dots i_{L}}={\partial\over\partial J_{i_1}}\dots {\partial\over\partial J_{i_{L}}} \log Z[J]\Big|_{J=0}\,.
\end{align}
It depends on the set of non-negative integers $i_1,\dots,i_L$ and it is different from zero only for even 
$i_1+\dots+i_{L}$. At large $N$, the correlation function admits an expansion \re{G-gen} in powers of $1/N^2$. 

The expressions for the correlation functions \re{Gc} are different for even and odd indices. For instance, for $L=2$ the correlation functions $\Q^+_{i_1,i_2}=G_{2i_1,2i_2}$ and $\Q^-_{i_1,i_2}=G_{2i_1+1,2i_2+1}$ are given by \re{Q2}. For $L=4$ we have
\begin{align}\notag\label{Q-4ind}
& \Q_{ijkl}^{+}=G_{2i,2j,2k,2l}={4\over N^2} \beta_i^+ \beta_j^+ \beta_k^+ \beta_l^+ (i+j+k+l-1)+O(1/N^4)\,,
\\\notag
& \Q_{ijkl}^{-}=G_{2i+1,2j+1,2k+1,2l+1}={4\over N^2} \beta_i^- \beta_j^- \beta_k^- \beta_l^- (i+j+k+l+4)+O(1/N^4)\,,
\\
& \Q_{ijkl}^{+-}=G_{2i,2j,2k+1,2l+1}={4\over N^2} \beta_i^+ \beta_j^+ \beta_k^- \beta_l^- (i+j+k+l)+O(1/N^4)\,,
\end{align}
where $\beta^\pm_i$ are defined in \re{beta}.
The relations \re{Q2} and \re{Q-4ind} are sufficient to compute the $O(1/N^4)$ corrections to the free energy in the $\SA$ and $\QQ$ models, see Eqs.~\re{3.9} and \re{dF-sum}. 

To find the $O(1/N^6)$ correction to \re{3.9}, we also need subleading corrections to $L=2$ and $L=4$ correlators as well as the leading order expression for $L=6$ correlator, all with odd indices. 
They are given by 
\begin{align} \notag \label{Q2Q4Q6}
& \Q^-_{i_1i_2} =2\beta^-_{i_1}\beta^-_{i_2}\left[{1\over e_1+1}+ \frac{1}{12 N^2}(e_1^2-5 e_1-e_2-13)\right. + {1\over 1440 N^4}(5 e_1^5-10 e_2 e_1^3+9 e_2^2 e_1-72 e_1^4  
\\ \notag & \qqqquad \qquad
+75 e_2 e_1^2-18 e_2^2+93 e_1^3 +223 e_2 e_1+906 e_1^2-906 e_2-164
   e_1-888) +O(1/N^6) \bigg],
\\\notag
& \Q^-_{i_1i_2i_3i_4} ={4\over N^2}\beta^-_{i_1}\beta^-_{i_2}\beta^-_{i_3}\beta^-_{i_4}\Big[e_1+4 + \frac{1}{12 N^2} \left(e_1^4-e_2 e_1^2-e_3 e_1-2 e_4+e_1^3-9 e_2 e_1 \right.
 \\\notag
{}& \qqqquad \qquad
 \left.
-49 e_1^2  +20 e_2 -94 e_1-6\right)+ O(1/N^4)\Big],
\\[2mm]
& \Q^-_{i_1i_2i_3i_4i_5i_6} = {8\over N^4}\beta^-_{i_1}\beta^-_{i_2}\beta^-_{i_3}\beta^-_{i_4}\beta^-_{i_5}\beta^-_{i_6}\Big[e_1^3+15 e_1^2-6 e_2+44 e_1+30+ O(1/N^2)\Big],
\end{align}
where $e_k$ (with $k=1,\dots,L$) are symmetric polynomials in $L$ variables $i_1,\dots, i_L$ 
\begin{align}\label{e}
e_1=\sum_{1\le p\le L} i_p\,,\qquad
e_2=\sum_{1\le p_1< p_2 \le L} i_{p_1}i_{p_2}\,,\qquad
\dots\,, \qquad e_L={i_1}\dots {i_L}
\,.
\end{align}
Notice that the coefficients of powers of $1/N^2$ in \re{Q2Q4Q6} are given by multi-linear combinations of the symmetric polynomials whose total degree is correlated with the power of $1/N^2$. 

Computing the free energy in the double scaling limit \re{dsl}, we encountered the correlators \re{Q2} evaluated for large values of indices $i_p=O(N^{2/3})$, or equivalently $e_k=O(N^{2k/3})$. In this limit, the coefficients of $1/N^2$ in \re{Q2Q4Q6} grow as powers of $N$ in such a way that all terms inside the brackets in \re{Q2Q4Q6} have the same behaviour for $N\to\infty$. In addition, as can be seen from \re{Q2} and \re{Q-4ind}, the correlators with even and odd indices, $\Q^+_{i_1i_2\dots}$ and $\Q^-_{i_1i_2\dots}$, are given  for large $i_p$  by the same function.  

Indeed, it is well-known that for $i_p=O(N^{2/3})$ with $p=1,\dots,L$, the correlators \re{G-gen} take the following universal form 
\begin{align}\label{A.27}
\vev{\OO_{i_1} \dots \OO_{i_L}}_c = {\beta_{i_1} \dots \beta_{i_L}\over N^{L-2}}\Big[ c_0 A_{0,L} + {c_1\over N^2} A_{1,L} + {c_2\over N^4} A_{2,L} + \dots \Big],
\end{align}
where $\beta_i = 2^{i/2-1} \sqrt {i/\pi}$ arises from the large $i$ limit of the functions \re{beta} and
the normalization factors $c_p$ with $p\ge 0$ are 
\begin{align}
c_p={2^{-L/2+3}\over 96^p \, p!} \,.
\end{align}
The functions $A_{g,L}(i_1,\dots,i_L)$ are independent 
of the parity of indices $i_p$. They describe genus $g$ contribution to the correlator.  

In the integral representation of the free energy \re{V-repr}, the correlation functions \re{A.27} play the role of the coupling constants $\Q_{i_1,i_2\dots } = \vev{\OO_{2i_1} \OO_{2i_2}\dots }_c$ defining 
the interaction potential \re{V} in the double scaling limit \re{dsl}. A nontrivial scaling behaviour of the correlation functions \re{A.27} in the  Gaussian matrix model for $i_p=O(N^{2/3})$ and that of the free energy for $g=O(N^{2/3})$ are in one-to-one correspondence with each other. 

The explicit expressions for $A_{g,L}$ for arbitrary $L$ and $g\le 3$ were derived in \cite{Eynard:2021zcj}
\begin{align}\notag\label{As}
A_{0,L} &=e_1^{L-3},
\\\notag
A_{1,L} &= e_1^L -\sum_{k=2}^L (k-2)! e_k e_1^{L-k},
\\\notag
A_{2,L} &=e_1^{L+3}-2 e_2 e_1^{L+1}-\frac{18}{5}
   e_3 e_1^L -\sum _{k=4}^{L} \frac{1}{30} \Big(k^3+21 k^2-70 k+96\Big) (k-3)! e_k 
   e_1^{L+3-k}
\\
&+\frac{9}{5} e_2^2 e_1^{L-1}    +\frac{18}{5} e_2 e_3
   e_1^{L-2} 
+\sum _{k=4}^{L} \frac{(k+16) (k-1)!}{10} e_2 
   e_k  e_1^{-k+L+1}-\sum _{k=3}^L \frac{k!}{10} e_3 
   e_k e_1^{L-k} ,
\end{align}
where symmetric polynomials $e_k$ are defined in \re{e}.  The expression for $A_{3,L}$ is more cumbersome and can be found in \cite{Eynard:2021zcj}.
One can verify that for $e_k=O(N^{2k/3})$ the relations \re{Q2Q4Q6} are in agreement with \re{A.27}.

For lowest values of $L$ the correlation function (\ref{A.27}) is known to any order in $1/N^2$, see \cite{Witten:1990hr,Dijkgraaf:1990qw}
\begin{align}\notag\label{L=1,2}
\vev{\OO_{i_1}} &=  4\sqrt 2 N {\beta_{i_1}}   e^{i_1^3\over 96 N^2}i_1^{-2},
\\
\vev{\OO_{i_1}\OO_{i_2}}_c &= 4\beta_{i_1} \beta_{i_2}  e^{i_1^3+i_2^3\over 96 N^2}\sum_{k=0}^\infty 
{1\over (2k+1)!!}{e_1^{k-1} e_2^k \over (4N)^{2k}}\,,
\end{align}
where $e_1=i_1+i_2$ and $e_2=i_1i_2$. Here the first relation holds for even $i_1$. For odd $i_1$ the correlator $\vev{\OO_{i_1}}$ vanishes.

The relations \re{As} take the form $A_{g,L} = e_1^{L+3(g-1)}+ \dots$ where dots denote terms with smaller power of $e_1=i_1+\dots+i_L$.  
Such terms can be neglected by considering the limit $i_1=O(N^{2/3})$ and $i_p\ll i_1$ with $p=2,\dots,L$. As follows from \re{A.27}, the correlation function is given in this limit by 
\begin{align}\label{A.13}
{\vev{\OO_{i_1} \dots \OO_{i_{L}}}_c} \stackrel{i_1\gg i_p}{=} {\beta_{i_1}\dots\beta_{i_{L}}\over N^{L-2}}
\times 2^{3-L/2}  i_1^{L-3} e^{\frac{i_1^3}{96 N^2}}\,.
\end{align}
In distinction 
to \re{L=1,2} this relation holds for $i_1\gg i_p$ with $p=2,\dots,L$.

Notice the presence of a universal factor $e^{i_1^3/( 96 N^2)}$ in  \re{L=1,2} and \re{A.13}. 
Its origin can be traced back to the universal behaviour of correlators in matrix models near a critical edge (for a review see e.g. \cite{Eynard:2015aea}). As an example, consider the two-point correlation function $\vev{\OO_{i_1}\OO_{i_2}}_c$.  It is well-known that at large $N$ the eigenvalues of the matrix $A/N^{1/2}$ in the $SU(N)$ Gaussian unitary ensemble \re{app:Z} 
condense on the interval $[-2,2]$ and their density is described by the Wigner semicircle distribution.
Defining the two-point connected distribution density of eigenvalues  
\begin{align}\label{rho2}
\rho(x_1,x_2) {}& = {1\over N^2} \VEV{\tr \delta \Big(x_1-{A\over  \sqrt N}\Big) \tr \delta \Big(x_2-{A\over  \sqrt N}\Big)}_c \,, 
\end{align}
we can express the two-point correlation function of $\OO_i=\tr\lr{A/ \sqrt N}^i$  as an integral over eigenvalues of the matrix $A/\sqrt N$ 
\begin{align}\label{OO-int}
\vev{\OO_{i_1}\OO_{i_2}}_c = \int_{-2}^2 dx_1 dx_2 \, \rho(x_1,x_2) x_1^{i_1} x_2^{i_2}\,.
\end{align}
It follows from this representation that for large $i_1$ and $i_2$ the integral receives a dominant contribution from integration in the vicinity of the end points $x_i=\pm 2$, or equivalently near the edges of the distribution of eigenvalues. Because the two edges provide the same contribution, we concentrate on  the right edge $x_i=2$.
Rescaling the integration variable in this region as~\footnote{In general, the scaling behaviour near the regular edge is parameterized by two integers $p$ and $q$, so that $x_i=  2  + \xi_i N^{-q/(p+q)}$. We encounter the special case $p/q=1/2$.}
\begin{align}
x_i=  2  + \xi_i N^{-2/3},
\end{align} 
one finds that at large $N$ the distribution density \re{rho2} is given by a remarkably simple expression (for a review, see e.g. \cite{Eynard:2015aea})
\begin{align}\notag\label{bar-rho}
&
\lim_{N\to\infty} \rho(x_1,x_2) =-[ K_{\rm Ai}(\xi_1,\xi_2)]^2\,,
\\
& K_{\rm Ai}(x,y)={{\rm Ai}(x) {\rm Ai}'(y) -{\rm Ai}(x) {\rm Ai}'(y)\over x-y}\,,
\end{align}
where ${\rm Ai}(x)$ is the Airy function. Substituting \re{bar-rho} into \re{OO-int} we get
\begin{align}\notag
\vev{\OO_{i_1}\OO_{i_2}}_c  
& \sim  \int_0^\infty d\xi_1 d\xi_2 \, e^{\frac12(\xi_1i_1 +\xi_2i_2) N^{-2/3}} K_{\rm Ai}^2(\xi_1,\xi_2)
\\
& \sim \int_0^\infty d\xi_1 d\xi_2 \, e^{\frac12(\xi_1i_1 +\xi_2i_2) N^{-2/3}-\frac43(\xi_1^{3/2}+\xi_2^{3/2})}  \sim e^{\frac{i_1^3+i_2^3}{96N^{2}}},
\end{align} 
where in the first relation we substituted $x^{i_p} = (2+\xi N^{-2/3})^{i_p}\sim 2^i e^{\frac12 \xi i_p N^{-2/3}}$ and
in the second relation replaced the Airy function ${\rm Ai}(\xi)$ by its leading behaviour at large $\xi$.  The above analysis can be generalized  to $L-$point correlators \re{A.13}.
  
\section{Auxiliary matrices}\label{app:mat}
 
In this appendix  we describe properties of various matrices that enter the calculation of non-planar corrections. 

In a Gaussian matrix model, the two-point correlators  of single traces
$\OO_i=\tr((A/\sqrt N)^{i/2})$ are given by the matrices $\Q_{ij}^+ = \vev{\OO_{2i}\OO_{2j}}_c$ and $\Q_{ij}^-=\vev{\OO_{2i+1}\OO_{2j+1}}$ defined in \re{Q2}. Taking linear combinations of 
$\OO_n$ 
\begin{align}\label{ortho}
\widehat \OO_{2i} = (U^+)_{ij}^{-1} \OO_{2j}\,,\qqqquad
\widehat \OO_{2i+1} = (U^-)_{ij}^{-1} \OO_{2j+1}\,, 
\end{align}
we can construct a basis of orthonormal traces satisfying $\vev{\widehat \OO_{2i}\widehat \OO_{2j}}=\vev{\widehat \OO_{2i+1}\widehat \OO_{2j+1}}=\delta_{ij} + O(1/N^2)$.
Here $U^\pm$ are lower triangular matrices, $U^\pm_{ij} =0$ for $i\le j-1$, satisfying \re{Q-UU}. Their explicit expressions are
\begin{align}\notag\label{U-mat}
& U^-_{ij} = \frac{ \sqrt{2 j+1}\, \Gamma (2 i+2)}{ 2^{i+1/2}\Gamma (i-j+1) \Gamma
   (i+j+2)}\,,
\\
& U^+_{ij} = \frac{\sqrt{2 j}\, \Gamma (2 i+1)}{2^{i}\Gamma (i-j+1) \Gamma
   (i+j+1)}\,.
\end{align}
The inverse matrices are given by
\begin{align}\notag
& (U^-)^{-1}_{ji} = (-1)^{i+j}\frac{ \,2^{i+1/2} \sqrt{2 j+1}\, \Gamma (i+j+1)}{\Gamma (2 i+2) \Gamma (-i+j+1)} \,,
\\
&  (U^+)^{-1}_{ji} = (-1)^{i+j}\frac{ \,2^{i} \sqrt{2 j}\, \Gamma (i+j)}{\Gamma (2 i+1) \Gamma (-i+j+1)} \,.
\end{align}
We can use these matrices to define the infinite-dimensional vectors
\begin{align}
 (R_n^\pm )_i = \sum_{j\ge 1} (U^\pm)_{ij}^{-1} j^n\beta_j^\pm\,,
\end{align}
where $\beta_j^\pm$ are given by \re{beta}.
Going through a calculation we find that
\begin{align} \label{U-P}  
 (R_n^- )_i = \sqrt{i+\tfrac12} \, P^-_{n}(i)\,,\qqqquad
 (R_n^+ )_i = \sqrt{i} \, P^+_{n}(i)\,,
 \end{align}
where $P^\pm_{n}(i)$ are polynomials in $i$ of degree $2n$. For lowest values of $n$ they are given by
\begin{align}\notag\label{P's}
& P^+_0=1\,,&& P^+_1=i^2\,,&& P^-_2=\frac12i^4 + \frac12 i^2\,,
\\
& P^-_0=1\,,&& P^-_1=i(i+1)-1\,,&& P^-_2=\frac12(i(i+1))^2 - i(i+1)+1\,.
\end{align}
For arbitrary $n$ they look as
\begin{align}\notag\label{PP}
& P_n^+(i)= \sum_{l=1}^i {(-1)^{i+l}\Gamma(i+l)\over \Gamma(1+i-l) \Gamma^2(l)}l^{n-1}\,,
\\
& P_n^-(i)= \sum_{l=1}^i {(-1)^{i+l}\Gamma(i+l+1)\over \Gamma(1+i-l) \Gamma(l)\Gamma(l+2)}l^{n}\,.
\end{align}
Applying the relations \re{phi-n}, \re{psi} and \re{U-P}, we get the following representation of the functions $\phi^\pm_n(x)$
\begin{align}\notag
& \phi_n^+  (x) = {1\over \sqrt{2x}} \sum_{i\ge 1} (-1)^i  \, 2i \, P^+_{n}(i) J_{2 i}(\sqrt{x})\,,
\\
& \phi_n^-  (x) = {1\over \sqrt{2x}} \sum_{i\ge 1} (-1)^i  (2i+1) \, P^-_{n}(i) J_{2 i+1}(\sqrt{x})\,.
\end{align}
Plugging in the expressions \re{PP} for the polynomials $P_n^\pm$, both sums can be evaluated  
leading to \re{phi-J}.  
  
\section{Method of differential equations}\label{app:method}  

In this appendix  we describe a technique that allows us to compute the matrix elements 
\begin{align}\label{w-nm1}
w_{nm}= \vev{\phi_n|\bm{\chi} {1\over 1-\bm K_\ell} |\phi_m} 
\end{align}
of the resolvent of the Bessel operator \re{Bes}
over the states $\phi_n=(x\partial_x)^n J_\ell(\sqrt x)$ with $n\ge 0$. 

\subsection*{Functional relations}

Expanding \re{w-nm1} in powers of $\bm{K}_\ell$ and using the definition \re{Bes} of the Bessel kernel, one can show that $w_{nm}$ is symmetric in indices, 
$
w_{nm} = w_{mn}
$.
The kernel of the Bessel operator \re{Bes} depends on the function $\chi(\sqrt x/(2g))$. It satisfies 
\begin{align}\label{chi-eq}
\lr{x\partial_x + \frac12 g\partial_g} \chi\lr{\sqrt x\over 2g} = 0\,.
\end{align}
One can use this relation together with the definition  of the Bessel operator \re{Bes} to show that its resolvent satisfies the following
operator identity \cite{Belitsky:2019fan}
\begin{align}\label{TW-id}
\left[x\partial_x + \frac12 g\partial_g, {1\over 1-\bm{K}_\ell} \right] = \frac14 {1\over 1-\bm{K}_\ell} \ket{\phi_0} \bra{\phi_0} \bm \chi {1\over 1-\bm{K}_\ell}\,,
\end{align}
where $\phi_0(x)=J_\ell(\sqrt x)$ and $\bm \chi$ is a diagonal operator with the kernel $\delta(x-y) \chi(\sqrt x/(2g))$. For special choice of the symbol $\chi(x)=\theta(1-x)$ the relation \re{TW-id} coincides with the identity derived by Tracy and Widom in \cite{Tracy:1993xj}.

Evaluating the matrix elements of both sides of \re{TW-id} over the states $\bra{\phi_n}\bm \chi$ and $\ket{\phi_m}$ and making use of \re{chi-eq}, we get
\begin{align}\label{Id}\notag
\frac12 g\partial_g\vev{\phi_n|\bm{\chi} {1\over 1-\bm K_\ell} |\phi_m} -
\vev{\partial_x  (x \phi_n)|\bm{\chi} {1\over 1-\bm K_\ell} |\phi_m}
-\vev{ \phi_n|\bm{\chi} {1\over 1-\bm K_\ell} |x\partial_x \phi_m}
\\
 = \frac14 \vev{\phi_n| \bm{\chi} {1\over 1-\bm K_\ell}|\phi_0}\vev{\phi_0|\bm{\chi} {1\over 1-\bm K_\ell}|\phi_m}\,.
\end{align}
Taking into account that $x\partial_x \phi_m=\phi_{m+1}$ and $\partial_x  (x \phi_n)=\phi_n+\phi_{n+1}$ we can cast 
\re{Id} into a functional relation for the matrix elements \re{w-nm} 
\begin{align}\label{w-fun1}
\lr{\frac12 g\partial_g-1}w_{nm} = \frac14w_{0n} w_{0m} +w_{n+1,m} + w_{n,m+1}\,,
\end{align}
where $n,m\ge 0$. Applying this relation, we can express $w_{nm}$ for arbitrary $n$ and $m$ in terms of the minimal set of independent matrix elements $w_{00}, w_{02}, w_{04},\dots$. For instance,
\begin{align}\label{5.12}\notag
& w_{01}=-\frac12\lr{1+\frac14 w_{00} -\frac12 g\partial_g} w_{00}\,,
\\
& w_{11} = -\lr{1+\frac14 w_{00} -\frac12 g\partial_g} w_{01}-w_{02} \, .
\end{align}

The dependence of the matrix elements $w_{nm}$ on the coupling constant $g$ enters through the symbol function $\chi(\sqrt x/(2g))$. Under a  variation of this function, $\chi\to\chi+\delta\chi$,  the matrix elements \re{w-nm1} change as
\begin{align}\notag\label{delta-w}
\delta w_{nm} {}&= \vev{\phi_n|\delta \bm{\chi} {1\over 1-\bm K_\ell} |\phi_m}+ 
\vev{\phi_n| \bm{\chi} {1\over 1-\bm K_\ell} \bm K_\ell \bm{\chi} ^{-1} \delta \bm{\chi}{1\over 1-\bm K_\ell} |\phi_m}
\\\notag
{}& 
=\vev{\phi_n|  {1\over 1- \bm{\chi} \bm K_\ell  \bm{\chi} ^{-1} } \delta \bm{\chi}{1\over 1-\bm K_\ell} |\phi_m}
\\
{}&
=\int_0^\infty dx \, Q_n(x) Q_m(x)\, \delta\chi \lr{\sqrt x\over 2g}\,,
\end{align}
where we introduced the notation 
\begin{align}\label{Qn}
Q_n(x) = \vev{x|{1\over 1-\bm K_\ell}|\phi_n} =\vev{\phi_n|  {1\over 1- \bm{\chi} \bm K_\ell  \bm{\chi} ^{-1} } |x} \,.
\end{align}
As above, the second relation can be verified by expanding the matrix element in powers of $\bm K_\ell$ and using the definition of the Bessel operator \re{Bes}. In the  special case when $\delta\chi=  \delta g \,\partial_g \chi$ the relation \re{delta-w} reduces to \re{dw}.

\subsection*{$\bm Q-$functions}

We can apply \re{TW-id} to show that the functions $Q_n(x)$ satisfy the  functional relation analogous to \re{Id}
\begin{align}\notag\label{5.6}
\lr{x\partial_x + \frac12 g \partial_g} Q_n(x)   
&=  \vev{x|{1\over 1-\bm{K}_\ell} |x\partial_x \phi_n}+\vev{x\Big |\left[x\partial_x + \frac12 g\partial_g, {1\over 1-\bm{K}_\ell} \right]  \Big | \phi_n}  
\\ \notag
& = Q_{n+1}(x) + \frac14 \vev{x\Big | {1\over 1-\bm K_\ell}\Big | \phi_0}
\vev{\phi_0\Big | \bm\chi{1\over 1-\bm K_\ell}\Big | \phi_n}   
\\
& = Q_{n+1}(x)  + \frac14 Q_0(x) w_{0n} \,.
\end{align}
It can be used to express $Q_n(x)$ for $n\ge 1$ in terms of $Q_0(x)$. For instance,
\begin{align}\notag\label{deqQ0}
& Q_1(x) = \lr{x\partial_x + \frac12 g \partial_g} Q_0(x)- \frac14 Q_0(x) w_{00}\,,
\\
& Q_2(x) = \lr{x\partial_x + \frac12 g \partial_g} Q_1(x)- \frac14 Q_0(x) w_{01}\,.
\end{align}
To find the function $Q_0(x)$, we take into account that  the Bessel function $\phi_0=J_\ell(\sqrt x)$ satisfies a differential equation
\begin{align}
\phi_2(x) = (x\partial_x)^2 \phi_0(x) =- \frac14(x-\ell^2)\phi_0(x)\,.
\end{align}
Together with \re{Qn} this leads to
\begin{align}\notag\label{Q2-Q0}
Q_2(x) {}&= \frac{\ell^2}4 Q_0(x) -\frac14 \vev{x|{1\over 1-\bm K_\ell}x|\phi_0} 
\\\notag
{}&=\frac14(\ell^2-x) Q_0(x) + \frac14 \vev{x|[x,{1\over 1-\bm K_\ell}]|\phi_0} 
\\
{}&= \frac14(\ell^2-x) Q_0(x) - \frac14 w_{01} Q_0(x) + \frac14 w_{00} Q_1(x) \,.
\end{align}
Here in the last relation we used the expression for the commutator from  \cite{Tracy:1993xj,Belitsky:2019fan}.

Combining  together  the relations \re{deqQ0} and \re{Q2-Q0}, we obtain the partial differential equation for $Q_0(x)$
\begin{align}\la{424}
\left[ (g\partial_g + 2 x\partial_x)^2 +x-\ell^2+(1-g \partial_g) w_{00}\right] Q_0(x)=0\,. 
\end{align}
This equation as well as the above relations hold for any coupling $g$. 
At weak coupling, we can expand \re{Qn} in powers of the Bessel operator to get
\begin{align}
Q_0(x) = \phi_0(x) + O(g^{2(\ell+1)})= J_\ell(\sqrt x)+ O(g^{2(\ell+1)})\ . 
\end{align}
This relation provides a boundary condition for the differential equation \re{424}.

\subsection*{Strong coupling expansion}

Let us apply \re{dw} and \re{424} to derive the strong coupling expansion of $w_{0n}$.  
Due to a  complicated form of $w_{00}(g)$ (see \re{w00-str}), the differential equation \re{424} can not be solved exactly for an arbitrary $g$. 
A significant simplification happens however at strong coupling. 

Because the symbol $\chi(x)$ vanishes rapidly at large $x$, the dominant contribution to the integral in \re{dw} comes from $\sqrt{x} =O(2g)$. This suggests the change of variables 
\begin{align}\label{change}
x= (2gz)^2\,,\qqqquad q_n(z,g) = Q_n((2gz)^2)\,.
\end{align}
Then  the relations \re{delta-w} and \re{5.6} can be rewritten as
\begin{align}\notag\label{q-rec}
& \partial_g w_{0n} = -8 g \int_0^\infty dz \, z^2 q_0(z) q_n(z) \partial_z \chi(z)\,,
\\
& q_{n+1}(z) = -\frac14 q_0(z) w_{0n} + \frac12  g\partial_g q_n(z)\,.
\end{align}
According to \re{424} the function $q_0(z)$ satisfies the  differential equation
\begin{align}\label{12.25}
& \Big[(g\partial_g)^2 + 4(gz)^2 -\ell^2 +(1 - g\partial_g)w_{00}  \Big]q_0(z)=0\,.
\end{align}
At strong coupling, the solution to this equation  was constructed using semiclassical methods  in \cite{Belitsky:2020qrm,Belitsky:2020qir}
\begin{align}\label{Q0}\notag
& q_0(z) = {f_0(z,g)\over \sqrt{1- \chi(z)}}\,,
\\
& f_0(z,g) = {1\over \sqrt{2\pi g z}}\Big[ a_0(z,g) \sin(2gz)+b_0(z,g)\cos(2g z)\Big],
\end{align}
where the functions $a_0(z,g)$ and $b_0(z,g)$ are given by series in $1/g$ 
\begin{align}\label{a-fun}
a_0(z,g) = 1+\sum_{k\ge 1} {a_{0,k}(z)\over g^k} \,,\qqqquad b_0(z,g) = 1+\sum_{k\ge 1} {b_{0,k}(z)\over g^k} \,.
\end{align}
The expansion coefficients can be found by substituting the ansatz \re{Q0} into \re{pde} and equating to zero the coefficients in front of the powers of $1/g$ and trigonometric functions. 
Replacing $w_{00}$ in \re{12.25} with its general expression at strong coupling  $w_{00}=A_0 g +A_1 + A_2/g+ O(1/g^2)$, we get 
\begin{align}
a_0(z,g) =b_0(-z,g)= 1-\frac{4 (\ell -2) \ell +3}{16 g z}-\frac{(2 \ell -5) (2 \ell -3) (2 \ell -1)
   (2 \ell +1)-128 A_2 z}{512 g^2 z^2}+O(1/g^3)\ . 
\end{align}
The expressions for the higher  order corrections in $1/g$ can be found in \cite{Belitsky:2020qrm,Belitsky:2020qir}. 

Combining together \re{Q0} and \re{q-rec} we can obtain the functions $q_n(z)$ for any $n$. They take the same form as \re{Q0} 
with the only difference that the coefficient functions $a_0(z,g)$ and $b_0(z,g)$ are replaced with $a_n(z,g)$ and $b_n(z,g)$, respectively.
They are fixed by the recurrence relations \re{Q-rec} in terms of the functions $a_0(z,g)$ and $b_0(z,g)$ and the matrix elements $w_{0m}(g)$ with $m\le n-1$. 

The functions $q_0(z)$ and $q_n(z)$ are given by the sum of two terms proportional to rapidly oscillating trigonometric functions $\sin(2gz)$ and $\cos(2gz)$. Substituting $q_0(z)$ and $q_n(z)$ into the first relation in \re{q-rec}, we replace the  rapidly oscillating trigonometric functions by their average values to get
\begin{align}\label{dw-as}
\partial_g w_{0n}= {2\over\pi}\int_0^\infty dz\, z \partial_z \log(1-\chi(z)) \big[a_0(z,g) a_n(z,g)+b_0(z,g) b_n(z,g)\big]\,.
\end{align}
The expression on the right-hand side depends on $w_{0m}(g)$ with $m\le n-1$. 
Solving \re{dw-as} recursively for $n=2,4,\dots$ we can determine the matrix elements $w_{0n}$ up to a few integration constants.
We recall that, in virtue of \re{fun}, $w_{0m}$ with odd $m$ are not independent. 

The expression inside the square brackets in \re{dw-as} is given by a double series in $1/g$ and $1/z$. 
Upon its  substitution into \re{dw-as}, the integral over $z$  can be expressed in terms of the so-called profile function
\begin{align}\label{In} 
I_{n} (\chi) =  \int_0^\infty {dz\over \pi} z^{1-2n} \partial_z  \log (1-  \chi(z))\,.
\end{align}
For the symbol $\chi(z)$ that is smooth at the origin, 
the integral on the right-hand side is well-defined for $n\le 1/2$. For $n>1/2$ the integral is understood through an analytical continuation. 
Replacing $\chi(z)$ in \re{In} with its expression \re{chi} we get
\begin{align}\label{In1} 
I_{n} (\chi)= 2 (-1)^{n-1} \left(1-2^{2-2 n}\right) \pi ^{1-2 n} \zeta (2 n-1)\,,
\end{align}
where $\zeta(z)$ is the  Riemann zeta-function. 

Going through the calculation we find from \re{dw-as}
\begin{align}\notag\label{w-I}
w_{00} ={}& 4 g I_0 + C_{00}-\frac{(2 \ell -3) (2 \ell -1)I_1}{8 g}-\frac{(2 \ell -3)
   (2 \ell -1) I_1^2}{16 g^2}+O(1/g^3)\,,
\\[2mm]\notag
w_{02} ={}& \frac{2}{3} g^3 \left(I_0^3-2 I_{-1}\right)+\frac{1}{2} g^2  (2 \ell -1)I_0^2 +g \left(I_0
   \ell ^2-\frac{1}{16} I_0^2 I_1 (2 \ell -3) (2 \ell -1) \right)
\\
&   +\left(C_{02}-\frac{1}{32} I_0 I_1 (2 \ell -3) (2 \ell -1)
   \left(I_0 I_1+2 \ell +1\right)\right)+O(1/g)\,,
\end{align}
where the integration constants $C_{00}$ and $C_{02}$ are independent of the coupling $g$. 

The value of $C_{00}$ can be determined by comparing the first relation in \re{w-I} with \re{w00-str}.  Replacing $I_n=I_n(\chi)$ with its expression \re{In}, we observe that the two relations coincide provided that 
\begin{align}\label{C00}
C_{00}=2\ell-1\,.
\end{align}
Writing down the second relation in \re{w-I}, we already took into account  \rf{C00}. 
In the last line in the expression for $w_{02}$ in \re{w-I}
 we used an ambiguity in defining the integration constant $C_{02}$ to insert an 
  additional term depending on $I_n$'s. The reason for this is that, as we show in Appendix~\ref{app:constant}, the constant $C_{02}$ defined in such a way does not depend on the choice of the symbol $\chi(z)$. 

We exploit this property  in Appendix~\ref{app:constant} to determine $C_{02}$. Namely, we show there that for a special choice of the symbol  $ \chi_0(z) = -{4/ z^2}$, the matrix elements $w_{0n}$ can be found exactly for any coupling, see \re{w-ex}. Expanding them at strong coupling and matching on to \re{w-I}, we can identify the integration constant as 
\begin{align}\label{C02}
C_{02}= \frac{1}{32} (2 \ell -1) \left(4 \ell ^2+4 \ell +3\right).
\end{align}
The same procedure  works for the integration constants $C_{04}, C_{06},\dots$.
 
\subsection*{Leading asymptotics at strong coupling}  

According to \re{w0-dsl} and \re{w1-dsl}, the matrix elements $w_{nm}$ scale at strong coupling as a power of the coupling constant
\begin{align}\label{w-LO}
w_{nm}=\omega_{nm}g^{n+m+1}+O(g^{n+m})\ . 
\end{align}
We can use this property to determine the leading coefficients $\omega_{nm}$. 

To find the leading asymptotics of $w_{0n}$, we apply \re{dw-as} and neglect $O(1/g)$ corrections to the functions $a_n(z,g)$ and 
$b_n(z,g)$. In application to \re{Q0}, this amounts to replacing $a_0(z,g)$ and $b_0(z,g)$ with $1$
\begin{align}\label{Q0-as} 
& q_0(z) = {\sin(2gz)+\cos(2g z)\over \sqrt{2\pi g z(1- \chi(z)) }}+\dots\ , 
\end{align}
where dots denote subleading corrections. Being combined with \re{q-rec}, this relation allows us to determine the remaining functions $q_n(z)$. For instance,
\begin{align}\notag\label{q2}
& q_1(z) = \frac{1}{2} g \partial_g q_0(z)-\frac{1}{4} q_0(z) w_{00},
\qquad  
\\
& q_2(z) = \frac{1}{4} ( g\partial_g)^2 q_0(z) -\frac{1}{8} g  \partial_g q_0(z)w_{00} -\frac{1}{8} q_0(z) \lr{g \partial_g w
   _{00}+ 2w_{01}} \,.
   \end{align}
Notice that $w_{00}=O(g)$ at strong coupling and, therefore, the first term in the expression for $q_1(z)$ is subleading. 
In a similar manner, one finds, using $w_{01}=O(g^2)$, that the second term in the expression for $q_2(z)$ 
is subleading and $w_{01} \gg g\partial_g w_{00}$. In addition, the first term in \re{q2} can be simplified with the  help of (\ref{12.25}) as $-(gz)^2q_0(z)+\dots$. In this way we get
\begin{align}\notag\label{q2-as}
& q_1(z) = -\frac{1}{4} q_0(z) w_{00} +\dots\,,
\qquad  
\\
& q_2(z) = -(gz)^2q_0(z)-\frac{1}{4} q_0(z) w_{01} +\dots \,.
\end{align}
As  above, we substitute \re{Q0-as} and \re{q2-as} into \re{dw-as}, replace trigonometric functions  by their average values, \eg,
\begin{align}
q_0^2(z) \to  {1\over  2\pi g z(1- \chi(z))}+\dots\,,
\end{align}
and express the resulting integrals in terms of the  functions $I_n$ defined in \re{In}. 
Taking into account \re{w-LO} we obtain a system of equations for the leading coefficients $\omega_{0n}$
\begin{align}\notag
& \omega_{00}= 4I_0\,,
\\[1.5mm]\notag
& \omega_{01}=-\frac12  I_0 \omega _{00} = -2 I_0^2 \,,
\\
& \omega_{02}=-\frac13 (I_0 \omega _{01}+4 I_{-1}) = \frac13 (2 I_0^3-4 I_{-1}) \,,\quad\dots
\end{align}
Replacing  $I_n$ with their expressions (\ref{In1}) we get 
\begin{align} 
\omega _{00}=-2\pi, &&  
\omega _{01}= -\frac{\pi ^2}{2}, && \omega _{02}= \frac{\pi ^3}{4}, && \omega
   _{03}= \frac{5 \pi ^4}{32}, 
   \qquad \dots
\end{align}
These relations are in agreement with \re{w00-str} and  \re{w04-as}. 
It is straightforward to compute $\omega_{0n}$ for any finite $n$. It proves convenient to introduce the generating function \re{gen-fun}.
Examining the resulting expressions for $\omega_{0n}$ we found that it has a remarkably simple form \re{G(x)}.

To find the remaining coefficients $w_{nm}$ we apply \re{w-fun1} and \re{w-LO} and take into account that the expression on the left-hand side of \re{w-fun1} is suppressed by the factor of $1/g$ as compared to the right-hand side. This leads to a functional equation
\begin{align}
\omega_{n+1,m}+\omega_{n,m+1}=- \frac14 \omega_{0n}\omega_{0m}\,,
\end{align}
where $n,m\ge 0$.
Going to the generating function \re{gen-fun}, we arrive at the relation \re{G(x,y)}. At large $x$ and $y$ the generating function \re{G(x,y)} 
admits the expansion
\begin{align}\notag
G(x,y) ={}& -2 \pi +\Big(-\frac{\pi ^2}{2 y }-\frac{\pi ^2}{2 x}\Big)   +\Big(\frac{\pi
   ^3}{4 y^2}-\frac{\pi ^3}{2 x y}+\frac{\pi ^3}{4 x^2}\Big)  
   +\Big(\frac{5 \pi ^4}{32 y^3}-\frac{\pi ^4}{32 y^2 x}-\frac{\pi ^4}{32 y
   x^2}+\frac{5 \pi ^4}{32 x^3}\Big) 
\\\notag &  
   +\Big(-\frac{23 \pi ^5}{64
   y^4}+\frac{7 \pi ^5}{16 y^3 x}-\frac{13 \pi ^5}{32 y^2 x^2}+\frac{7 \pi
   ^5}{16 y x^3}-\frac{23 \pi ^5}{64 x^4}\Big)  
   \\  & 
  +\Big(-\frac{53 \pi
    ^6}{256 y^5}+\frac{7 \pi ^6}{256 y^4 x}-\frac{\pi ^6}{128 y^3 x^2}-\frac{\pi
   ^6}{128 y^2 x^3}+\frac{7 \pi ^6}{256 y x^4}-\frac{53 \pi ^6}{256 x^5}\Big)
 + \dots .
\end{align}
By definition, $\omega_{nm}$ can be read off as coefficients in front of $1/(x^n y^m)$, 
\begin{align}
\omega_{13} = \frac{7 \pi ^5}{16}\,,\qqquad \omega_{22}=-\frac{13 \pi ^5}{32}\,,\qqquad
\omega_{23} = -{\pi^6\over 128}\,,\qqquad \omega_{14}=\frac{7 \pi ^6}{256 }\,,\ \ \ \ \  \textit{etc.} 
\end{align}

\section{Integration constants}
\la{app:constant}

The differential equation \re{q-rec} allows us to determine the matrix elements $w_{0n}$ up to integration constants $C_{0n}$. Similar to  $w_{0n}$, these integration constants are not independent and can be expressed in terms of an independent set of constants $C_{00},C_{02},C_{04},\dots$. 

In general, the matrix elements $w_{0n}$ and integration constants $C_{0n}$ depend on the symbol $\chi(z)$ in a nontrivial way. It follows from \re{delta-w} and \re{change}, that under the variation of the symbol, 
\begin{align}
\delta w_{0n} = 8g^2 \int_0^\infty dz\, z\, q_0(z) q_n(z) \delta\chi(z)\,.
\end{align} 
Replacing $q_0(z)$ with its asymptotic expressions \re{Q0} and doing 
the same for $q_n(z)$, we can repeat the calculation of the integral on the right-hand side to obtain the expression for $\delta w_{0n}$ in terms of functions $I_n$ defined in \re{In} and their variation $\delta I_n$. Comparing it with the expression for $w_{0n}$ (see \re{w-I}), we can obtain the  expression  for $\delta C_{0n}$. In this way, one finds from \re{w-I} that  
\begin{align}
\delta C_{00}=\delta C_{02}=0\,.
\end{align}
Thus, the integration constants $C_{00}$ and $C_{02}$ are independent of the symbol. 

We can exploit this property to choose $\chi(z)$ to our convenience, e.g., 
\begin{align}\label{chi-simp}
\chi_0(z) = -{4\over z^2}\,.
\end{align}
It coincides with the symbol \re{chi} at small $z$ but has different behaviour at infinity. A distinguished feature of the symbol \re{chi-simp} is that, as we show below, the corresponding matrix elements $w_{nm}(\chi_0)$ can be found exactly. For instance, 
\begin{align}\notag\label{w-ex}
& w_{00}(\chi_0) =-8 g \frac{I_\ell(4 g)}{I_{\ell-1}(4 g)}\,,
\\\notag
& w_{02}(\chi_0) = - 2 g \left(8 g^2+\ell ^2\right) \frac{ I_{\ell }(4 g)}{I_{\ell -1}(4 g)}\,,
\\
& w_{04}(\chi_0) =-g\left(48g^4 +8(\ell^2-2\ell+2)g^2 +\tfrac12 \ell^4\right) \frac{ I_{\ell }(4 g)}{I_{\ell -1}(4 g)}-48 g^4
\,,
\end{align}
where $I_\ell(4g)$ is the Bessel function (not to be confused with the profile function $I_n(\chi_0)$).
These relations hold for an arbitrary coupling constant. 

At large $g$ the relations \re{w-ex} have to match \re{w-I} after we replace the profile functions $I_n$ in \re{w-I} with their expressions \re{In} evaluated for the symbol \re{chi-simp},
\begin{align}
I_n(\chi_0)=(-1)^{n-1} 2^{-2n+1}\,.
\end{align}
Indeed, expanding \re{w-ex} at large $g$ and neglecting exponentially small $O(e^{-8g})$ corrections, we reproduce the first few terms on the right-hand side of \re{w-I}
and determine the integration constants \re{C00} and \re{C02}. 
 
\subsection*{Exact solution} 

As was mentioned above, for the symbol $\chi(z)$ of the form \re{chi-simp}, the Fredholm determinant of the Bessel operator \re{Bes} 
and 
the matrix elements \re{w-nm1} can be computed exactly.

Replacing $\chi(x)$ with its expression \re{chi-simp}, we find that the matrix \re{K-B} has nonzero elements on the main diagonal and two adjacent subdiagonals
\begin{align}
(K_\ell)_{ij}={}& -{4 g^2(-1)^{i+j} \over \sqrt{\ell+2 i-1}\sqrt{\ell+2 j-1}}\bigg[ {\delta_{ij}+\delta_{j,i+1}\over \ell+2i}+{\delta_{ij}+\delta_{i,j+1} \over \ell+2(i-1)} \bigg]\,,
\end{align} 
where $i,j \ge 1$.  
Let us examine the 
 eigenvalue problem for this matrix
\begin{align}
(K_\ell)_{ij}\Psi_j = \Lambda\,\Psi_i \,.
\end{align}
We can use the properties of Bessel functions to verify that eigenfunctions are given by
\begin{align}\label{Psi}
\Psi_i =  (-1)^i \sqrt{2i+\ell-1} {J_{2i+\ell-1}(\sigma) \over \sigma}\,,\qqqquad \Lambda = -{(4 g)^2\over \sigma^2}\ , 
\end{align}
where $i\ge 1$. Notice that $\Psi_i$ coincides with the function $\psi_i(x)$ in \re{psi}
 evaluated at $x=\sigma^2$. 
A quantization condition for the eigenvalues $\Lambda$ follows from the requirement $\Psi_0(y)=0$. Together with \re{Psi} this implies that 
the eigenvalues of the matrix $K_\ell$ are related to zeros of the Bessel function
\begin{align}
J_{\ell-1}(\sigma_k) = 0 \,,\qqqquad \Lambda_k=-{(4g)^2\over \sigma_k^2}\,,
\end{align}
where $k\ge 1$ enumerates the zeros.
As a consequence, the  Fredholm determinant of the Bessel operator with the symbol \re{chi-simp} is given by
\begin{align}\notag\label{Fred-simp}
\det(1-\bm K_\ell) {}& = \det\Big[\delta_{ij} - (K_\ell)_{ij}\Big]\Big|_{_{i,j\ge 1} }
\\
{}& = \prod_{k=1}^\infty (1-\Lambda_k) 
= \prod_{k=1}^\infty\lr{1+ {(4g)^2\over \sigma_k^2}} =  \Gamma(\ell) (2g)^{1-\ell} \rI_{\ell-1} (4g)\,,
\end{align}
where in the last relation we used a well-known representation of the Bessel function $\rI_{\ell-1}(4g)$ as a product involving its zeros. 
It is interesting to note that up to a  rescaling of  the coupling constant, $g\to \pi g$, the expression on the right-hand side coincides with 
the correlation function of the product of circular Wilson loop and half-BPS operator of dimension $\Delta=\ell-1$ in planar $\mathcal N=4$ SYM theory  \cite{Semenoff:2001xp}.  

Let us now examine the matrix elements \re{w-nm1}. 
Taking into account \re{w00} and \re{Fred-simp}, we can determine $w_{00}$ as
\begin{align}\label{w00-sim}
w_{00}(\chi_0) = - 2 g \partial_g \log \det(1-\bm K_\ell) 
=-8 g \frac{\rI_\ell(4 g)}{\rI_{\ell-1}(4 g)}\,.
\end{align}
Substituting this expression into \re{5.12} we get
\begin{align}
w_{01}(\chi_0) = -8g^2 + 4 g\ell \frac{\rI_\ell(4 g)}{\rI_{\ell-1}(4 g)}\,.
\end{align}
These relations hold for an arbitrary coupling.

Notice that $2w_{01}+\ell w_{00} = -16 g^2$. This relation is a particular case of a general relation stating that
$2w_{1n}+\ell w_{0n}$ is a polynomial in $g^2$ of degree $\lfloor n/2 \rfloor+1$. For instance,
\begin{align}\notag\label{mag}
& 2w_{11}+\ell w_{01} = 8\ell g^2\,, 
\\\notag
& 2w_{12}+\ell w_{02} =-16g^4-4\ell^2g^2 \,, 
\\\notag
& 2w_{13}+\ell w_{03} =8(\ell-2)g^4+2\ell^3g^2 \,, 
\\
& 2w_{14}+\ell w_{04} =-32 g^6-8 g^4 \left(\ell ^2-2 \ell +2\right)-g^2 \ell ^4\,,\qquad \dots
\end{align}
Combining these relations with \re{w-fun1} and \re{5.12} and taking into account \re{w00-sim}, we can determine all matrix elements $w_{nm}$ and reproduce \re{w-ex}. 

The underlying reason for simplicity of \re{mag}  is that the sum of matrix elements $2w_{1n}+\ell w_{0n}$ involves the state $2\phi_1(x)+\ell\phi_0(x)$. Replacing $\phi_0(x)$ and 
$\phi_1(x)$ with their expressions \re{phi}, we find that it is proportional to the function $\psi_0(y)$ defined in  \re{psi},
\begin{align}\label{Id-B}
\ell\phi_0(y) + 2 \phi_1(y) =  {y \psi_0(y)\over \sqrt{\ell-1}}\,.
\end{align}
Recall that the states $\psi_i(x)$ form the orthonormal basis (see \re{psi}). Applying the Bessel operator \re{Bes} to both sides of \re{Id-B} we find that for the  symbol \re{chi-simp}
\begin{align}\label{Id-B1}
\bm K_\ell (\ell\ket{\phi_0} + 2 \ket{\phi_1}) \sim \sum_{i\ge 1} \ket{\psi_i}\vev{\psi_i|\psi_0}= 0\,,
\end{align}
where in the second relation we used \re{K-sum}. We can apply this relation to get
\begin{align}\notag\label{qm}
{}&  \ell w_{n0} + 2 w_{n1}  = \vev{\phi_n|\bm{\chi}_0{1\over 1-\bm K_\ell}| \ell \,\phi_0 + 2\phi_1} 
\\
{}& \qquad \stackrel{?}{=}  \vev{\phi_n|\bm{\chi}_0|\ell  \phi_0 +2 \phi_1} =  - 16 g^2 \int_0^\infty {dx\over \sqrt x}J_{\ell-1}(\sqrt x)  (x\partial_x)^n J_\ell(\sqrt x)  
=-16 g^2 (-\ell/2)^n\,,
\end{align}
where in the second line we expanded the matrix elements in powers of $\bm K_\ell$. For $n=0,1$ this relation is in agreement with \re{mag}. For $n\ge 2$ 
it correctly reproduces $O(g^2)$ terms in \re{mag} but fails to reproduce terms with higher power of $g^2$.

 The reason for this is that the expansion of \re{qm} in powers of  $\bm K_\ell$ is not well-defined because the matrix elements $\vev{\phi_n|\bm{\chi}_0(\bm K_\ell)^p | \ell \,\phi_0 + 2\phi_1}$ give rise to divergent integrals.
Indeed, one can check that the integral of the Bessel functions on the second line of \re{qm} diverges at large $x$ for $n\ge 1$. The expression on the right-hand side of \re{qm} was obtained by
inserting the cut-off factor $x^{-\epsilon}$ inside the integral and sending $\epsilon\to 0$ afterwards. Carefully regularizing the integrals in $\vev{\phi_n|\bm{\chi}_0(\bm K_\ell)^p | \ell \,\phi_0 + 2\phi_1}$ and going through the calculation one arrives at \re{mag}. 
 
\providecommand{\href}[2]{#2}\begingroup\raggedright\endgroup

\end{document}